%% file: resubmit.tex
\documentclass[12pt]{article}
\usepackage{geometry}
\usepackage{amsmath}
\usepackage{graphicx}
\usepackage{comment}
\usepackage{natbib}
\usepackage{lmodern}
\usepackage{url}
\usepackage{appendix}
\usepackage{amsmath,amssymb,amsthm,setspace,paralist}
\usepackage[colorlinks,citecolor=blue,urlcolor=blue]{hyperref}
\usepackage{epstopdf}
\usepackage[normalem]{ulem}
\usepackage{multirow}
\usepackage{float}
\usepackage[inline]{enumitem}
\usepackage{bbm}
\usepackage{subcaption}
\usepackage{rotating}
\usepackage{standalone}

\theoremstyle{plain}
\newtheorem{lemma}{Lemma}[section]
\newtheorem{theorem}{Theorem}[section]
\newtheorem{prop}[theorem]{Proposition}
\newtheorem{corollary}[theorem]{Corollary}
\def\m{\mathcal}
\def\mb{\mathbb}
\newcommand{\+}[1]{\ensuremath{\boldsymbol {#1}}}

\newcommand{\be}{\begin{equs}}
	\newcommand{\ee}{\end{equs}}
\newcommand{\bpm}{\begin{pmatrix}}
	\newcommand{\epm}{\end{pmatrix}}
\DeclareMathOperator{\E}{\mathbf E}

\newcommand{\ind}{\mathbbm 1}
\newcommand{\indm}{\mathbbm I}

\DeclareMathOperator*{\argmin}{argmin}

\usepackage{accents}
\usepackage{bibunits}
\usepackage{tikz}
\usepackage{animate}
\usetikzlibrary{shapes,bayesnet}
\defaultbibliographystyle{apalike}
\defaultbibliography{DNAMP}
\newcommand{\blind}{1}

\begin{document}

\def\spacingset#1{\renewcommand{\baselinestretch}%
{#1}\small\normalsize} \spacingset{1}
\newtheorem{assumption}{Assumption}


\if1\blind
{
  \title{ Factorized Fusion Shrinkage for Dynamic Relational Data}
  \author{Peng Zhao \\ Department of Applied Economics and Statistics, University of Delaware\\
  \vspace{0.1cm}
  \\
  Anirban Bhattacharya,
    Debdeep Pati and Bani K. Mallick \\
    Department of Statistics, Texas A\&M University}
  \maketitle
} \fi

\if0\blind
{
  \bigskip
  \bigskip
  \bigskip
  
    \title{\bf Factorized Fusion Shrinkage for Dynamic Relational Data}

  \medskip
} \fi

\bigskip

	\graphicspath{{figures/}}
	
\maketitle
	\bigskip
	\begin{abstract}
    Modern data science applications often involve complex relational data with dynamic structures. In systems that experience regime changes, such as changes in alliances between nations after a war or air transportation networks in the wake of the COVID-19 pandemic, abrupt alterations in the relational dynamics of such data are commonly observed. To address this scenario, we propose a Factorized Fusion Shrinkage model, which involves dynamically shrinking each decomposed factor of the low-rank approximation of the data. To achieve the dynamic shrinkage, we use global-local shrinkage priors applied to successive differences of the decomposed factors. The adopted prior preserves both the separability of clusters and the long-range properties of latent factor dynamics, facilitating post-processing such as cluster analysis and change-point detection. Under specific conditions, we prove that the associated fractional posterior attains the minimax optimal rate up to logarithmic factors. 
    For efficient computation, we introduce a structured mean-field variational inference algorithm that balances optimal posterior inference with computational scalability.  Our framework is versatile and can accommodate a wide range of models, including latent space models for networks, dynamic matrix factorization, and low-rank tensor models. The efficacy of our methodology is tested through extensive simulations and real-world data analysis.
	\end{abstract}
	\noindent%
	\vfill

	\newpage

  \spacingset{1.3} 
   
   \begin{bibunit}
  
        \section{Introduction}
      Relational data describes the relationship between two or more sets of variables and is typically observed as matrices. One of the objectives of relational data analysis is to explain the variation in the entries of a relational array through unobserved explanatory factors. For example, given matrix-valued data $\+Y \in \mathbb{R}^{n \times p}$, the static low-rank plus noise model of the form $\+Y=\+U\+V'+ \+E$, where $\+A'$ is the transpose for a matrix $\+A$, $\max\{\mbox{rank}(\+U), \mbox{rank}(\+V)\} \ll \min\{n, p\}$ and $\+E$ is an error term, has been extensively studied; see \cite{hoff2007model,chatterjee2015matrix,gavish2014optimal,donoho2020screenot} for a flavor and connections with truncated singular value decompositions.   In recent decades, there has been a rapid growth in the interest in analyzing dynamic, complex data sets. Models for dynamic relational data have found widespread application in
         dynamic social network analysis \citep{sarkar2005dynamic,zhu2016scalable},  subspace tracking \citep{doukopoulos2008fast}, traffic prediction \citep{tan2016short}, recommendation system \citep{zhang2021dynamic} among others. Compared to static relational data, additional challenges are posed when the observation matrices possess a dynamic structure. In particular,  modeling the evolution of $\+Y_t, t =1,...,T$  over time has been of particular interest. 
         {For example, the classical vector autoregressive (VAR) model \citep{stock2016dynamic,lutkepohl2013vector} has been employed to model the evolution of the observations directly  and \cite{hoff2015multilinear} considered a bi-linear form of the autoregressive model; 
        \cite{hoff2011hierarchical} and  \cite{friel2016interlocking} parameterized time-varying latent factors in terms of static factors with time-varying weights or coefficients for bipartite network models; \cite{sarkar2005dynamic} used an AR(1) type of evolution of the latent factors in the context of latent space models for dynamic networks. }
        
      Abrupt changes are important examples of non-stationarity, typically observed in systems that undergo regime changes due to an intervention, such as alliances between nations before/after a war \citep{gibler2008international}, voting records before and after an election \citep{lee2004voters}, or the impact of protein networks after a treatment \citep{hegde2008dynamic}.  This article presents a novel shrinkage process for dynamic relational data to handle such abrupt changes: 
      given the dynamic likelihood $\+Y_t=\+U_t\+V_t'+\+E_t$, $t =1,...,T$, we introduce time-dependence by shrinking
      the successive differences  (e.g., $\+u_{it}-\+u_{i(t-1)}$) between the row vectors $\{\+u_{it}\},\{\+v_{it}\}$ of $\+U_t$, $\+V_t$ to group-wise sparse vectors.  In particular, we propose a factorized fusion shrinkage (FFS) prior, where group-wise global-local shrinkage priors on all the transitions of latent factors are applied to promote group-wise fusion structures. The adopted global-local prior has a sufficient mass around zero, effectively shrinking transitions to zeros while keeping large enough transitions, promoting fusion structures. In addition, all components of the transitions share the same local scales from the prior, leading to interpretable group-wise shrinkage. The proposed model attempts to account for the dynamic change and dependence across $\+Y_t$ by reflecting the piecewise constant changes among  $\+U_t$ and $\+V_t$.
      We also propose a symmetric version of the model which is useful for modeling adjacency matrices in network models.

The proposed FFS priors differ substantially from those in literature analyzing dynamic relational data, like normal transition priors with a common variance and Gaussian process priors \citep{sarkar2005dynamic,sun2014collaborative,sewell2015latent,durante2014nonparametric,    sewell2017latent,zhang2018local}.  While the above commonly used dynamic priors tend to introduce the smoothness effect among dynamic transitions, FFS priors are designed to introduce the \textit{stopping} and \textit{separation} effects so that the transitions are either closed to zeros or large. In particular, the stopping effect of FFS priors can significantly enhance the interpretation of the dynamics of latent vectors.  Furthermore, the separation effect of FFS priors contributes to the detection of clusters among subjects after estimating the latent factors. The detected cluster results provide information on which subjects have similar effects in generating the relational data.
        When performing clustering algorithms such as $K$-means, the cluster separation  $\delta_t = \min_{i \ne j} \|\+u_{it}-\+u_{jt}\|_2 $ for any $\+u_{it},\+u_{jt}$ not in the same cluster plays a crucial role in determining the difficulty level of the problem (e.g., eigengap for spectral clustering, see \cite{ng2001spectral}). Generally, the larger $\delta_t$ is, the easier it is to detect clusters at time $t$. The separation effect of FFS priors is justified by encouraging more separation among clusters:
        suppose at time $t$ we have $\+u_{it}=\+u_{jt}$ for subjects $i,j$, and at time $t+1$, subject $i$ moves to another cluster while subject $j$ stays in the same cluster. Then the cluster separation at time $t+1$ satisfies $\delta_{t+1} \leq  \|\+u_{j(t+1)}-\+u_{i(t+1)}\|_2 = \|\+u_{j(t+1)}-\+u_{jt}\|_2=\|D\+u_{jt}\|_2 $,  which highlights the importance of shrinking towards larger transitions. In contrast, priors that introduce smoothness to the transitions tend to impede the separation of the clusters. Overall, the model explicitly links the transitions of latent factors with the changing of cluster memberships for each subject: a zero transition implies no change in cluster membership, while a change in cluster membership implies a large transition.

        Dynamic shrinkage priors have been studied in a wide range of literature \citep{fruhwirth2010stochastic,chan2012time,nakajima2013bayesian,kalli2014time,kowal2019dynamic}. 
        In this article, we consider matrix-valued responses with complex dependence structures using a structural dynamic shrinkage model for the transition of the latent factors and illustrate its applicability across a wide range of problems.
         Additionally, we provide an in-depth theoretical analysis of the proposed prior. 
        We first derive an informative-theoretic lower bound for the model under an appropriate parameter space, incorporating both the initial estimation errors of the matrix and the selection errors due to the sparsity of the transitions. 
         Using the proposed prior concentration {and a fractional posterior device}, we demonstrate that the fractional posterior under the proposed prior can achieve minimax rates up to logarithmic factors. To our knowledge, both the lower bound and fractional posterior concentration are the first results reported in the related literature. {Note that additional works are needed to prove similar results for the usual posterior.} In addition, it has been established that a frequentist approach cannot achieve this fusion type of minimax optimal rate when $\ell_1$-regularization is used \citep{fan2018approximate}. Therefore, such near-optimal convergence rates improve upon related models via $\ell_1$ penalized approaches. 
         
         Finally, using dynamic network models as an example, {we offer theoretical support for the proposed post-processing technique. While posterior post-processing has been increasingly popular in the Bayesian literature, theoretical validations are comparatively rare; see \cite{lee2021post} for a recent example in a different context.} {In particular, dynamic comparisons, cluster analysis, and change point detection are considered.} We show that the optimal estimation of the proposed method can help achieve better performance in these post-processing tasks.


From a computational aspect, we present posterior approximations based on variational inference for scalability and computational efficiency while noting that MCMC approaches can also be readily developed.
We consider a structured mean-field (SMF) variational inference framework where the temporal dependence  is taken into account. A corresponding coordinate ascent variational inference (\cite{bishop2006pattern}, CAVI) algorithm is developed to incorporate the temporal dependence into the variational inference where simple closed-form updatings can be achieved. The proposed algorithm is more efficient than gradient descent or other first-order algorithm types with little increase in the complexity per iteration since the computation can utilize the banded (or block tri-diagonal) structure of the second-order moments, which incurs 
a $O(Td^3)$ cost for matrix inversion. The overall complexity is $O(npTd^3)$, which is similar to the complexity described in the related literature, for example, in \cite{matias2017statistical}. However, the existing literature discusses cases where the number of time points $T$ is small (e.g., in \cite{matias2017statistical}, the real data contains around 5 time points), whereas our algorithm is specifically designed for scenarios with a large number of time points (e.g., $T=200$). Finally, we extend our SMF variational inference framework to tensor data by utilizing a CP type of low-rank factorization in the appendix.

      	{\bf Notation.} For a vector $\+x$, we use $\|\+x\|_2$, $\|\+x\|_1$, $\|\+x\|_\infty$ to represent its $\ell_2$, $\ell_1$ and $\ell_\infty$ norms and $\+x'$ as its transpose.  For a matrix $\+A$, let $\|\+A\|_F$ be its Frobenius norm.  We use $\indm$  and $\+1$  to denote the identity matrix and vector with all ones. Suppose $P$ and $Q$ are probability measures on a common probability space with density $p$ and $q $. We use $D_{KL} \left\{ p \mid \mid q \right\} = \int p \log (p/q) d\mu $ to denote the KL divergence and $D_{\alpha} \left\{ p\mid \mid q \right\} = \log \int p^{\alpha} q^{1-\alpha} d\mu $ to denote the  R\'{e}nyi divergence of order $\alpha$.  Given sequences $a_n$ and $b_n$, we denote $a_n = O(b_n)$ or
	$a_n \lesssim b_n$ if there exists a constant $C>0$ such that $a_n \leq C b_n$ for all large enough $n$. Similarly, we
	define $a_n \gtrsim b_n$. In addition, let $a_n=o(b_n)$ to be $\lim_{n \rightarrow \infty} a_n/b_n = 0$. Let $P_X$ denote a probability distribution with parameter $X$, and $p_X$ denote its density. Denote $\E_{X} $ as the expectation taken with respect to a variable $x$. Let $\mathcal{N}(\mu,\sigma)$ be the normal distribution with mean $\mu$ and variance $\sigma$ while $N(x;\mu,\sigma)$ be the  density at $x$. 
\section{Methodology}\label{sec:method}
\subsection{Factorized Fusion Shrinkage}
        First, we lay down the factorized fusion shrinkage (FFS) approach for dynamic matrix-valued data. Specifically, let $\m Y = \{\+Y_t\}_{t=1}^T$ be the observed data, where $\+Y_t \in \mathbb{R}^{n \times p}$ is an matrix-valued observation corresponding to the $t$th time point. For example, in the context of daily air transportation networks providing flights between cities, where the entries of the matrices represent whether the airlines provide flights between two cities on a given date,   $\+Y_t$ is a matrix with rows and columns representing the city of origin, city of destination and the time-index $t$ runs over days.    For such data, we consider the following dynamic model:
        \begin{equation}\label{eq:matrix_data_generate}  
                \+Y_t \sim   p(\+ M_t;\beta), \quad \m M_t =  \+U_t\+V'_t, \quad t \in [T]:\,= \{1, \ldots, T\}, 
        \end{equation}
        where  $\+u_{it}, \+v_{jt}  \in \mathbb{R}^{d}$ for each $t \in [T]$, $i \in [n]$ and $j \in [p]$ are the row vectors of $\+U_t$ and $\+V_t$ respectively; $\mb E_p (\+Y_t \mid \+ M_t) = g(\+ M_t)$ for some link function $g$ which operates elementwise on a matrix; and $\beta$ represents additional parameters (e.g., variance, subject-specific effects, etc.).

        Even though the Singular value decomposition (SVD) vastly reduces the effective number of parameters for static matrix factorization models, additional structural assumptions are necessitated in the dynamic setting to reduce model complexity. We achieve this by proposing a parsimonious yet flexible evolution of {\em latent factor} vectors. To that end, we first introduce some notation. For each define an $n  \times d$ matrix $\+U_t = [\+u_{1t},...,\+u_{n t}]'$ with $\+u_{it} \in \mathbb{R}^{d}$ for each $i \in [n]$ and similarly $\+V_t = [\+v_{1t},...,\+v_{p t}]'$ with $\+v_{jt} \in \mathbb{R}^{d}$ for each $j \in [p]$. One may interpret $\+u_{it}$ as  $d$-dimensional vector of latent factors for the $i$th data unit of the column of the tmatrix at time $t$. For example, in the air transportation network setting, $\+u_{it}$ represents latent factors corresponding to the $i$th city of origin at time $t$. We consider the following \textit{group-wise} fusion structure on the evolution of the latent factors: 
        \begin{equation}\label{eq:fusion} 
            \begin{aligned}
                        \sum_{t=2}^{T}\sum_{i=1}^{n} \ind \{D\+u_{it} \ne \+0 \} \leq s_{u}, \quad \mbox{and} \quad \sum_{t=2}^{T}\sum_{j=1}^{p} \ind \{D\+v_{jt} \ne \+0 \} \leq s_{v},
                \end{aligned} 
        \end{equation}  
        where $D\+u_{it} = \+u_{it}-\+u_{i(t-1)}$ and $ \+a\ne \+0_d$ means that $\+a \in \mb R^d$ is different from a zero vector. In particular, when $D\+u_{it}= \+0_d$, the entire effect of subject $i$  remains unchanged from time point $t$ to $t+1$. Therefore, when $s_u\ll n$ and $s_v \ll p$, the proposed dynamic fusion structure in equation~\eqref{eq:fusion} sparsely constrains the transitions over time in a group-wise manner and significantly constrains the active number of parameters across all time points.

        We operate in a Bayesian framework and adopt a continuous shrinkage framework to shrink towards the fusion structure in \eqref{eq:fusion}. Specifically, we propose a factorized fusion shrinkage (FFS) prior, which employs \textit{group-wise} global-local shrinkage priors to model the transition of the latent factors:   
        \begin{align}\label{eq:prior_adaptive}
                \begin{aligned}
                        &    \+u_{i(t+1)}\mid \+u_{it} \sim \m N( \+u_{it},\lambda^{(u)2}_{it}\tau_i^{(u)2}\mb I_d),\quad \+v_{j(t+1)}\mid \+v_{jt} \sim \m N( \+v_{jt},\lambda^{(v)2}_{jt}\tau_i^{(v)2}\mb I_d). \\
                        &   \lambda^{(u)}_{it},\lambda^{(v)}_{jt} \overset{ind.}\sim \mbox{Ca}^+(0,1), \quad \tau^{(u)}_i,\tau^{(v)}_j \overset{ind.}\sim g, \quad i\in[n], j\in[p], t \in [T-1].
                \end{aligned} 
        \end{align}
The proposed priors for both $\+u_{it}$ and $\+v_{jt}$ are similarly structured, so let's take $\+u_{it}$ as an example to discuss.  Observe that under the FFS prior, $D\+u_{i(t+1)} \mid \+u_{it} \sim \m N( \+0_d,\lambda^{(u)2}_{it}\tau_i^{(u)2}\mb I_d)$, i.e., the latent factor transitions are conditionally mean zero Gaussian vectors with conditional variance given by $\lambda^{(u)2}_{it}\tau_i^{(u)2}$. The FFS prior adopts a {\em group-wise} global-local parameterization for the transition variances, which simultaneously shrinks all the $d$ components of $D\+u_{i(t+1)}$ to zero, as they all share the same local scale $\lambda^{(u)}_{it}$ from the prior.  We place independent half-Cauchy priors \citep{carvalho2009handling} on the local transition scales, $ \lambda^{(u)}_{it} \overset{ind.}\sim \mbox{Ca}^+(0,1)$, and the global prior $g$ is chosen to ensure that it places sufficient mass around zero. This structure ensures strong shrinkage of the entire vector $D\+u_{i(t+1)}$ towards the origin, while at the same time, the Cauchy tails allow $D\+u_{i(t+1)}$ to have large magnitude when warranted, allowing the prior to capture sharp changes.   For example, the global military alliance networks among nations in the last two centuries, analyzed in Section~\ref{sec:formal_alliance}, underwent dramatic changes in the late 1940s due to the end of World War II, which the proposed prior adequately captures.    Finally, the prior specification is completed by letting $\+u_{i1} \sim \m N(\+0,\sigma^{(u)2}_{0i}\mb I_d)$, with $ \sigma^{(u)2}_{0i} \sim \mbox{IG}(a_{\sigma_0},b_{\sigma_0})$ for the first time point $t=1$, independently for each $i$. The same structure is also introduced on the set of vectors of $\{\+v_{jt}\}_{j\in [p],t\in [T]}$. We assume the latent dimension $d$ is fixed across time. {Given the strong shrinkage employed, over-specifying the number of factors should not lead to a substantial loss in estimation. We also note that at the cost of additional computational burden, more elaborate modeling of the local scales is possible that allows automatic factor selection (e.g.,  \cite{bhattacharya2011sparse,legramanti2020bayesian,schiavon2022generalized,fruhwirth2023generalized}).}

    Two special cases of FFS are worth discussing independently. When assuming an additive Gaussian model for $p$, equation~\eqref{eq:tensor_data_generate} reduces to a matrix factorization model 
        \begin{equation}\label{eq:dynamic}
     Y_{ij,t} \overset{ind.} \sim  \m N(\+u_{it}'\+v_{jt}, \sigma^2), \quad i \in [n], j \in [p], t \in [T], 
        \end{equation}
   where $Y_{ij,t}$ is the $(i,j)$-th component of observed data matrix $\+Y_t \in \mathbb{R}^{n \times p}$ for $t \in [T]$. 
   Our proposed FFS prior~\eqref{eq:prior_adaptive} shrinks the row vectors of $\+U_t$, $\+V_t$ towards the following two-sided group-wise fusion structure~\eqref{eq:fusion},
        where  $s_u, s_v$ are much less than $n(T-1)$ and $p(T-1)$ respectively. 
         A prominent example related to~\eqref{eq:dynamic} is latent factor models, where $\+Y_t$ is a data matrix with rows corresponding to individuals and columns corresponding to variables. The $i$-th row of $\+U_t$ is a vector of latent factors for the $i$-th observation, and $\+V_t$ corresponds to the factor loadings matrix. There is substantial literature on dynamic factor models \citep{stock2011dynamic,assmann2016bayesian,stock2016dynamic}. The FFS prior assumes both the latent factors and the factor loadings have transitions shrunk towards the above two-sided fusion structure.

   Second, the popular latent space model for network data \citep{hoff2002latent} can be realized as a special case of our general modeling framework. Suppose $\{\+Y_t\}_{t=1}^T$ is a collection of time-varying binary networks representing  networks of $n$ individuals observed over $T$ time points, where each $\+Y_t \in \{0, 1\}^{n \times n}$. For $1 \le i, j \le n$, denote $Y_{ij,t} \in \{0, 1\}$ as the absence/presence of an edge between nodes $i$ and $j$ at time $t$. The latent space modeling then posits 
        \begin{equation}\label{eq:model_lsm} 
                Y_{ij,t} \sim \mbox{Bernoulli}(\mbox{logistic}(\+u_{it}'\+u_{jt}+\beta)),
        \end{equation} where $Y_{ij,t}$ is the $(i,j)$-th component of $\+Y_t \in \mathbb{R}^{n \times n}$ and $\+U_t=[\+u_{1t},...\+u_{nt}]' \in \mathbb{R}^{n \times d}$ for $t \in [T]$. Here $\mbox{logistic}(a)=1/(1+\exp(-a))$ is the standard logistic link function and $\+u_{it} \in \mathbb{R}^d$ denotes the $d$-dimension latent Euclidean position of node $i$ at time $t$. Then the prior~\eqref{eq:prior_adaptive} is applied to shrink the parameters towards structures $\sum_{t=2}^{T}\sum_{i=1}^{n} \ind \{D\+u_{it} \ne \+0 \} \leq s$, which indicates that the total count of transitions of nodes'  latent factors is upper bounded by $s$.  
        
     
 \subsection{Posterior Post-processings}\label{sec:post}
 
We consider three post-processing tasks with the model: Procrustes rotations, cluster analysis of estimated latent factors at each time point, and change point detection for dynamic systems.

\noindent \textbf{Procrustes rotations:}
%
When considering the latent space model or matrix factorization model, the non-identifiability in our model is more pronounced than in other dynamic relational models, as it can disrupt the structure of sparse transitions in the truth. For instance, the row-wise cluster structure of the truth $[\+U^{*'}_{t},\+U^{*'}_{t+1}]'$, which also reflects sparsity in transitions, is compromised if different orthogonal transformations are applied. The identity rows in $[\+U^{*'}_{t},\+U^{*'}_{t+1}]'$ may no longer be the same in $[(\+U^*_{t}\+O_{t})',(\+U^*_{t+1}\+O_{t+1})']'$ for different orthonormal matrices $\+O_{t}$ and $\+O_{t+1}$, even though the likelihood remains invariant to such orthogonal transformations. To address this issue, Procrustes rotation is the most commonly used approach \citep{sarkar2005dynamic, assmann2016bayesian, papastamoulis2022identifiability}. Specifically, we adopt the following optimization objective: 
\begin{equation}\label{eq:sequential_Procrustes_rotation}
\begin{aligned}
\hat{\+O}_{t,(t-1)} &=  \argmin_{\+O \in \mathbb{O}^{d \times d}} \|\hat{\+U}_{t-1} \hat{\+O}_{(t-1),(t-2)}-\hat{\+U}_t \+O\|_F,  \quad t=2,...,T,
\end{aligned} 
\end{equation}
with $\+O_{1,0}=\+I$. We then define the final estimator of latent vectors as  $\hat{\+U}^{o}_1=\hat{\+U}_1,\,\hat{\+U}^{o}_2=\hat{\+U}_2\hat{\+O}_{2,1}, \ldots , \hat{\+U}^{o}_T =$ $\hat{\+U}_T\hat{\+O}_{T,(T-1)}$. The optimization objective~\eqref{eq:sequential_Procrustes_rotation}  aims to rotate an estimated latent factor to maximize its similarity to the estimation at its previous time point by minimizing the sum of squared differences. This can be solved in a closed form using singular value decomposition (SVD) \citep{gower2004procrustes}. To interpret the estimator obtained via~\eqref{eq:sequential_Procrustes_rotation}, we evaluate the transformed estimators $\hat{\+U}^o_{t+1},\ldots, \hat{\+U}^o_{t+k}$ using the following criterion for a given value of $k$: 
\begin{equation}\label{eq:loss_dynamical}
\inf_{\+O \in \mathbb{O}^{d \times d}} \sum_{k_0=1}^{k}\|\hat{\+U}^o_{t+k} -\+U^*_{t+k}\+O\|_F^2, 
\end{equation}
where a \textit{common} orthogonal transformation is considered for the truth $\+U^{*}_{t+1} \ldots \+U^{*}_{t+k}$. When consistency holds under the loss function~\eqref{eq:loss_dynamical}, the rotated estimators $\hat{\+U}^o_{t+1}, \ldots, \hat{\+U}^o_{t+k}$ can exhibit the same sparse transition property as the truth because the common orthogonal transformation preserves the identity rows in $[\+U^{*'}_{t+1},\ldots, \+U^{*'}_{t+k}]'$. For exploratory purposes, the user should provide $k$, which can be interpreted as the length of time the sparse transition property of the latent factors is expected to persist. The consistency under the loss function~\eqref{eq:loss_dynamical} can be considered a long-range property of the proposed model, influenced by the choice of $k$ and the sparsity $s$ of the truth. In the theoretical section, we discuss how these factors impact this long-range property. 

\begin{figure}[ht!]
\centering
\includegraphics[width=0.8\textwidth]{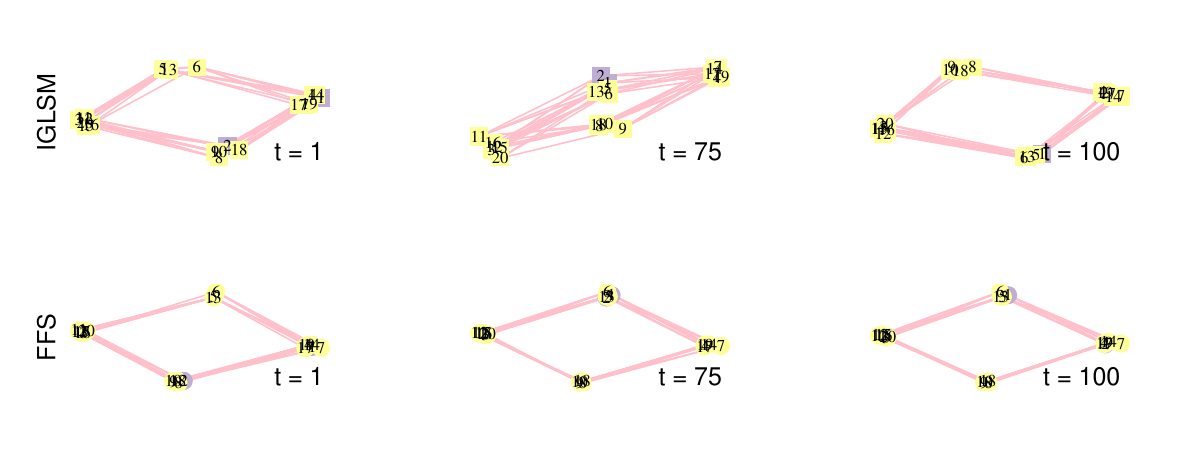}
\caption{\it Snapshots of the estimated latent space for time points $1,75,100$, where only nodes $1,2$ transit and the rest stay static over time. Top row: IGLSM; bottom row: FFS. The estimated latent spaces for FFS are consistent across all time points, illustrating the long-range property of the proposed model.}
\label{fig:Case12}
\end{figure}

To demonstrate the long-range property discussed above, we use the simulation setting as in subsection~\ref{sec:simu_fig12} in the supplement to compare the recovered latent positions of our model to IGLSM (where the same Procrustes rotation~\eqref{eq:sequential_Procrustes_rotation} is also applied), only changing the sample size to $n=20$. It is important to note that in the true data-generating process, only nodes $1,2$ transit, while the rest remain static over time. As shown in Figure~\ref{fig:Case12}, for IGLSM, several nodes, including $5,6,13$, experience a significant shift in their estimated locations from time point $1$ to $100$. Since these nodes also move over time, it becomes difficult to determine whether the movement of the node positions is due to random error or an intrinsic property of the truth. However, with the proposed FFS prior, the property of the truth, where all nodes except $1,2$ remain static, can be recovered, which aligns with the long-range property discussed earlier for $k=T$.

\noindent \textbf{Cluster analysis:} Moreover, inferring discrete structures from a continuous model through post-processing is a common objective in the literature (e.g., variable selection \citep{hahn2015decoupling, bashir2019post} and rank estimation \citep{chakraborty2020bayesian}). Here, we consider a cluster analysis, which assigns cluster labels to the subjects $i=1,\ldots,n$, enabling the automatic grouping of subjects at a given time point $t$. In this article, we obtain cluster assignments using $K_t$-means after estimating the latent vectors, similar to spectral clustering, where $K$ means are obtained after acquiring latent vectors through spectral decomposition \citep{ng2001spectral}. Assume that the true latent vectors have only $K_t$ distinct rows. Then, given any obtained $\hat{\+U}$, we can perform a $K_t$-means analysis on the row vectors of $\hat{\+U}_t$: 
	\begin{equation*}
	(\hat{\+\Xi}_t, \hat{\+X}_t) =\argmin_{\+\Xi \in \mathbb{M}_{n,K_t},\+X \in \mathbb{R}^{K_t \times d}}\|\+\Xi \+X - \hat{\+U}_t\|_F^2, 
	\end{equation*}
where $\mathbb{M}_{n,K_t}$ is the collection of membership matrices, each of which has exactly one $1$ and $K_t-1$ zeros in every row. The membership matrix $\hat{\+\Xi}_t$ reveals the cluster assignments of subjects $i \in [n]$ at time $t$, and the row vectors of $\hat{\+X}_t$ represent the estimated unique rows of latent factors at time $t$. Variants of $K_t$-means can also be applied, such as performing $K_t$-means after normalizing all row vectors \citep{von2007tutorial}. The separation between the distinct rows of the truth significantly affects the performance of $K_t$-means.

\noindent \textbf{Change point detection:}
 Furthermore, the proposed approach holds promise for change point detection of dynamic networks. Change point detection, a widely investigated
 problem in statistical analysis for networks (e.g., \cite{zou2017modeling,hewapathirana2020change,wang2021optimal}), involves identifying abrupt shifts in the underlying properties or behaviors of a dynamic network system. Let $\{\+Y_t\}_{t=1}^T \subset\{0,1\}^{n \times n}$ be the observed dynamic network generated from a sequence of distributions $\left\{\mathcal{L}_t\right\}_{t=1}^T$. Then the sequence $\left\{\eta_k\right\}_{k=1}^K \subset\{2, \ldots, T\}$ with $1=\eta_0<\eta_1<\ldots<\eta_K \leq T<\eta_{K+1}=T+1$ are defined as change points when
 $\mathcal{L}_{t-1} \neq \mathcal{L}_t$ if and only if $t \in\left\{\eta_1, \ldots, \eta_K\right\}$.
 The goal is to estimate the change points $\left\{\eta_k\right\}_{k=1}^K$.
 {\cite{padilla2022change} proposed a change point detection process for dependent dynamic networks employing the random dot product model (RDP) and a similar inner product discrepancy measure for latent positions. }
  The proposed change point detection process for dynamic networks in \cite{padilla2022change} consists of the following steps: 1. Obtain denoised estimators 
  of all connecting probabilities for all networks at all time points. 2.  Vectorize the denoised estimations of all connecting probabilities and utilize wild binary segmentation  \citep{padilla2021optimal} for change point detection of multivariate time series.    Our approach when used for change-point detection has a key improvement in \cite{padilla2022change}. We take into account the dependence of latent positions across time, which allows us to achieve near-optimal statistical rates when estimating the denoised connecting probabilities.
  {While their approach constructs test statistics by estimating each network separately, we aim to estimate a bulk of adjacency matrices at once to achieve near-optimal statistical accuracy, 
  which addresses the issue discussed in \cite{padilla2022change} (e.g., see their discussion paragraph on page 15).}
  
        \section{Theoretical Analysis}\label{sec:3}

        \subsection{Posterior contraction of $\alpha$-fractional Posterior}
   
        Denote $\varTheta=[\m U',\m V']',\m U =[\+U_1,...,\+U_T], \m V=[\+V_1,...,\+V_T],  \+\Theta_t =[\+U_t',\+V_t']'$ and $\+\theta_{it}= \+u_{it}$ if $i \in [n]$; $\+\theta_{it}= \+v_{(i-n)t}$ if $i =n+1,...,n+p$. Let $\+\theta_{i\cdot}=[\+\theta_{i1}',...,\+\theta'_{iT}]'$. Without ambiguity, we use the symbol $\varTheta, \+\Theta_t, \+\theta_{it},\+\theta_{i\cdot}$ for the symmetric case in equation~\eqref{eq:model_lsm} interchangeably with $\m U, \+U_t, \+u_{it},\+u_{i\cdot}$.   We
first consider the following parameter space (Double-sided fusion, DSF):
        
    \noindent    \textbf{Double-sided Fusion (DSF):} 
        \begin{equation}\label{truth:GS}
                \begin{aligned}
                        \mbox{DSF}(s_u,s_v) :=\left\{\varTheta^*:  \sum_{t=2}^{T}\sum_{i=1}^{n} \ind \{D\+u^*_{it} \ne \+0 \} \leq s_u, \,  \sum_{t=2}^{T}\sum_{j=1}^{p} \ind \{D\+v^*_{jt} \ne \+0 \} \leq s_v, \right.\\
                        \left. \max_{i=1}^{n}\max_{t=1}^{T} \|\+u_{it}\|_2 \lesssim 1, \quad \max_{j=1}^{p}\max_{t=1}^{T} \|\+v_{jt}\|_2 \lesssim 1        \right\}.
                \end{aligned} 
        \end{equation}

The sparsity constraint mentioned above implies that out of the total $n \times (T-1)$ transitions for the subjects of $\+U$ and $p \times (T-1)$ transitions for the subjects of $\+V$, only $s_u$ and $s_v$ transitions, respectively, are nonzero, while in the remaining cases, the latent vectors stay unchanged. Additionally, under the boundness assumption provided, for the binary likelihood, all the probabilities induced by the inner product $p_{u^*_{it},v^*_{jt}}:= 1/\{1+\exp(-\+u_{it}^{*'}\+v^*_{jt})\}$ are bounded away from $0$ and $1$, which is only valid for dense networks. We provide the extension of theoretical results of sparse networks in Section~\ref{sec:sparse_network}. Sparsity levels $s_u$ and $s_v$ can be expressed as functions of $n, T$ and $p, T$, respectively. Next, we propose regularity conditions to further refine our understanding of the problem properties and behaviors.

        \begin{assumption}[KL divergence regularity]\label{asm:KL}
                For any $\varTheta^a$, $\varTheta^b \in \mbox{DSF}(s_u,s_v) $, we assume the likelihood induced by $\+\Theta_t^a$, $\+\Theta_t^b$ for all $t \in [T]$ satisfies:
                $ 
                        \max \left\{D_{KL}(p_{\+\Theta_t^a},p_{\+\Theta_t^b}), V_2(p_{\+\Theta_t^a},p_{\+\Theta_t^b})\right\} \lesssim \|\+U_t^a\+V_t^{a'}-\+U_t^b\+V_t^{b'}\|_F^2,
               $
                where $D_{KL}(p_{\+\Theta_t^a},p_{\+\Theta_t^b})$ is the Kullback–Leibler (KL) divergence and  $V_2(p_{\+\Theta_t^a},p_{\+\Theta_t^b})=$\\
                $ \int p_{\+\Theta_t^a} \{\log(p_{\+\Theta_t^a}/p_{\+\Theta_t^b})^2\} d p_{\+\Theta_t^a}$ is the second order moment of log-likelihood ratio between $p_{\+\Theta_t^a}$ and $p_{\+\Theta_t^b}$. 
        \end{assumption}
        The assumption holds for many commonly used likelihood functions, like Gaussian and binary with true probabilities bounded away from $0$ and $1$.    Given an estimator $\hat{\varTheta}$ of $\varTheta^*$, we  consider the squared loss 
        $\sum_{t=1}^{T}\|\hat{\+U}_t\hat{\+V}_t'-\+U^*_t\+V^{*'}_t\|_F^2/(npT) $
        to formulate the minimax lower bound. Since $\+U_t,\+V_t$ can only be identified up to rotation and scaling, the loss function is formulated in terms of the matrix products, which are rotation and scaling invariants. Then the following statement holds for the minimax lower bound:
        \begin{theorem}\label{thm:lower_minimax}
                Suppose the data generating process follows equation~\eqref{eq:dynamic} and Assumption~\ref{asm:KL} holds, then suppose that $d$ is fixed. For large enough $n,p,T$, we have:
                \begin{align*}\label{eq:ST_lower}
                        \begin{aligned}
                                \inf_{\hat{\varTheta}}\sup_{ \varTheta\in \mbox{DSF}(s_u,s_v)} \E_{\varTheta} \left[\frac{\sum_{t=1}^{T}\|\hat{\+U}_t\hat{\+V}_t'-\+U^*_t\+V^{*'}_t\|_F^2}{Tnp}  \right] 
                                \gtrsim \frac{ {s_u \log \left(Tn/s_u\right)}+s_v \log \left(Tp/{s_v}\right)+ n+p}{npT}.
                        \end{aligned}
                \end{align*}
        \end{theorem}

Theorem~\ref{thm:lower_minimax} presents a novel result concerning the estimation of low-rank structured DSF matrices with an information-theoretic lower bound. As far as we know, there are no similar results in the literature. We explicitly ignore the latent dimension $d$ in the bound by assuming it remains a fixed constant across time. The terms $s_u \log ({Tn/s_u})$ and $s_v \log(Tp/s_v)$ represent the selection errors for the fusion structure of $\{\+U_t\}_t$ and $\{\+V_t\}_t$ respectively. The term $(n+p)/(npT)$ identifies the initial estimation errors. Even in the extreme case that $s_u=s_v=0$, where all matrices are equal, it is still necessary to estimate $\+U_1$ and $\+V_1$. This matches the minimax lower bound $1+s\log(T/s)$ of the linear fused model \citep{fan2018approximate} with fusion sparsity $s$. On the other hand, in the dense case where $s_u \geq cnT$ and $s_v \geq cpT$ for some constant $c>0$, the lower bound is then $(n+p)/(np)$, which equates to estimating each low-rank mean matrix individually. If $n$ and $p$ are approximately equal, then both $\{\+U_t\}_t$ and $\{\+V_t\}_t$ must be structurally fused to deliver a measurable improvement in the lower bound. However, in the case of only a one-sided fusion structure, the lower bound may not even be improved from the error rate of the static case $(n+p)/(np)$. This highlights the importance of having two-sided fusion structures for better estimation performance.

        
                To facilitate the theoretical analysis of the proposed model, we adopt the fractional posterior \citep{walker2001bayesian} framework, where the usual likelihood $P(\m Y\mid \m U,\beta)$ is raised to a power $\alpha \in (0, 1)$ to form a pseudo-likelihood $P_\alpha(\m Y \mid \m U, \beta):= \left[P(\m Y\mid \m U,\beta)\right]^\alpha$, leading to a fractional posterior $P_\alpha(\m U,  \beta \mid \m Y) \propto P_\alpha(\m Y \mid \m U, \beta) \, p(\m U) \, p(\beta)$. Such adaptation only requires minor changes in computation, while the theoretical analysis requires fewer conditions than the usual posterior \citep{bhattacharya2019bayesian}. Similarly to the usual posterior,  the optimal convergence of a fractional posterior can imply a rate-optimal point estimator derived from the fractional posterior.   We then need the following assumptions to establish results for the fractional posterior convergence:
        \begin{assumption}[Likelihood regularity]\label{asm:likelihood}
                For any $0<\alpha <1$, and $\varTheta^a \, , \varTheta^b \in \mbox{DSF}(s_u,s_v) $, the $\alpha$-divergence induced by $\varTheta^a$, $\varTheta^b$ for all $t \in [T]$ satisfies:
                $\label{eq:alpha_divergence}
                        D_{\alpha}(p_{\varTheta^a},p_{\varTheta^b})\gtrsim \|\+U_t^a\+V_t^{a'}-\+U_t^b\+V_t^{b'}\|_F^2,
               $
           {where we recall  $D_{\alpha} \left\{ p_x\mid \mid p_{x_0} \right\} = \log \int p_x^{\alpha} p_{x_0}^{1-\alpha} d\mu $ to denote the  R\'{e}nyi divergence of order $\alpha$ between the density $p_x$ and $p_{x_0}$.}
        \end{assumption}

        \begin{assumption}[Global prior]\label{asm:global_prior}
                For any $s_i \in [\max\{n,p\}]$, with $\log T =o (np),$ the global prior $g$ satisfies:
             $  
                        \log    \left\{g(\tau_i^\ast<\tau<2\tau_i^\ast)  \right\}\gtrsim -s_i\log(npT)$ with $\tau_i^\ast = s_i^{\frac{1}{2}}(np)^{-\frac{3}{2}}T^{-\frac{5}{2}}(\log (npT))^{\frac{1}{2}}.
         $      
        \end{assumption}         Similar to Assumption~\ref{asm:KL}, Assumption~\ref{asm:likelihood} also holds for Gaussian and binary cases with true probabilities bounded away from $0$ and $1$, see \cite{gil2013renyi} for some detailed calculations.   Assumption~\ref{asm:global_prior} requires that the global prior has a sufficient mass around proper small values. It can be satisfied by the priors $\tau \sim \mbox{Ca}^+(0,1)$ or $\tau^2 \sim \mbox{Gamma}(a_\tau,b_\tau)$ with constants $a_\tau,b_\tau$ as shown Proposition~\ref{proposition_lemma_8} in the appendix. Then we have the following main theorem:
        
        \begin{theorem}\label{thm:posterior}
                Suppose the data generating process follows equation~\eqref{eq:dynamic} and Assumptions~\ref{asm:KL}, \ref{asm:likelihood} and~\ref{asm:global_prior} hold. Then under the prior~\eqref{eq:prior_adaptive}, if $\varTheta^* \in \mbox{DSF}(s_u,s_v)$ with $\log T = o(np)$ and $d$ is fixed, denote $\epsilon_{n,p,T} = M\sqrt{(s_u+s_v+n+p) \log(npT)/npT}$ for some constant $M>0$,
                then for large enough $n,p,T$, any $D\geq 2$ and $\eta>0$, with probability at least $1-2/\{(D-1+\eta)^2npT \epsilon_{n,p,T}^2\}$, we have 
                \begin{equation*}\label{eq:result}
                        \Pi_{\alpha}\left\{\frac{1}{Tnp} \sum_{t=1}^{T}\|\+U_t\+V_t'-\+U^*_t\+V^{*'}_t\|_F^2 \geq \frac{D+3\eta}{1-\alpha} \epsilon_{n,p,T}^2\mid \m Y\right\} \leq e^{-\eta npT \epsilon_{n,p,T}^2}.
                \end{equation*}
        \end{theorem}
        
The above theorem indicates that the fractional posterior of the mean estimation error can achieve the minimax lower bound with only a logarithmic factor when the model is correctly specified. In the worst case, such a fusion type of lower bound cannot be matched by $\ell_1$ penalized regressions \citep{fan2018approximate}. Therefore,  the proposed method is better than using $\ell_1$ regularization to introduce fusion structures. The proof of Theorem~\ref{thm:posterior} is based on an $\ell_1$ type of prior concentration for the shrinkage prior shown in the Lemma~\ref{lem:horse_net} in the supplement. The above theorem also demonstrates an advantage in comparison to the conditions required with the fusion model with only one variable. When there is only one variable of time $T$ with fusion sparsity $s$, then $s=o(T/\log(T))$ is required to make sure that the targeted rate $s\log T/T$ converges to zero. However, while the fusion structure holds for $np$ variables, the error rate for a subject with sparsity $s_i$ is $s_i\log(npT)/(npT)=s_i \log(np)/(npT)+s_i \log(T)/(npT)$. As long as the mild assumption $\log T = o(np)$ holds, the above rate converges to zero even if $s_i=T$, which is considerably more flexible since the successive differences of some subjects are therefore permitted to be non-sparse.


\subsection{Post-processing for dynamic networks}
  For dynamic network models, the estimation error in the inner product in Theorem~\ref{thm:posterior} cannot directly imply the estimation of latent vectors that can be directly compared across time. To handle the issue, we study the recovery of latent vectors in the loss function~\eqref{eq:loss_dynamical}. In particular, we divide the time $1,..,T$ into $T/k$ sets, where each set contains $k$ consecutive time points: ${(t-1)k+1,...,(t-1)k}_{t=1}^{T/k}$. We assume $\bar t = T/k$ is an integer for simplicity. We consider the comparison for latent space within each set of periods so that the maximal time gap is $k$ for each comparison. Then the following corollary characterizes the optimal estimation error of the estimator after sequential Procrustes rotations~\eqref{eq:sequential_Procrustes_rotation}:

	\begin{corollary}\label{thm:dynamical_comparison}
		Suppose the Assumptions in Theorem~\ref{thm:posterior} hold. 	In addition, assume that the smallest singular value of $\+U^*_t$, denoted as $\lambda_{t}$, satisfies $\lambda_{t} = \Omega(\sqrt{n})$ for $t \in [T]$. For the fractional posterior means  estimated $\hat{\+U}_t$, after performing sequential Procrustes rotations~\eqref{eq:sequential_Procrustes_rotation} to obtain estimation $\hat{\+U}^{o}_1,...,\hat{\+U}^{o}_T$,	the average error for all windows satisfies 
		\begin{equation*}
		\begin{aligned}
		\frac{1}{nT}\sum_{t=1}^{\bar t}\inf_{\bar {\+O}_{t,k} \in \mathbb{O}^{d \times d}}\sum_{k_0=1}^{k}\|\hat{\+U}^o_{(t-1)k+k_0} -\+U^*_{(t-1)k+k_0} \bar{\+O}_{t,k}\|_F^2\lesssim \frac{  (s+n) \log(nT)}{n^2T}\\
		+\min \left\{\frac{ k(s+n) \log(nT)}{n^2T},  \frac{\log(nT)}{n} \right\}+\min\left\{\frac{s k (k-1)}{nT},\frac{s}{n},1 \right\}.
		\end{aligned} 
		\end{equation*}
	\end{corollary}

An additional assumption characterizes the lower bound of the order of the smallest eigenvalue of the true latent vectors. The lower bound of the order $\sqrt{n}$ holds for many matrices. For example, for an $n \times d$ random matrix whose entries are i.i.d. distributed with zero mean and unit variance, the famous Bai-Yin's law \citep{bai2008limit} states that its smallest singular value is approximately $\sqrt{n}-\sqrt{d}$ (we refer to \cite{vershynin2010introduction} for more matrices with the same rate). When $k=1$, the error rate becomes the optimal estimation error in Theorem~\ref{thm:posterior}. When $k$ is a constant not increasing with $n$ and $T$, the average error for all windows converges to zero at rate $(s+n) \log(nT)/(n^2T)+s/(nT)$ as $n, T \rightarrow \infty$, leading to consistency as long as $s=o(nT)$. If the transition is extremely sparse $s=o(n)$, then one can simultaneously compare all the latent space for all time points of length $k=O(T)$. For the challenging case, where $s=o(nT)$, only constant time latent positions can be compared across time. In some special cases, the above error rate is not optimal. In supplementary subsection~\ref{sec:improved_rate}, we provide an improvement under a more stringent assumption of the true transitions. To our best knowledge, although Procrustes rotations have been popularly employed, there is no similar theoretical result to Corollary~\ref{thm:dynamical_comparison} that tries to quantitatively understand the possibility and limitations of long-term comparison of the estimated latent factors. 

Once the latent vectors have been obtained, $K$-means can be used to cluster the subjects. 
For any membership matrix $\+\Xi$, the cluster
	membership of a subject $i$ is denoted by $g_i \in \{1, . . . , K\}$, which satisfies $\Xi_{i g_i}=1$. Let $G_k(\Xi)=\{1\leq i \leq n: g_i=k\}$. We consider the following loss function to evaluate clustering accuracy:
$
	L(\hat{\+\Xi},\+\Xi) =  \min_{\+J \in E_K} \|\hat{\+\Xi} \+J-\+\Xi \|_{2,0},
$
	which represents the number of misclustering subjects, where $E_K$ is the set of all $K \times K$ permutation matrix. Our next result concerns performing separate community detection for each network while improving the overall misclustering rate based on the network dependence structure.
	
	\begin{theorem}\label{thm:single_cluster}
		Suppose the assumptions in Theorem~\ref{thm:posterior} hold. In addition, assume that for $t \in [T]$, the truth $\+U^*_t$ satisfies:
		\begin{enumerate*}
			\item $\+U^*_t$ has $K_t$ distinct rows: $\+U^*_t= \+\Xi^*_t \+X_t$,  $\+\Xi^*_t \in \mathbb{M}_{n,K_t}$ for some $K_t>0$ and $\+U^*_t \in \mathbb{R}^{K_t \times d}$ is full rank;
			\item The smallest singular value of $\+U^*_t$, denoted as $\lambda_{t}$, satisfies $\lambda_{t} = \Omega(\sqrt{n})$;
			\item Cluster separation: $\delta \leq \min_{i \ne j} \|\+u^*_{it}-\+u^*_{jt}\|_2 $ for any $\+u^*_{it} \ne \+u^*_{jt}$, where $\delta>0$;
			\item The minimal block size satisfies $\min_{i} |G_i(\+\Xi^*_t)| = \omega( \log (n)/\delta^2)$.
		\end{enumerate*}	Then after performing $K_t$-means for fractional posterior means $\hat{\+U}_t$ to obtain estimation of membership matrix $\hat{\+\Xi}_t$, as $n,T \rightarrow \infty$, we have
		$
		\sum_{t=1}^{T} L(\hat{\+\Xi}_t,\+\Xi^*_t)/T \lesssim (s+n) \log(nT)/(n T  \delta^2) .
		$
	\end{theorem}

  Regarding assumptions in Theorem~\ref{thm:single_cluster}: Assumption 1 is prevalent in the literature on community detection for networks. {Similar to the assumption in Corollary~\ref{thm:dynamical_comparison}, Assumption 2 requires the lower bound of the order of the smallest eigenvalue of the matrix formed by true latent vectors.}
  Assumption 3 represents a minimum separation between different clusters, which can be a constant for many cases. For example, $\delta=1$ for a random matrix whose entries are i.i.d. binary distributed. Assumption 4 requires a minimum block size of the clusters. The usage of FFS is justified by cluster separation. Suppose at time $t$ we have $\+u^*_{it}=\+u^*_{jt}$ for subjects $i,j$. Then for time $t+1$, the subject $i$ moves to another cluster while subject $j$ stays the same. Under the cluster separation condition, we have $\delta_{t+1} \leq  \|\+u^*_{j(t+1)}-\+u^*_{i(t+1)}\|_2 = \|\+u^*_{j(t+1)}-\+u^*_{jt}\|_2 $, which requires the transitions of the subjects to be heterogeneous: either the transition is zero, or the transition is large. On the contrary, when smoothness-induced priors (e.g., Gaussian process priors) are adopted on the transitions, one assumes that the true difference satisfies some smoothness condition, which cannot adapt to the above heterogeneous requirement.  
   
        \subsection{Theory for sparse dynamic networks}\label{sec:sparse_network}

        In the previous section, when considering the latent space model, all the theoretical results have been presented in dense network settings where the connection probabilities are bounded away from $0$ and $1$ due to Assumption~\ref{asm:likelihood}. In this section, we consider sparse binary networks where the connection  probabilities can converge to zero as $n$ and $T$ approach infinity. Theoretical results for latent space models of the single sparse network from a frequentist perspective can be found in \cite{ma2020universal},  \cite{zhang2022directed}, and \cite{zhang2022joint}. However, the results for sparse dynamic networks from Bayesian settings are relatively new. For the sparse pairwise networks, we consider the data-generating process as 

\begin{equation}  
\begin{aligned}
\label{likelihood:beta_unknwon}
    	Y_{ijt} \stackrel{ind.}\sim  \mbox{Bernoulli}\left[1/\{1+\exp(-\beta^*-\+u_{it}^{*'}\+u_{jt}^{*})\}\right], \, \,\,\mbox{s.t.,} \quad\sum_{t=2}^{T}\sum_{i=1}^{n} \ind \{D\+u^*_{it} \ne \+0 \}\leq s_u,
     \end{aligned} 
\end{equation}
 for $i, j \in [n], t\in [T]$, where the parameter $\beta^*$ controls the sparsity level of all networks and is not known and needs to be estimated. 

  \begin{assumption}[Sparsity level]\label{asm:sparsity_level} Suppose all the $\ell_2$ norms of latent positions $\+u^*_{it}$ are unformly bounded by constant. In addition,  denote $\epsilon_{n,T} = M_0 \sqrt{(s_u+n) \log(nT)/(n^2T)}$ for some constant $M_0>0$, we assume that the sparsity level of the connection  probabilities satisfies
$
    e^{\beta^*} \gg \epsilon_{n,T}^2. $

\end{assumption}
 The literature commonly considers Assumption~\eqref{asm:sparsity_level} as the smallest sparsity level for which the estimated inner product plus intercept can be consistently recovered. For static networks, this sparsity level is $\sqrt{\log(n)/n}$. The rate $\epsilon_{n,T}^2$ has improvement over $\log(n)/n$ as long as $s_u \ll nT$, which shows the gain of FFS on the proposed structure. 
For the prior setting, in addition to FFS~\eqref{eq:prior_adaptive}, we also have the following assumptions on the sparsity parameter $\beta$:
 \begin{assumption}[Prior for the sparsity parameter]\label{asm:sparsity_prior} We use prior $ \Pi(\beta)$ such that \\
$
\Pi\left( |\beta  - \beta^*| \leq  c_4 \epsilon_{n,T}\right) \gtrsim  e^{-n^2T \epsilon_{n,T}^2}
$  for constant $c_4>0$.
\end{assumption}

The above assumption means that the prior $\Pi(\beta)$  has a sufficient mass around $\beta^*$ for $\beta^*$ goes to negative infinity at a specific rate. We show later that the assumption can be satisfied by the commonly used normal prior with mean zero and constant scales. We also have a technical assumption for the prior: consider the event $ B_p = \{  \|\+u_{it}\|_2 \leq C_5 , \forall  i,j \in [n], t \in [T]\} $ for a large enough constant $C_5>0$.  
\begin{equation}\label{eq:prior_prob}
   \mbox{We consider the prior restricted on event }  B_p(\m U), \,\, \tilde \Pi := \Pi(\cdot \cap B_p(\m U))/\Pi(B_p(\m U)),  
\end{equation} 
to replace the original prior such that $\tilde \Pi(B_p^c)=0$.
Without ambiguity, we still use $\Pi(\cdot)$ to denote $\tilde \Pi()$. {The assumption assumes the usage of a truncated prior, which guarantees that the latent positions estimated by the fractional posterior are bounded. As a result, the order of the estimated connection probabilities is determined solely by $\beta$ as $\beta$ dominates  $\+u_{it}'\+u_{jt}$ for all $i,j,t$.} 
In this paper, the assumption is only for technical use to prove the theorem while not used in the algorithm, as adopting such prior will only result in a negligible difference between using the original prior.  A similar phenomenon is also reported in Remark 2 in \cite{ma2020universal}.  Then, we have the following theorem about the fractional posterior convergence in different loss measures.
       \begin{theorem}\label{thm:posterior_sparse_new}
                Suppose the data generating process follows equation~\eqref{likelihood:beta_unknwon} and Assumption~ \ref{asm:sparsity_level} hold. Then under the FFS prior~\eqref{eq:prior_adaptive} 
                with Assumptions~\ref{asm:global_prior} and~\ref{asm:sparsity_prior}, 
                if $\log T = o(n)$ and $d$ is fixed,
                then for large enough $n,T$, any $D\geq 2$ and $\eta>0$, with probability at least $1-2/\{(D-1+\eta)^2n^2T \epsilon_{n,T}^2\}$, we have 
                \begin{equation*}
    \Pi_{\alpha}\left(\frac{1}{n^2T} D_{\alpha}\left( p_{\m U,\beta},  p_{\m U^*,\beta^*}\right) \geq  \frac{D+3\eta}{1-\alpha} \epsilon_{n,T}^2 \mid \m Y\right) \rightarrow 0.
\end{equation*}
  In addition, if technical assumption~\eqref{eq:prior_prob} also holds, we also have for large enough constant $M>0$, 
                \begin{equation*}\label{eq:result3}
                        \Pi_{\alpha}\left\{\frac{1}{n^2T} \sum_{t=1}^{T}\sum_{i,j=1}^{n}(\+u_{it}'\+u_{jt}+\beta-\+u^{*'}_{it}\+u^{*}_{jt}-\beta^*)^2 \geq M\frac{D+3\eta}{1-\alpha}  e^{-\beta^*}\epsilon_{n,T}^2\mid \m Y\right\} \leq e^{-\eta n^2T \epsilon_{n,T}^2}. 
                \end{equation*} 
        \end{theorem}

{Compared to the convergence rate of fractional posterior under the $\alpha$ divergence $ D_{\alpha}( p_{\m U,\beta},  p_{\m U^*,\beta^*})$, there is an additional $e^{-\beta^*}$ multiplier for the convergence of the squared $\ell_2$ loss for the mean parameter $\+u'_{it}\+u_{jt}+\beta$. Note that the logistic function has a flat curvature near zero probabilities causing the mean parameter to approach negative infinity.  Therefore, when the connection probabilities are close to zero, even if they can be estimated accurately, the mean parameter will have an additional error factor that is determined by the sparsity level.
The results align with those reported in the static latent space models as in \cite{ma2020universal}, \cite{zhang2022directed} and \cite{zhang2022joint}.} The proposed framework can also handle dynamic networks with different sparsity levels, which is left for future research.

        \section{Computation}\label{sec:2}
We present an efficient variational inference algorithm to approximate the fractional posterior distribution under the FFS prior. The algorithm uses closed-form iterations of a coordinate-ascent method, which relies on the conditionally conjugate nature of the FFS prior as in the related literature \citep{loyal2021eigenmodel,zhao2022structured}. Our variational inference algorithm targets the fractional posterior $p_{\alpha}(\m U, \beta \mid \mathbf{Y})$ and aims to find the best approximation (in terms of KL divergence) from a structured mean-field variational (SMF) family $\Gamma$: 
        \begin{equation}\label{eq:SMF}
                q(\m U,\beta) = \left\{\prod_{m=1}^{M}\prod_{i=1}^{n}q(\+u^{(m)}_{i\cdot}) \right\}q(\beta), 
        \end{equation}
        where   $\+u^{(m)}_{i\cdot}=[\+u^{(m)'}_{i1},...,\+u^{(m)'}_{iT}]'$. The adopted SMF family accommodates the special structures of the $Td \times Td$ covariance matrix for $\+u^{(m)}_{i\cdot}$, which includes the temporal dependence across time $t \in [T]$ and the group-wise dependence among components for the same subject $i$. The covariance of $\+u^{(m)}_{i\cdot}$ has a block tri-diagonal structure, which can be inverted efficiently (in $O(Td^3)$ operations) using the Kalman smoothing framework.


        The goal of the variational inference is  
        \begin{align*}
                \begin{aligned}
                        \hat{q}(\varTheta, \+\beta) &=\argmin_{q(\varTheta, \+\beta) \in \Gamma} D_{KL}  \left\{ q(\varTheta, \+\beta) \mid \mid p_\alpha(\varTheta,\+\beta \mid \m Y) \right\} =\argmin_{q(\varTheta,\+\beta) \in \Gamma} -\E_q \left\{\log \left(\frac{p_\alpha(\m Y,\varTheta,\+\beta)}{q(\varTheta,\beta)}\right)\right\},
                \end{aligned} 
        \end{align*}
        where the term $\E_q\{\log(p_\alpha(\m Y,\varTheta,\beta)/q(\varTheta,\beta))\}$ is the evidence-lower bound (ELBO) and  $p_\alpha(\varTheta,\+\beta \mid \m Y)$ is the marginal fractional posterior $p_\alpha(\varTheta,\+\beta \mid \m Y)= \int p_\alpha(\varTheta,\+\beta, \+\Lambda, \+\sigma_0 \mid \m Y) d  \+\Lambda \, d \+\sigma_0$. Here we denote $\varTheta=[\m U',\m V']', \+\Theta_t =[\+U_t',\+V_t']'$ and $\+\theta_{it}= \+u_{it}$ if $i \in [n]$; $\+\theta_{it}= \+v_{(i-n)t}$ if $i =n+1,...,n+p$ for notation simplicity.  Furthermore, we adopt the variable augmentation that the square of the half-Cauchy distribution can be expressed as a mixture of Inverse-Gamma distributions \citep{neville2014mean}: for $\lambda_{it} \sim \mbox{Ca}^+(0,1)$, we can write $  \lambda_{it}^2 \mid \eta_{it} \sim \mbox{IG}(1/2,1/\nu_{it}),\quad \eta_{it} \sim \mbox{IG}(1/2,1). $  Denote $\+H=\{\eta_{it}\}_{i,t}$. Then the objective of KL minimization is as follows: 
        \begin{equation*} 
                \begin{aligned}
                        \hat{q}(\varTheta, \beta, \+H,\+\Lambda,\+\tau,\+\sigma_0) 
                        &=\argmin_{q( \varTheta, \beta, \+H,\+\Lambda,\+\tau, \+\sigma_0) \in \Gamma} -\E_q \left\{\log \left(\frac{p_\alpha(\m Y, \varTheta, \beta, \+H,\+\Lambda,\+\tau, \+\sigma_0)}{q(\varTheta, \beta, \+H,\+\Lambda,\+\tau,\+\sigma_0)}\right)\right\},
                \end{aligned}  
        \end{equation*}
        where now the SMF family is defined as:
        \begin{equation}\label{eq:SMF_joint}
                q(\varTheta, \beta, \+H,\+\Lambda,\+\tau, \+\sigma_0) = \prod_{i=1}^{n+p} \left[q(\+\theta_{i\cdot})q(\tau_i)\prod_{t=1}^{T-1}\left\{q(\lambda_{it})q(\eta_{it}) \right\}q(\sigma_{0i}) \right]q(\beta).
        \end{equation}
        The variational family defined in equation~\eqref{eq:SMF_joint} allows for updating $q(\eta_{it})$, $q(\lambda^2_{it})$, and $q(\sigma_{0i} ^{2})$ in the inverse-Gamma conjugate family. Furthermore, $q(\+\tau)$ can be updated with a closed-form expression when the recommended half-Cauchy or Gamma prior is used. For likelihoods that are Gaussian or binary, updating $q(\beta)$ and $q(\+\theta_{i\cdot})$ can be efficiently obtained through a message-passing framework. The derivations for these updates are provided in Section~\ref{sec:CAVI} of the supplementary material for completeness.

{
When applying a CAVI algorithm to models with sparsity structures, the final accuracy depends heavily on the order of component-wise updating. Updating in a naive cyclical manner can lead to error accumulation, as noted in \cite{huang2016variational}. To prevent this, \cite{ray2021variational} proposed a prioritized updating approach for sparse regression. Inspired by their approach, we partition the subject-related components into different blocks based on the subject index, ${q(\+\theta_{i\cdot}),q(\sigma_{0i}),q(\+\lambda_{i\cdot}),q(\tau_i),q(\+\eta_{i\cdot})}, {i \in [n+p]}$, to avoid error accumulation. We update each block ${q(\+\theta_{i\cdot}),q(\sigma_{0i}),q(\+\lambda_{i\cdot}),q(\tau_i),q(\+\eta_{i\cdot})}$ until convergence, and then move onto the next block. This block-wise updating strategy reduces cumulative error and converges quickly because the variational family~\eqref{eq:SMF_joint} captures temporal dependence. In Section~\ref{sec:convergence} of the supplement, we compare this algorithm with proximal gradient descent using an $\ell_1$ penalized objective function for each block (node fixed), and we extend the algorithm to tensor data in Section~\ref{sec:tensor_computation} of the supplement.
}

        \section{Data Analysis}\label{sec:4}

         \subsection{Simulations}\label{sec:simu}
      \noindent \textbf{Matrix factorization model:} First, we perform simulations to show that the simultaneous low-rank and fusion structures can not be fully captured by the methods that only consider low-rank or fusion structures. Simulation cases are considered to compare with the following approaches: 1. SVD1: perform SVD to obtain the best rank $d$ approximation for matrices at each time; 2. SVD2:  combine the observed matrices at the neighbor time together to perform a low-rank approximation: let $\tilde{\+Y}_t=[\+Y_{t-1}, \+Y_t, \+Y_{t+1}]$ ($\tilde{\+Y}_1=[\+Y_1,\+Y_2]$ and $\tilde{\+Y}_T=[\+Y_{T-1},\+Y_T]$). We then perform low-rank approximation on $\tilde{\+Y}_t$ where the rank is obtained by optimal hard thresholding \citep{gavish2014optimal}, and the estimator $\hat{\+Y}_t$ is the corresponding sub-matrix of the low-rank estimator of  $\tilde{\+Y}_t$; 3. Flasso1: perform fused lasso to estimate each component over time separately; 4. Flasso2: perform fused lasso to estimate each component over time separately, then apply SVD to obtain the best rank $d$ approximation for each estimated matrix.
        SVD2 and Flasso2 are some modified approaches based on SVD and fused lasso to consider both low-rank and fusion structures in the final estimation.
        Throughout all simulations, we fix the fractional power for the fractional likelihood  as $\alpha=0.95$, and hyperparameters $a_{\sigma_0}=b_{\sigma_0}=1/2$.  First, $25$ replicated data sets are generated from the following case:   
         $n,p=5,10,20$ with $n \geq p$, $\+u^{(m)}_{i1} \sim \m N(\+0,\indm),  \quad i=1,...,n_m, m=1,2$, $D\+u^{(m)}_{it} = (0,0)' \mbox{  with probability $\rho$ }$, $D\+u^{(m)}_{it} =(-1,-1)' \,\,\mbox{  with probability $(1-\rho)/2$}$, $D\+u^{(m)}_{it} =(1,1)' \,\,\mbox{  with probability $(1-\rho)/2$} $, and $Y_{ijt} \sim \m N(\+u^{(1)'}_{it}\+u^{(2)}_{jt}, 0.3^2) \quad i \in [n_1]; j \in [n_2]; t \in [T].$  The above simulation corresponds to the example~\eqref{eq:dynamic} introduced in the methodology section. The supplement provides two additional cases about the binary network and tensor models in subsection \ref{sec:additional_simu}. We have $T=100, \,d=2, \, \rho=0.5$, $0.8$, $0.85$, $0.9$,  $0.95$, $0.99$ across all cases.  The expected number of effective parameters in the above simulations is $(n+p)(1-\rho) Td$ for the Gaussian factorization model. In the Gaussian case, iterations are stopped when the difference between predictive root mean squared errors (RMSEs) in two consecutive cycles is less than $10^{-4}$.   The estimated RMSE ( $\sqrt{  \sum_{i \in [n], j\in [p],t \in [T]} (\hat{\+u}_{it}'\hat{\+v}_{jt}-\+u_{it}^{*'}\+v_{jt}^*)^2 /(npT)}$ ) is also used to measure the discrepancy between the estimated and true latent distances. For the fused lasso, we use cross-validation to tune the hyperparameters. 
         
\begin{figure}
            \centering
            \includegraphics[width=0.8\linewidth]{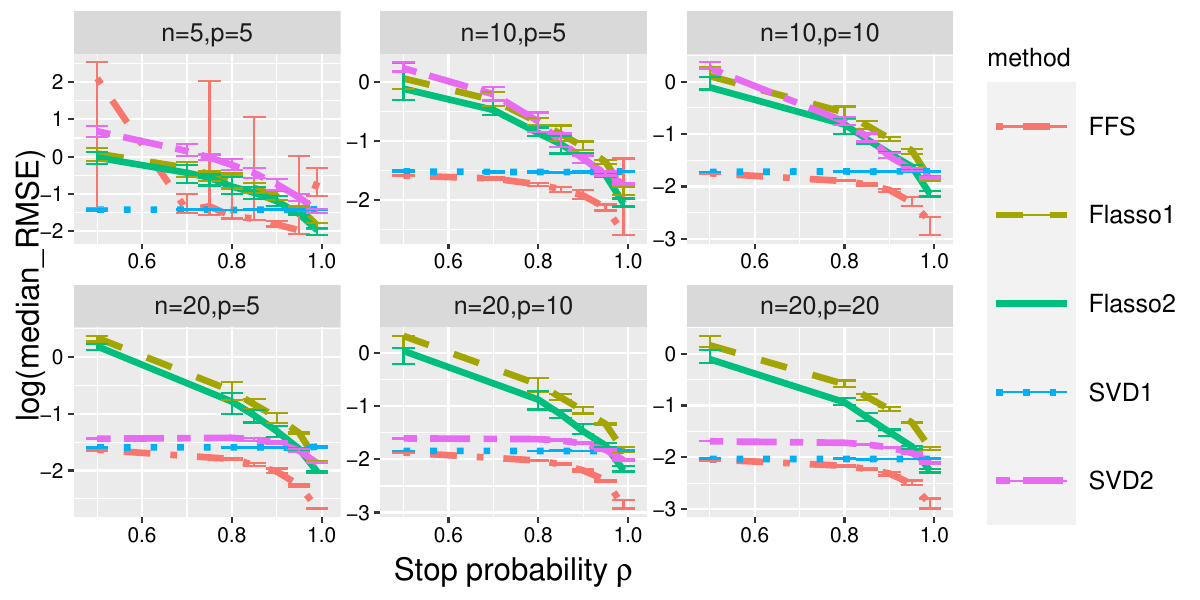}
            \caption{\it Performance comparison for Case 1 between FFS, SVD and Flasso. The measure is the median RMSE for the mean estimation of $\m M$ across $25$ replicates, and the lower and upper error bars are $25\%$ and $75\%$ quantiles. All numbers are converted to logarithm scales. Stop probability $\rho$: the probability of remaining static for each time point and observation. FFS achieves lower RMSE than all other approaches when the number of observations is sufficient (except in the case $n=5,p=5$). }
            \label{tab:2}
        \end{figure}

Figure~\ref{tab:2}  provides numerical support for our theoretical results in multiple aspects. First, as $\rho$ increases, the estimation errors of FFS, Flasso1, Flasso2 and SVD2 decrease, whereas that of SVD barely changes. The reason is that SVD estimates observation matrices separately without considering temporal dependence, whereas other approaches can take advantage of dynamic dependence in the model. It can be seen that even though SVD2 can take into account partial temporal dependence, the exact fusion structure can not be captured, leading to larger estimation errors than the proposed approach. Moreover, by comparing Flasso and Flasso 2, we can see that the simultaneously low-rank and fusion structure should be considered if the true data-generating process has such a structure. When the number of observations is sufficient, FFS consistently achieves lower RMSE than all other approaches. The more accurate estimation of FFS over Flasso and Flasso2 matches the theoretical results that our final estimators are near minimax optimal while the $\ell_1$ type of regularizations is not.      

\begin{figure*}[t!]
    \centering
    \begin{subfigure}[t]{0.48\textwidth}
       \includegraphics[width=\textwidth]{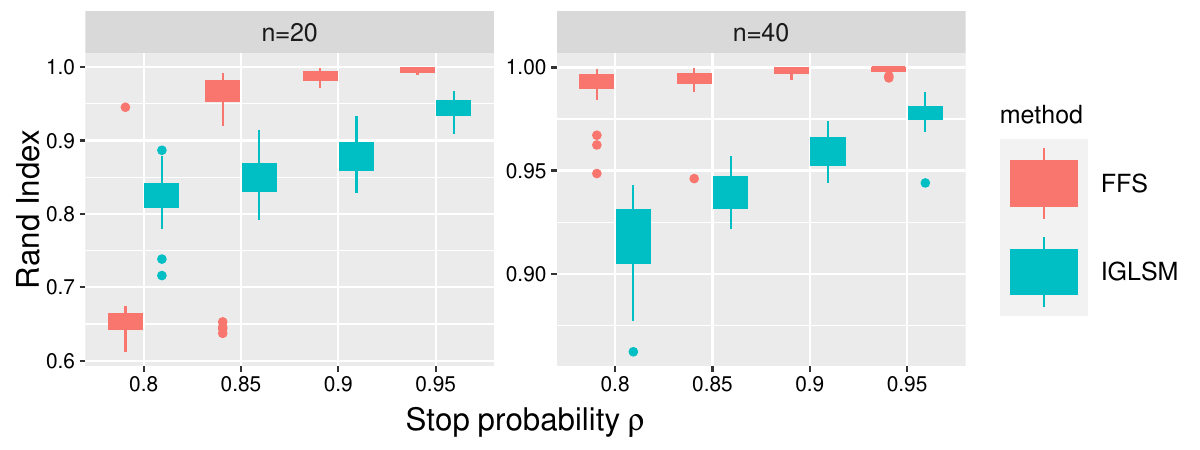}
            \caption{\it Boxplot comparing the performance of community detection between FFS and IGLSM. }
            
            \label{fig:cluster}
    \end{subfigure}%
    ~ 
    \begin{subfigure}[t]{0.48\textwidth}
    \includegraphics[width=\linewidth]{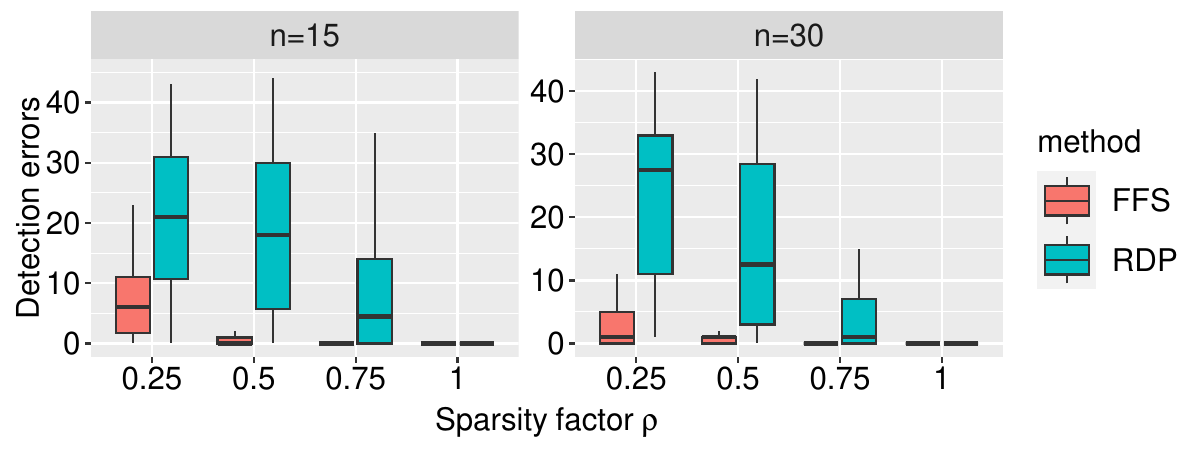}
    \caption{\it Boxplot comparing the performance of change point detection between FFS and RDP. }
    \label{fig:change_point}
    \end{subfigure}
    \caption{\it Boxplot comparing the performance ofcommunity detection and change point detection between FFS and other approaches.}
\end{figure*}

\noindent \textbf{Community detection:}     In addition, we perform simulations in the context of the dynamic latent space model~\eqref{eq:model_lsm} for dynamic networks introduced in the methodology section to demonstrate the effectiveness of FFS in clustering in the presence of sparse changes. We compare FFS priors to IGLSM, where we implement IGLSM via the same likelihood and variational inference framework and only change the prior for the transition variance.  Let $d=2,n=20,40,T=100$. The initial distribution of each component of the true latent vectors is uniformly sampled from $\{-2,2\}$, so that there are $2^d = 4$ true clusters across all time: $(-2,-2),(-2,2),(2,-2),(2,2)$. Then we let all subjects transit with a probability $\rho$ each time, and switch uniformly to the other $3$ locations with a probability of $(1-\rho)/3$. Thus, the expected number of changes in cluster membership is $nT(1-p)$.  We generate data according to the model $Y_{ij,t} \sim  \mbox{Ber(logistic}(-2+\+u_{it}'\+u_{jt})), i>j;  Y_{ij,t}=Y_{ji,t}, t \in [T],$ where the intercept $-2$ is to make sure the connection probabilities between cluster $(-2,2)$ and $(2,-2)$ are small enough. After obtaining the estimate of the latent vectors via variational means, we normalized all the latent subjects, performed Procrustes rotations~\eqref{eq:sequential_Procrustes_rotation}  and then applied $K$-means with $4$ centers at each time independently. Finally, the average rand index over time between the true cluster and the estimated cluster determined by $K$-means is used to assess the accuracy of the clustering. Simulations are repeated $25$ times; the results are shown in Figure~\ref{fig:cluster}. According to the figure, FFS performs better than IGLSM except in the case $\rho=0.8, n=20$, in which the sparsity assumption is
somehow violated. Given the weak signals where the number of subjects $n$ is relatively small, the adaptive patterns between stopping and transition are hard to capture. In general, the more sparse the transitions, the better the performance in clustering, both for FFS and IGLSM, since sparsity introduces more dependence across time.  In summary, FFS better captures the dependence introduced by sparsity in transitions than IGLSM.
        
\noindent \textbf{Change point detection:}        To demonstrate the advantage of our methods in change point detection of dynamic networks, we perform the simulations to compare our approach with \cite{padilla2022change} (RDP), implemented in \textit{changepoints} (\cite{xuchangepoints2022}) package.  We employ a structured approach to simulate network data using stochastic block models (SBMs) to make sure both our model and \cite{padilla2022change} are not exactly specified. Our simulation constructs a network comprising $n=15, 30$ nodes,  
partitioned into three distinct groups $k=3$. We construct two different network connecting patterns. For each segment, we generate a robust sample size of $50$ instances, indicating the true change point is $t_0^* =50$. The detailed generating process of the  two segments is listed below
\begin{equation*}
 \mbox{first segment: }  \rho \times \begin{pmatrix}
    0 & 1 & 0 \\
    1 & 0 & 0 \\
    0 & 0 & 1 \end{pmatrix} \quad \mbox{second segment: }  \rho \times\begin{pmatrix}
    1 & 0 & 0 \\
    0 & 0 & 1 \\
    0 & 1 & 0 \end{pmatrix},
\end{equation*}
where $\rho =0.25,0.5,0.75,1$ stands for the sparsity factor. Nodes are randomly assigned to groups. We utilized the proposed FFS method to estimate the connecting probabilities for each time and vectorized them to perform change-point detection for multivariate time series using \cite{padilla2021optimal}. We repeat the simulation 100 times and use the absolute difference between the estimated change point and the truth, denoted as $|\hat{t_0}-t_0^*|$, as the measure of errors. Figure~\ref{fig:change_point} illustrates the comparison between FFS and RDP. FFS has less error than RDP in all cases as it has near-optimal estimation, where the dependence of networks across all time is taken into account in the estimation step.


   \subsection{Analysis for Formal alliances data}\label{sec:formal_alliance}

       Our methodology is applied to the formal alliances data set (\cite{gibler2008international}, v4.1), which is a dynamic network data set recording all formal alliances (e.g., mutual defense pacts, non-aggression treaties, and ententes) among different nations between 1816 and 2012.  An undirected edge between the two nations indicates that they formed a formal alliance in that year. We use a dynamic latent space model~\eqref{eq:model_lsm} under our methodology to analyze the data set. Studying the global military alliance networks can provide insight into the evolution of geopolitics and international relations across nations over the past two centuries. Visualizing the dynamic networks without a proper statistical model is challenging due to a large number of subjects ($n=180$ nations) and time points ($T=197$ years). For example, \cite{park2020detecting} analyzed the data set with only selected years and subjects, both less than $10$. Our analysis aims to identify the significant historical events and countries that impacted the global alliance structure and provide a visual summary of these changes over time. While many territories changed during the period, we kept all the nations from the beginning to the end as long as they had at least one alliance at any given time. Therefore, the data set contains all subjects and all time points.     Similar to the simulation subsection, we fix the fractional power $\alpha=0.95$ and hyperparameters $a_{\sigma_0}=b_{\sigma_0}=1/2$. We also choose $d=2$ for visualization purposes. Stopping criteria are taken as a difference between two consecutive training AUCs of not more than $0.01$.  The estimate of the latent vectors is performed via variational means with Procrustes rotation~\eqref{eq:sequential_Procrustes_rotation}.

\begin{figure*}[t!]
    \centering
    \begin{subfigure}[t]{0.48\textwidth}
        \centering
\includegraphics[width=\textwidth]{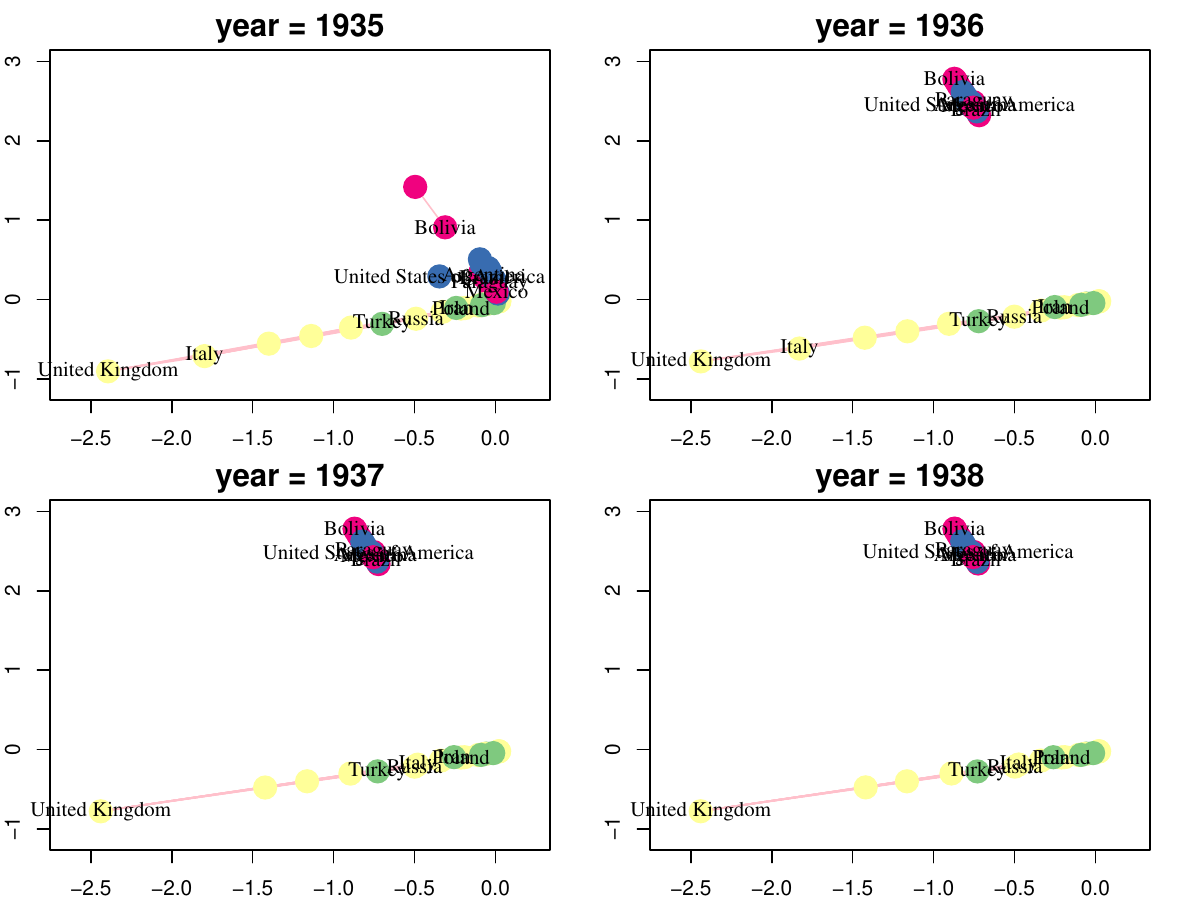}
            \caption{\it Latent space of world's formal alliances for 1935 to 1938. }
            \label{fig:milltary}
    \end{subfigure}%
    ~ 
    \begin{subfigure}[t]{0.48\textwidth}
\includegraphics[width=\textwidth]{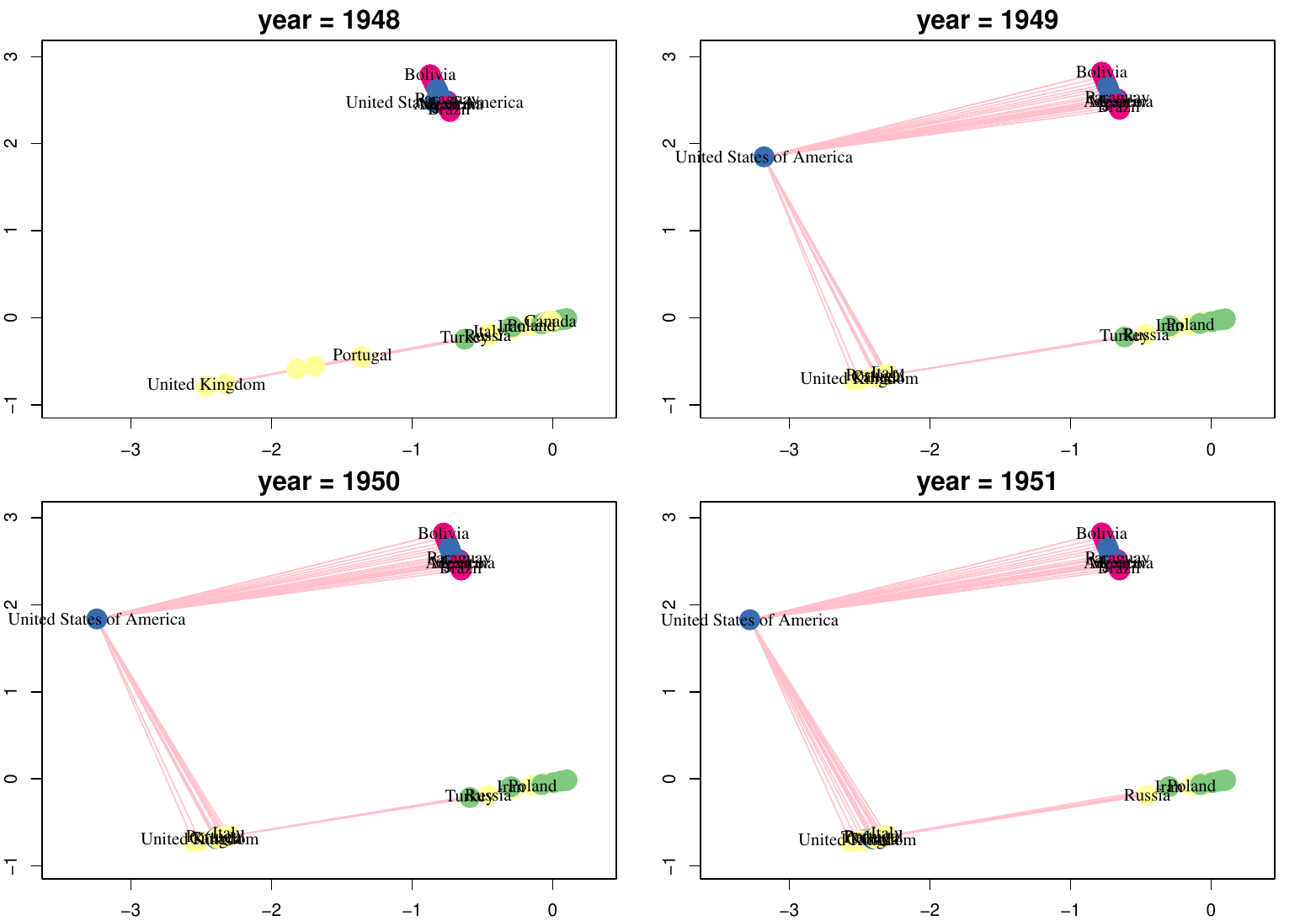}
                \caption{\it Latent space of world's formal alliances for 1948 - 1951. }
                \label{fig:US1948}
    \end{subfigure}
    \caption{The locations of countries' names represent the corresponding subjects' estimated locations after performing Procrustes rotation~\eqref{eq:sequential_Procrustes_rotation}. Some selected labels are shown. Isolated subjects are not shown. Yellow, European countries; Red, South American countries; Blue, North American countries; Green, Asian countries; Green,  African countries; Orange, Australian countries.}
\end{figure*}
       

After obtaining the estimated connecting probabilities, we vectorized them to perform change-point detection for multivariate time series using \cite{padilla2021optimal} to find out the most significant time points for changes in the networks. We successfully detected two time points with significant changes, the year 1935 and the year 1948. We then visualize the estimated latent positions of all the nodes for both time points after performing Procrustes rotation~\eqref{eq:sequential_Procrustes_rotation}.
Figure~\ref{fig:milltary}  compares the estimated latent space between 1935 and 1938. From 1935 to 1936, significant moves came from Latin American countries. This demonstrates that with the Montevideo Convention created in 1933 and registered in the League of Nations Treaty Series in 1936 \citep{montevideo1933}, Latin American countries signed treaties exclusively with each other, separate from European countries. Figure~\ref{fig:milltary} clearly describes the dynamics that these Latin American countries moved far away from European countries and formed a new community.  Figure~\ref{fig:US1948} displays the initial formation of the North Atlantic Treaty Organization (NATO). In 1948, the United States allied with several South American countries, as part of the Inter-American Treaty of Reciprocal Assistance. The graph shows that the United States was closely connected to these countries at that time. Then with the formation of NATO in 1949, the United States shifted its position and gained connections with NATO's European members, demonstrating its strong influence in both America and Europe. Notably, from 1950 to 1951, Turkey also underwent a significant shift, moving away from its proximity to Russia, indicating its participation in NATO during 1951-1952.

\section{Discussions}
{A theoretical framework for Bayesian sparse (dynamic) networks from a latent space model perspective is presented. A natural next step is to allow varying sparsity for different networks in dynamic models.
It is also important to determine the latent dimensionality $d$ in a model. Our model assigns structure to the latent factors, which means that it is necessary to calculate the direct differences across different time points. In this case, the latent dimension $d$ is fixed across time. However, if the dimension is allowed to vary across time, then the new latent dimension considered in our model should be defined as the maximum value of $d$ across all time points. In such a case, models that allow automatic factor selection (e.g., \cite{bhattacharya2011sparse}) would be helpful. In addition, extending the proposed framework to heterogeneous networks containing many isolated nodes during some time points will be interesting to explore. Finally, considering a stochastic version of the proposed CAVI algorithm as in \cite{loyal2024fast} can help in handling large-scale networks from a computational perspective.}

%

	\section{Supplementary material}
Supplementary material covers various topics: extension to a tensor model in Section~\ref{sec:tensor}, additional simulations in Section~\ref{sec:additional_simu}, which includes a comparison between SMF and MCMC, simulation of FFS under smooth transitions and sensitivity analysis with the latent dimension and fractional power, algorithm details for SMF variational inference in Section~\ref{sec:CAVI}, comparison of convergence for our algorithm within each block with others in Section~\ref{sec:convergence}, the algorithm for tensor data in Section~\ref{sec:tensor_computation}, and proofs for the main theorems in Sections~\ref{proof_thm:lower_minimax} to~\ref{sec:proof_thm:single_cluster}. Reproducible examples for both simulation and real data analyses can be found in the attached files
or GitHub at \url{https://github.com/pengzhaostat/Factorized-Fusion-Shrinkage}.

{\putbib}
\end{bibunit}

	\newpage
	
	\section{Appendix}
	\spacingset{1.25} 

 \begin{bibunit}

  \subsection{Extension to dynamic tensor model}\label{sec:tensor}

\begin{figure}[ht]
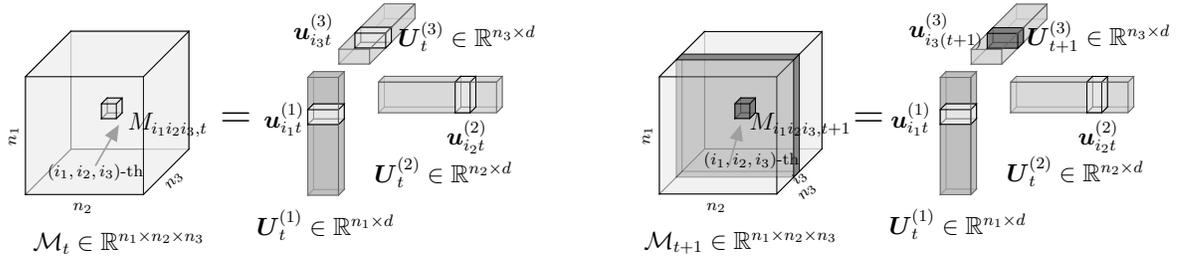

            \centering
            \includestandalone[width=0.47\linewidth]{CP_standalone}
            \includestandalone[width=0.47\linewidth]{CP2_standalone}
            \caption{\it A schematic illustration of the dynamic fusion structure in equation~\eqref{eq:fusion} for a three-way tensor decomposition. 
            From time $t$ to $t+1$, $\+U^{(1)}_{t} $ and $\+U^{(2)}_{t}$ remain unchanged, while only the $i_3$-th row of $\+U^{(3)}_{t}$ changes,  thereby only changing the entire frontal plane indexed by $i_3$ of $\m M_t$.}
            \label{fig:CP_decomposition}
        \end{figure}

        We extend the FFS approach in its most general form for dynamic multi-way arrays. Specifically, let $\m Y = \{\m Y_t\}_{t=1}^T$ be the observed data, where $\m Y_t \in \mathbb{R}^{n_1\times...\times n_M}$ is an $M$-way tensor corresponding to the $t$-th time point. 
         For such data, we consider the following dynamic low-rank tensor model: 
        \begin{equation}\label{eq:tensor_data_generate}
                \m Y_t \sim   p(\m M_t;\beta), \quad \m M_t = \sum_{l=1}^d \+u^{(1)}_{t,l} \otimes ... \otimes \+u^{(M)}_{t,l}, \quad t \in [T], 
        \end{equation}
        where $\otimes$ is the vector outer product\footnote{As an example, given vectors $u, v, w$, the outer-product $A :\,= u \otimes v \otimes w$ is a three-way tensor with $A_{ijk} = u_i v_j w_k$. Naturally extends to more than three arms.}; $\+u_{t,l}^{(m)} \in \mathbb{R}^{n_m}$ for each $t \in [T]$ and $l \in [d]$; $\mb E_{p}(\m Y_t \mid \m M_t) = g(\m M_t)$ for some link function $g$ which operates elementwise on a tensor; and $\beta$ represents additional parameters. 
        The decomposition of $\m M_t$ in equation~\eqref{eq:tensor_data_generate} is based on the CP decomposition \citep{kolda2009tensor}, which expresses a tensor as a sum of $d$ rank-one tensors, where a rank-one tensor is an outer product of vectors. The CP decomposition provides a natural extension to the matrix singular value decomposition (SVD) and model~\eqref{eq:tensor_data_generate} is a corresponding extension of model~\eqref{eq:matrix_data_generate}.
        When $d$ is small compared to  $\prod_{m=1}^M n_m$, the CP decomposition is highly parsimonious as it reduces the number of bits of information in $\m M_t$ from $\prod_{m=1}^M n_m$ to $d \sum_{m=1}^M n_m$.   
         For each $m \in [M]$ representing an arm of the tensor, define an $n_m \times d$ matrix $\+U_t^{(m)}=[\+u^{( m  )}_{t,1},...,\+u^{( m  )}_{t,d}]$. Let us also represent $\+U_t^{(m)}$ in terms of its rows as $\+U_t^{(m)} = [\+u^{( m  )}_{1t},...,\+u^{( m  )}_{n_m t}]'$ with $\+u_{it}^{(m)} \in \mathbb{R}^{d}$ for each $i \in [n_m]$. One may interpret $\+u_{it}^{(m)}$ as a $d$-dimensional vector of latent factors for the $i$th data unit in the $m$th arm of the tensor at time $t$. 
         Similar with structure~\eqref{eq:fusion}, we impose the following group-wise fusion structure on the evolution of the latent factors: $ 
                \sum_{t=2}^T \sum_{i=1}^{n_m} \ind \{D\+u^{(m)}_{it}\ne \+0_d\} \leq  s^{(m)},  m \in [M].$       
         In particular, when $D\+u^{(m)}_{it}= \+0_d$, the entire effect of subject $i$ of arm $m$ remains unchanged from time point $t$ to $t+1$.
         Figure~\ref{fig:CP_decomposition} provides a schematic illustration of this dynamic fusion structure in the case of a $3$-way tensor. When the latent factor $\+u^{(3)}_{i_3t}$ corresponding to the third arm of the tensor changes to $\+u^{(3)}_{i_3(t+1)}$ while $\+U^{(1)}_{t} $, $\+U^{(2)}_{t}$ and all other row vectors of $\+U^{(3)}_{t}$ remain unchanged similar with the matrix case, only the frontal plane indexed by $i_3$ of $\m M_t$ changes, leaving the rest of the tensor intact.           
          We employ group-wise global-local shrinkage priors to model the transition of the latent factors:   
        \begin{align}\label{eq:prior_adaptive_tensor} 
                \begin{aligned}
                        &    \+u^{(m)}_{i(t+1)}\mid \+u^{(m)}_{it} \sim \m N( \+u^{(m)}_{it},\lambda^{(m)2}_{it}\tau_i^{(m)2}\mb I_d),\quad  t \in [T-1]. \\
                        &   \lambda^{(m)}_{it} \overset{ind.}\sim \mbox{Ca}^+(0,1), \quad \tau^{(m)}_i \overset{ind.}\sim g, \quad  t \in [T-1],
                \end{aligned} 
        \end{align}
        independently for $i \in [n_m],\,m \in [M]$. This structure ensures strong shrinkage of the entire vector $D\+u^{(m)}_{i(t+1)}$ towards the origin, while at the same time, the Cauchy tails allow $D\+u^{(m)}_{i(t+1)}$ to have large magnitude when warranted, allowing the prior to capture sharp changes.     For $m>2$, since the CP-type of low-rank tensor models is invariant only up to scaling and permutations (e.g., see Section 3.2 in \citep{kolda2009tensor}), but not for orthogonal transformations, we cannot perform Procrustes rotations on $\+U_t^{(m)},m \in [M]$ to their previous time points without altering the values of $\mathcal{M}_t$.

		\subsection{Additional simulations}\label{sec:additional_simu}
	 \subsection{Additional simulations of FFS prior}\label{sec:additional_FFS}
\begin{figure}[h!]
                \centering
                \includegraphics[width=0.7\linewidth]{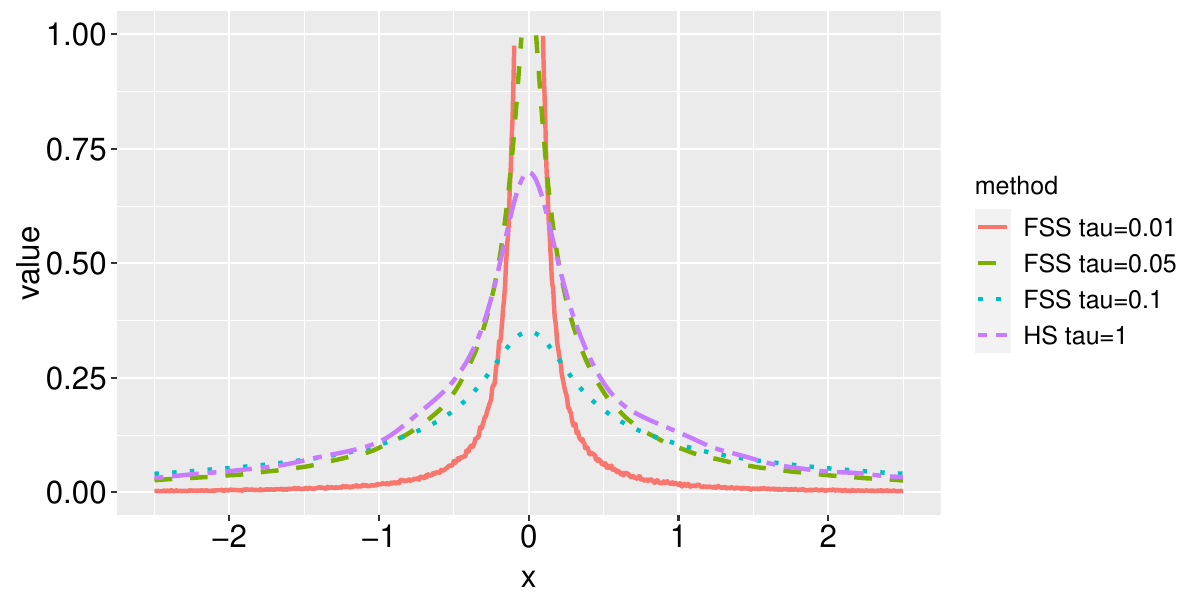}
                \caption{\it Density plot comparisons among marginal priors of $\+u_{i(t+1)}'\+v_{j(t+1)}-\+u'_{it}\+v_{jt}$ with different global parameters  $\tau$ for FFS priors to the horseshoe prior (HS) component-wise on $D\m M_t$ with the global parameter fixed at $\tau =1 $. Value: values of fitted densities; $x$: $\+u_{i(t+1)}'\+v_{j(t+1)}-\+u'_{it}\+v_{jt}$.
                Small choices of $\tau$ in FFS priors can lead to similar marginal distributions with the HS prior applied directly to the component-wise differences.}\label{fig:prior_density}
        \end{figure}

     In this section, we remark on the shrinkage effect of the FFS prior on the transitions at the scale of $\m M_t$. Focusing on the Gaussian matrix factorization model with $d = 2$, we examine the induced prior on the components of $D\m M_{t} = \+U_{t+1}\+V_{t+1}'-\+U_{t}\+V_{t}'$ for different choices of the global parameter $\tau$. Given that $D\+u_{it}$ and $D\+v_{jt}$ are assigned group-wise shrinkage priors under FFS, the induced prior on the components of $D \m M_t$ conditional on $\+u_{i(t+1)},\+v_{jt}$ can be expressed as a weighted sum of horseshoe-like priors. The summation and marginalization slightly decrease the mass assigned near the origin, which nevertheless can be adjusted by setting the global scale $\tau$ to a smaller value. To illustrate this, Figure~\ref{fig:prior_density} compares the shape of the FFS prior for different $\tau$ with the horseshoe prior on components of $D\m M_t$ with $\tau = 1$. According to the figure, the shrinkage effect towards $D\m M_t$ is less pronounced than applying the usual horseshoe prior on elements of $D\m M_t$ with the same $\tau$. We can, however, resolve this problem by using a smaller value of $\tau$. With $\tau=0.05$, the mass around zero is almost the same as that of the horseshoe prior applied directly to $D\m M_t$ with $\tau=1$, while $\tau=0.01$ has even more mass around zero. Therefore, FFS priors can achieve a similar component-wise fusion shrinkage on mean matrices if the global prior on $\tau$ has a sufficient mass around zero. In particular, based on the details described in Assumption~\ref{asm:global_prior} in Section~\ref{sec:3}, the half-Cauchy prior $\tau^{(m)}_i \sim \mbox{Ca}^+(0,1)$ and Gamma prior $\tau_i^{(m)2} \sim \Gamma(a_\tau,b_\tau)$ can be adopted. As noted earlier, the application of the shrinkage on the latent factors instead of the components of $\m M_t$ directly leads to a substantial reduction in the effective number of parameters.     

        We then provide some additional simulations in the binary network model and tensor model.  The expected number of effective parameters is $n^2Td(1-\rho)$ for the binary network model, and $(n_1+n_2+n_3)Td(1-\rho)$ for the tensor model. The piecewise change of binary and tensor cases is small to ensure that the absolute value of all latent vectors is bounded.   We use the sample Pearson's correlation coefficient (PCC) in the binary case to measure the discrepancy between the true and estimated probabilities. The stopping criterion is the difference between the training AUC (area under the curve) in two consecutive cycles not exceeding $0.01$. 
        
	 {\bf Case 2, Binary network model:}
        $n=5,10,20,50$:
        \begin{align*}
                &\+u_{i1} \sim  0.5 \m N((1,0)',\indm) + 0.5 \m N((-1,0)',\indm),  \quad i \in [n]\\
                & D\+u_{it} = \begin{cases}
                        (0,0)' \,\, \mbox{  with probability $\rho$ } \\
                        (-0.25,-0.25)' \,\,\mbox{  with probability $(1-\rho)/2$} \\
                        (0.25,0.25)' \,\,\mbox{  with probability $(1-\rho)/2$}. 
                \end{cases}  \quad  i \in [n]; t=2,...,T\\
                &Y_{ijt} \sim  \mbox{Ber(logistic}(\+u_{it}'\+u_{jt}))  \quad i>j; \quad Y_{ijt}=Y_{jit}, t \in [T].
        \end{align*}
        
        {\bf Case 3, Gaussian Tensor model:}
        $M=3$, $n_1,n_2,n_3=5,10$, $n_1 \geq n_2 \geq n_3$:
        \begin{align*}
                &\+u^{(m)}_{i1} \sim \m N(\+0,\indm),  \quad i \in [n_m],m=1,2,3\\
                &D\+u^{(m)}_{it} = \begin{cases}
                        (0,0)' \,\, \mbox{  with probability $\rho$ }  \quad \quad \quad m=1,2,3\\
                        (-0.25,-0.25)' \,\,\mbox{  with probability $(1-\rho)/2$} \\
                        (0.25,0.25)' \,\,\mbox{  with probability $(1-\rho)/2$}, 
                \end{cases} \quad  i \in [n_m]; t=2,...,T\\
                &\m Y_{t} \sim \m N(\sum_{l=1}^{d} \+u^{(1)}_{t,l} \otimes  \+u^{(2)}_{t,l} \otimes  \+u^{(3)}_{t,l} , 0.3^2) \quad i \in [n_m], m=1,2,3; \, t \in [T].
        \end{align*}

	 According to the simulation results from Table~\ref{tab:1}, when $\rho \geq 0.85$ increases, FFS, SVD and SVD2 improve estimation accuracy. The phenomenon occurs because the Bernoulli distribution has a smaller variance if the probabilities are close to zero or one. In addition, the initial distribution is generated from a mixture distribution with separate components that generate true probabilities close to zero or one. Consequently, if $\rho$ becomes larger, then the true binary responses for $t=2,...,T$ also tend to have probabilities close to $0$ or $1$, making it easier to obtain higher accuracy for the entire problem. Nevertheless,  the proposed method still outperforms SVD and SVD2 in terms of PCC in almost all cases for $\rho \geq 0.8$. The results indicate that the proposed approach can benefit from the time dependence induced by fusion structure as $\rho$ increases, which cannot be ascribed to SVD and SVD2. When the sparsity assumption is violated with $\rho=0.5$, the proposed method is still comparable to SVD and SVD2.    Next, Table~\ref{tab:3} compares FFS with CP decomposition for tensor data generated in Case $3$. FFS performs better than CP across all simulation settings. Note that CP decomposition is optimized through alternating least squares, which is in a similar fashion to the CAVI algorithm. Therefore, both methods suffer from some bad local optimal induced by the alternating optimization mechanism. Therefore, the improvement in the estimation of FFS is simply because the fusion structure is taken into account.

        \begin{table}[ht!]
                \centering
                \caption{Performance comparison for binary cases between FFS and SVD. The measure is the median PCC for estimation of the connection  probabilities.}\label{tab:1}
                \begin{tabular}{rlrrrrrrrr}
                        \hline
                        $n$  & Method & \multicolumn{5}{c}{$\rho$} \\
                        \hline
                        &      &0.5  & 0.8 & 0.85 & 0.9 & 0.95 & 0.99 \\ 
                        \hline
                        5  & FFS & \textbf{ 0.909}  & \textbf{0.867} & \textbf{0.920} & \textbf{0.959} & \textbf{0.984} & \textbf{0.990} \\ 
                        &     SVD & 0.774 & 0.758 & 0.771 & 0.735 & 0.726 & 0.737\\ 
                        &    SVD2 &0.531  & 0.579 & 0.755 & 0.709 & 0.476 & 0.690 \\
                        \hline
                        
                        10  & FFS & \textbf{0.956}   &  \textbf{0.959} & \textbf{0.951} & \textbf{0.940} & \textbf{0.949} & \textbf{0.953} \\ 
                        &   SVD &0.864 & 0.833 & 0.833 & 0.829 & 0.876 & 0.889\\ 
                        &    SVD2 & 0.829  & 0.819 & 0.844 & 0.852 & 0.884 & 0.881 \\
                        \hline
                        20  &  FFS & \textbf{0.973} & \textbf{0.980} & \textbf{0.985} & \textbf{0.985} & \textbf{0.994} & \textbf{0.999}  \\ 
                        &   SVD  &0.919 & 0.941 & 0.947 & 0.959 & 0.957 & 0.950\\ 
                        &    SVD2 & 0.938  & 0.949 & 0.957 & 0.967 & 0.965 & 0.963 \\
                        \hline
                        50  &FFS &  0.970 &  \textbf{0.985} & \textbf{0.984} & \textbf{0.992} & \textbf{0.996} & \textbf{0.997} \\ 
                        &   SVD & 0.934 & 0.967 & 0.936 & 0.970 & 0.973 & 0.973 \\ 
                        &    SVD2 & \textbf{0.972} & 0.975 & 0.960  & 0.978 & 0.979 & 0.981 \\
                        \hline
                \end{tabular}
        \end{table}

        \begin{table}[ht!]
                \centering
                \caption{Performance comparison for tensor cases between FFS and CP decomposition.  The measure is the median RMSE for estimation of the $\m M$.}\label{tab:3}
                \begin{tabular}{llrrrrrrrr}
                        \hline
                        $(n_1, n_2, n_3)$ & Method & \multicolumn{5}{c}{$\rho$} \\
                        \hline
                        &    &0.5  & 0.8 & 0.85 & 0.9 & 0.95 &0.99 \\ 
                        \hline
                        (5 ,5 ,5)  & FFS & \textbf{0.132} & \textbf{0.119} & \textbf{0.119} &\textbf{0.121} & \textbf{0.127 }& \textbf{0.113} \\ 
                        &    CP  &0.187 & 0.147 & 0.153 & 0.145 & 0.148 & 0.143 \\ 
                        \hline
                        
                        (10 ,5 ,5)  & FFS& \textbf{0.114} & \textbf{0.107} & \textbf{0.102} & \textbf{0.106} & \textbf{0.103} & \textbf{0.104} \\ 
                        &    CP& 0.213 & 0.127 & 0.132 & 0.117 & 0.117 & 0.121\\ 
                        \hline
                        (10 ,10 ,5) &  FFS & \textbf{0.093} & \textbf{0.088} & \textbf{0.087} & \textbf{0.082} & \textbf{0.081} &  \textbf{0.074}\\ 
                        &    CP &  0.159 & 0.121 & 0.111 & 0.092 & 0.092 & 0.092 \\ 
                        \hline
                        (10 ,10 ,10)  &FFS & \textbf{0.073} & \textbf{0.068} & \textbf{0.069} & \textbf{0.067} & \textbf{0.067} & \textbf{0.065} \\ 
                        &    CP  & 0.192 & 0.072 & 0.073 & 0.071 & 0.071 & 0.072\\ 
                        \hline
                \end{tabular}
        \end{table}

		\subsection{Simulation settings for Figure~\ref{example_2move}}\label{sec:simu_fig12}
	   Let $d=2,n=10,T=100$. The initial distribution of each component of the true latent vectors is uniformly sampled from $\{(-1,-1)',(-1,1)',(1,-1)',(1,1)'\}$.  The subjects $1$ and $2$ transit with probability $0.05$ at each time point, while the rest of the subjects stay static across all time.  Data is then generated according to the model:  $Y_{ijt} \sim  \mbox{Ber(logistic}(2\+u_{it}'\+u_{jt})), i>j;  Y_{ijt}=Y_{jit}, t \in [T]$.

    \subsection{Comparison between SMF and MCMC}

We compared standard MCMC and SMF in terms of estimation accuracy and computation time for binary networks through the data generating Case 2, Binary network model in subsection~\ref{sec:additional_FFS}. We ran MCMC with  200, 1000 and 5000 iterations using a Gibbs Sampler algorithm, where each coefficient was sampled from its full conditional distribution. For the MCMC chain, we discarded the first half of iterations as burn-in and used the sample means from the last half of iterations to calculate the estimator. We then compared this accuracy with SMF using the PCC (Pearson Correlation Coefficient) with the true probabilities, as well as considering computation time. We set the probability of unchanging of latent positions, $\rho=0.9, 0.95, 0.99$, sample size $n=10, 20, 50$ and $p=n$. The simulations were repeated 25 times for each setting.

\begin{figure}[H]
\centering
\includegraphics[width=0.8\linewidth]{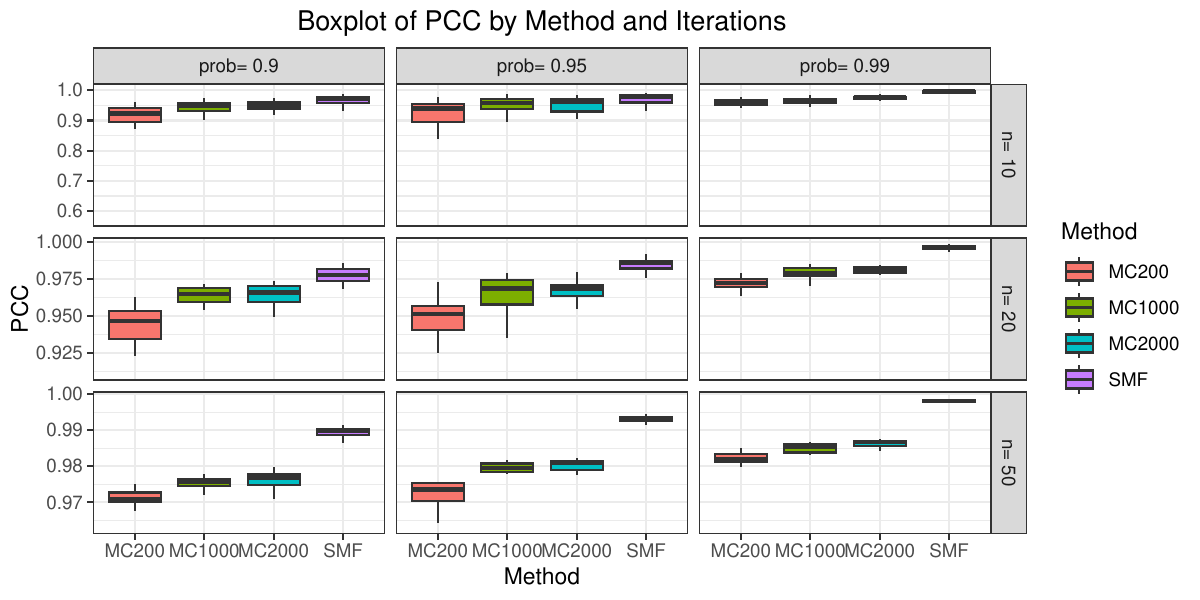}
\caption{\it {Boxplots comparing the estimation accuracy of Pearson Correlation Coefficient (PCC) between the estimated and true connection probabilities for SMF,  and various numbers of MCMC iterations. A higher PCC indicates better estimation performance for the corresponding method.   MC200, MC1000 and MC 2000 represent posterior means obtained after   200, 1000 and 2000 iterations of Gibbs samplers, respectively, with the first half of iterations discarded as burn-in. Among all cases, SMF achieves the best estimation accuracy.}}
\label{fig:MCMC_PCC}
\end{figure}

\begin{figure}[H]
\centering
\includegraphics[width=0.8\linewidth]{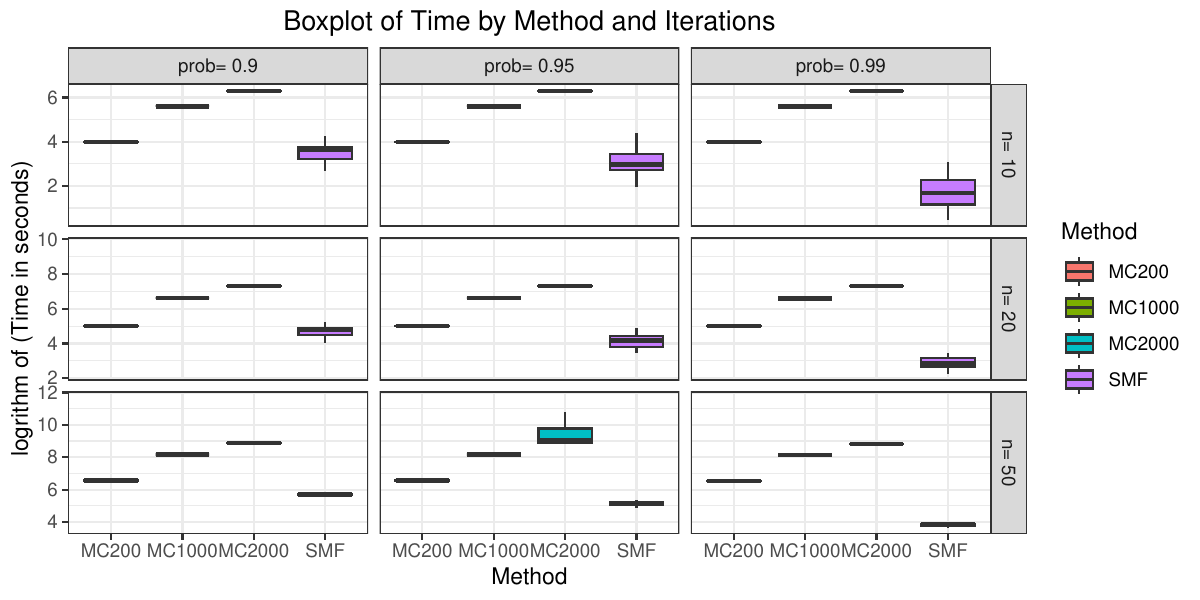}
\caption{\it Boxplots comparing the computation time for SMF,  and different numbers of MCMC iterations, as simulated in Figure~\ref{fig:MCMC_PCC}.  MC200, MC1000 and MC 2000 represent  posterior means obtained after   200, 1000 and 2000 iterations of Gibbs samplers. Remarkably, SMF exhibits the shortest computation time while achieving the highest level of estimation accuracy in Figure~\ref{fig:MCMC_PCC} across all cases.}
\label{fig:MCMC_Time}
\end{figure}

The results are presented in boxplot comparisons shown in Figure~\ref{fig:MCMC_PCC} and Figure~\ref{fig:MCMC_Time}. Overall, SMF requires less computation time than MCMC under the given settings while achieving the best estimation accuracy. This indicates that when the dependence across time is strong, SMF significantly improves computation efficiency.


\subsection{Performance of FFS under smooth transition}
\begin{figure}[H]
    \centering
    \includegraphics[width=0.5\linewidth]{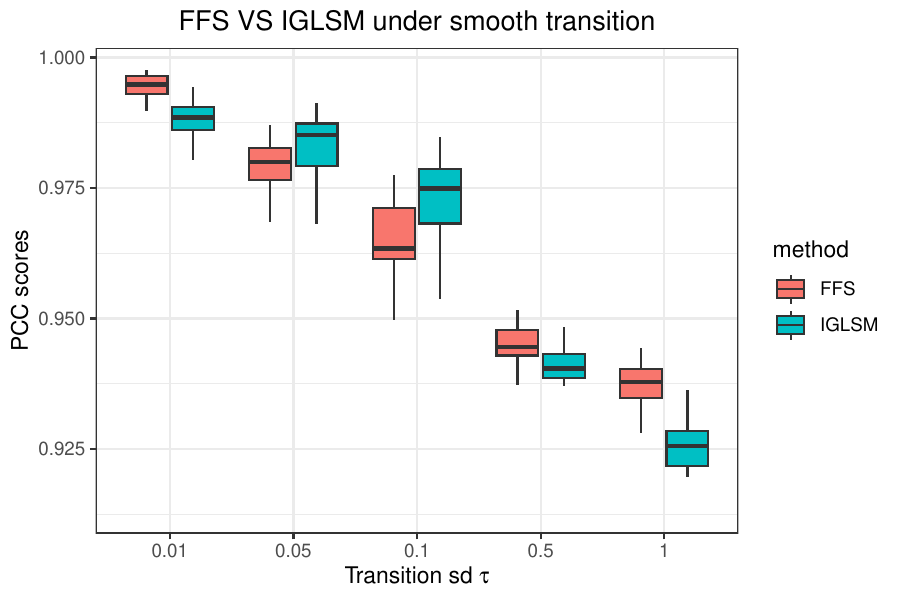}
    \caption{\it Boxplot comparing the performance of estimation error FFS and IGLSM under smooth data generating process.
The measure is the PCC scores between the estimated probabilities and truth, with 25 replications. }
    \label{fig:simu_smooth}
\end{figure}

Our new simulations demonstrate that FFS prior performs relatively well even when the model is misspecified. We consider smooth transition, we use $\tau$ as the true transition standard derivation to control the magnitude of changes: $\+u_{it} \mid \+u_{i(t-1)} \sim \m N(\+u_{i(t-1)},\tau^2 \+I_d) $. The other settings are similar: we use 25 replicated data sets are generated from
	$Y_{ijt} \sim \mbox{Bernoulli}(\+u_{it}'\+u_{jt})$ for $i \ne j=1,...,n$ and $t=1,...,T$ where $n=20, \,T=100, \,d=2$. Let $\+u_{i1} \sim \mathcal{N}((0,0)',  \indm)$.  We compare the estimation accuracy between the IGLSM and FFS across different values of $\tau$ such as $\tau=0.01,0.05,0.1,0.5,1$. Figure~\ref{fig:simu_smooth} illustrates the performance between them. If the transition standard deviation, $\tau$, is either too large ($\tau =0.5,0.1$) or too small ($\tau =0.01$), the FFS method outperforms the IGLSM. This is because the shrinkage prior in FFS encourages the transition to be either too large or close to zero, which aligns with the real data-generating cases. However, when $\tau$ is neither too large nor too small, the FFS method performs worse than IGLSM method because the adopted shrinkage prior has no correct preference for which direction to shrink the transitions towards.

\subsection{Sensitivity analysis with respect to latent dimensions and fractional power}
\begin{figure}[H]
		\centering
		\includegraphics[width=0.9\linewidth]{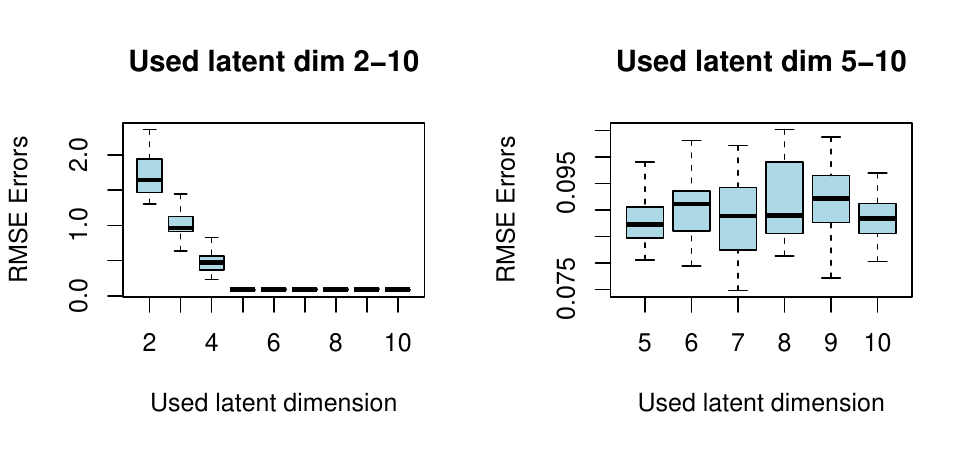}
		\caption{\it Performance comparison for Gaussian networks of estimation of SMF across different specifications of latent dimension $\hat{d}$ while the true latent dimension is $d=5$. Simulation settings are provided in the matrix factorization model in subsection~\ref{sec:simu}.}
  		\label{fig:error_vs_d}
	\end{figure}

{\bf Sensitivity analysis with respect to the choice of latent dimension:} When the $d$ is overspecified, where the using of $d$ is greater than the true $d^*$, we find out the misspecification doesn't provide too much damage. We have conducted a new simulation as in the matrix factorization model in subsection~\ref{sec:simu} with $n=20,p=10,T=100, \rho=0.99$ to demonstrate the above finding. For true latent dimension $d=5$, we use 9 grids of correct and misspecifications of used latent dimension $\hat{d}=2,3,4,5,6,7,8,9,10$. The simulation result is shown in Figure~\ref{fig:error_vs_d}, where we compared different effects of used latent dimension $\hat{d}$ in our model. The figure shows that if the latent dimension is under-specified, the estimation errors will increase significantly because the complexity of the model cannot be captured. However, if the latent dimension is overspecified, the outcome is not affected too much, as the estimation errors remain almost the same for $\hat{d}\geq 5$, with only minor variations due to stochastic errors.
	
\begin{figure}[H]
		\centering
		\includegraphics[width=0.6\linewidth]{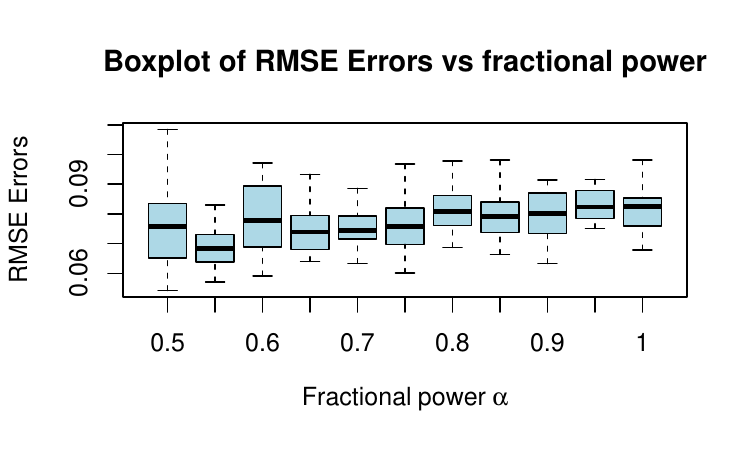}
		\caption{\it Performance comparison for Gaussian networks of estimation of SMF and MF across different $\alpha$. Simulation settings are provided in the matrix factorization model in subsection~\ref{sec:simu}.}
		\label{fig:error_vs_alpha}
	\end{figure}

{\bf Sensitivity analysis with respect to the choice of fractional power:}
We compared the effect of different $\alpha$ values in our model. 	we have conducted a new simulation as in the matrix factorization model in subsection~\ref{sec:simu} with $n=20,p=10,T=100, \rho=0.99$ to provide a numerical demonstration. We use 11 grids $0.5,0.55,0.6,0.65,0.7,0.75,0.8,0.85,0.9,0.95,1$ for $\alpha$ values. The simulation result is shown in Figure~\ref{fig:error_vs_alpha}, where we compared different effects of $\alpha$ values in our model. As can be seen from the figure, the results are consistent with previous findings in other papers and suggest that the choice of $\alpha$ does not affect the outcome.
	
	\subsection{CAVI algorithm}\label{sec:CAVI}

	First, given $q(\varTheta)$, the updating of $q(\beta)$ is as follows:
	\begin{align*}
	q(\beta) \propto \exp [\E_{-\beta}\{\log p_\alpha (\varTheta,\beta,\m Y )\}] \propto \exp [\E_{\varTheta}\{\log P_\alpha(\m Y \mid \varTheta,\beta )\}+ \log p(\beta)],
	\end{align*}  
	which has a closed-form expression when the likelihood and priors are in a Gaussian form.
	
	Given the above moment of  $\lambda_{it}$, $\tau_i$ and $\sigma_{0i}$, the updating of $q(\+\theta_{i\cdot})$  can be obtained through the message-passing framework:  suppose $q(\beta)$, $q(\tau)$,$q(\sigma_0)$ and $q_j(\+\theta_{j\cdot}), \, j \ne i$  are given and the target is to update $q_i(\+\theta_{i\cdot})$, we have:
	
	\begin{equation}\label{eq:variational_xit}
	\begin{aligned}
	q_i(\+\theta_{i\cdot}) \propto \exp \left[ \E_{-\+\theta_{i\cdot}}\left\{{\sum_{t=1}^{T}\sum_{j\neq i, j=1}^{n}  \{\log P_\alpha(Y_{ijt}\mid \+\theta_{it},\+\theta_{jt},\beta)\}} \right.\right. \\
	\left.\left.+ { \sum_{t=1}^{T-1}\log p(\+\theta_{i(t+1)}\mid \+\theta_{it},\lambda_{it}) + \log p(\+\theta_{i1} \mid \sigma_{0i})} \right\} \right],
	\end{aligned}
	\end{equation}
	where $\E_{-\+\theta_{it}}$ is the expectation taken with respect to the density $\big[\prod_{j \neq i} q_j(\+\theta_{j\cdot})\big] q(\beta)q(\+\Lambda)q(\+\sigma_0)$.
	
	 We have the following factorial property for fractional joint distribution $p_\alpha(\m Y, \varTheta, \beta, \+H,\+\Lambda,\+\tau, \+\sigma_0)$:
        \begin{align}\label{eq:lsm}
                \begin{aligned}
                        &       p_\alpha(\m Y, \varTheta, \beta, \+H,\+\Lambda,\+\tau, \+\sigma_0)\\
                        &\propto P_\alpha(\m Y \mid \varTheta ,\+\Lambda,\+\tau,\beta,\+\sigma_0,\+H)p(\varTheta \mid \+\Lambda,\+\tau,\+\sigma_0)p(\+\Lambda \mid \+H)p(\+H)p(\beta)p(\+\tau) \\
                        &=    \prod_{i=1}^{n} \left\{\prod_{t=1}^{T-1}p(\+\theta_{i(t+1)}\mid \+\theta_{it}, \lambda_{it},\tau_i)p(\lambda_{it} \mid \eta_{it})p(\eta_{it}) p(\+\theta_{i1} \mid \sigma_{0i})\right\} p(\beta)p(\tau_i) \\
                        &\times \prod_{t=1}^T \prod_{1\leq i \ne j \leq n} P_\alpha(Y_{ijt}\mid \+\theta_{it},\+\theta_{jt},\beta),
                \end{aligned}
        \end{align}
        with $p(\+\theta_{i(t+1)}\mid \+\theta_{it}, \lambda_{it},\tau_i) \propto \exp(-\|\+\theta_{i(t+1)}- \+\theta_{it}\|_2^2/(2\tau_i^2\lambda^2_{it}))$ for $t \in [T-1]$.

        By the scheme of CAVI, given the distributions $q(\varTheta),q(\beta)$, we have updatings of  $\lambda_{it},\eta_{it},\sigma_{0i}$:
        \begin{align*}
                \hat{q}(\lambda_{it},\eta_{it},\sigma_{0i}) &\propto \exp [\E_{-\lambda_{it},\eta_{it},\sigma_{0i}}\{\log p_\alpha(\m Y, \varTheta, \beta, \+H,\+\Lambda,\+\tau, \+\sigma_0)\}]\\
                & \propto \exp [\E_{-\lambda_{it},\eta_{it},\sigma_{0i}}\{ \log P(\varTheta \mid \+\sigma_{0}, \+\Lambda) +\log P(\+\Lambda \mid \+H) + \log P(\+H)+\log p(\+\sigma_{0})\}] \\
                & \propto \exp \left[- \frac{d+3}{2} \log (\lambda_{it}^2) -\E_{\+\theta_{i\cdot}}\left\{\frac{\|\+\theta_{it}-\+\theta_{i(t+1)}\|^2}{2\tau_i^2 \lambda_{it}^2}  \right\}  -2\log (\eta_{it}^2) -\frac{1}{\eta_{it} \lambda^2_{it}}  -\frac{1}{\eta_{it} }  \right.\\ 
                &\left.  -\E_{x_{i1}}\left(\frac{\|\+\theta_{i1}\|^2_2}{2\sigma^2_{0i}}\right) -\left(\frac{d}{2}+a_{\sigma_0}+1\right) \log(\sigma_0^2) -\frac{b_{\sigma_0}}{\sigma_{0i}^2}\right].
        \end{align*}  
        
        Therefore, we have
        \begin{equation}\label{eq:update_scale}
                \begin{aligned}
                        \eta_{it}^{(new)} \quad &\sim \mbox{IG}\left(1, 1+ \mu_{1/\lambda_{it}^2} \right); \\
                        \lambda_{it}^{2(new)}  \quad &\sim \mbox{IG}\left(\frac{d+1}{2},  \mu_{1/\eta_{it}}+\E_{\+\theta_{it},\+\theta_{i(t+1)}}\left[\frac{\|\+\theta_{it}-\+\theta_{i(t+1)}\|_2^{2}}{2 \tau_i^{2}} \right]\right); \\
                        \sigma_{0i} ^{2(new)} \quad &\sim \mbox{IG}\left(\frac{nd+a_{\sigma_0}}{2},\frac{\E_{q(\+\theta_{i1})}(\|\+\theta_{i1}\|_2^2)+2b_{\sigma_0}}{2} \right),
                \end{aligned}
        \end{equation}
        where $\mu_{1/\lambda_{it}^2} = \E_{q(\lambda_{it})}(1/\lambda_{it}^2)$ and $\mu_{1/\eta_{it}} = \E_{q(\eta_{it})}(1/\eta_{it})$.
        Note that the key moment has a closed-form expression in terms of the parameters: $a \sim \mbox{IG}(\alpha,\beta)$, $\E(1/a)=\alpha/\beta$. 
        
	Given the equation~\eqref{eq:variational_xit}, note that $q_i(\+\theta_{i\cdot})$ has the following form:
	\begin{equation}\label{eq:Markov}
	q_i(\+\theta_{i\cdot}) = q_{i1}(\+\theta_{i1})\prod_{t=1}^{T-1}q(\+\theta_{i(t+1)} \mid q(\+\theta_{it})) = \prod_{t=1}^{T-1}\frac{q_{it,i(t+1)}(\+\theta_{it},\+\theta_{i(t+1)})}{q_{it}(\+\theta_{it})q_{i(t+1)}(\+\theta_{i(t+1)})}\prod_{t=1}^T q_{it}(\+\theta_{it}).
	\end{equation}
	It follows that the graph of random variable $\+\theta_{i\cdot}$ is structured by a chain from $\+\theta_{i1}$ to $\+\theta_{iT}$.   Due to the above structure~\eqref{eq:Markov}, the computation of $q_i(\+\theta_{i\cdot})$ given the rest densities can be carried out efficiently in a message-passing manner. In particular, the message-passing algorithm involves computing all the unary marginals $\{q_{it}\}$ and binary marginals $\{q_{it, i(t+1)}\}$, which in turn help to update scales in closed forms \eqref{eq:update_scale}. 
	
	When Gaussian likelihood is adopted:
	\begin{align*}
	P_\alpha(\m Y\mid \varTheta,\beta) = \prod_{t=1}^T \prod_{1\leq i \leq n, 1 \ne j \leq p}  \frac{1}{\sqrt{2 \pi} \sigma} \exp \left[-\alpha\frac{\{Y_{ijt} -\+u_{it}'\+v_{jt}\}^2}{2\sigma^2}\right].
	\end{align*}
	where $\sigma$ is the $\beta$ in the previous general setting, the MP updating can be implemented in the framework of Gaussian belief propagation networks,  where the mean and covariance of the new update of variational distribution can be calculated by Gaussian conjugate and marginalization using the Schur complement. In addition, the updating of $\sigma$ can also be obtained in the closed forms via inverse-gamma conjugacy.
	
	For the Bernoulli likelihood:
	$$P_\alpha(Y_{ijt}\mid  \+u_{it},\+v_{jt})=\exp[\alpha Y_{ijt}(\+u_{it}'\+v_{jt})-\alpha\log \{1+\exp(\+u_{it}'\+v_{jt})\}].$$ The tangent transform approach proposed by \cite{jaakkola2000bayesian} is applied in the present context to obtain closed-form updates.

	By introducing $\Xi=\{\xi_{ijt}: i,j=1,...,n,t \in [T]\}$ with $A(\xi_{ijt})=-\mbox{tanh}(\xi_{ijt}/2)/(4 \xi_{ijt})$ and $C(\xi_{ijt})=\xi_{ijt}/2-\log(1+\exp(\xi_{ijt}))+\xi_{ijt} \mbox{tanh}(\xi_{ijt}/2)/(4 \xi_{ijt})$ for any $\xi_{ijt}$,  the following lower bound on $P_\alpha (Y_{ijt}\mid \+u_{it},\+u_{jt},\beta)$ holds:
	\begin{equation*}
	\begin{aligned}
	\underline{P}_\alpha(Y_{ijt}\mid  \+u_{it},\+v_{jt} ;\xi_{ijt}) = \exp \left[\alpha A(\xi_{ijt})(\+u_{it}'\+v_{jt})^2+\alpha \left(Y_{ijt}-\frac{1}{2} \right)(\+u_{it}'\+v_{jt})+\alpha C(\xi_{ijt}) \right].
	\end{aligned}
	\end{equation*}
	
	The likelihood $P_\alpha(Y_{ijt}\mid  \+u_{it},\+u_{jt})$ is replaced by its lower bound $\underline{P}_\alpha(Y_{ijt}\mid \+u_{it},\+v_{jt} ;\xi_{ijt})$, where the updating of $\varTheta$ in the Gaussian conjugate framework can be performed. After updating all the variational densities, $\xi_{ijt}$ is optimized based on EM algorithm according to \cite{jaakkola2000bayesian}: $\xi^{(new)}_{ijt} =\E_{q(\varTheta)}\{(\+u_{it}'\+v_{jt})^2\}$. To summarize, for Gaussian or binary likelihoods, the proposed variational framework allows all updating in the Gaussian conjugate paradigm by assuming only independent relationships between different subjects within the variational family. 
	
	\subsection{Comparison of the MP algorithm to proximal gradient for $\ell_1$ penalized trend filtering}\label{sec:convergence}
\begin{figure}[h]
		\centering
		\includegraphics[width=0.7\linewidth]{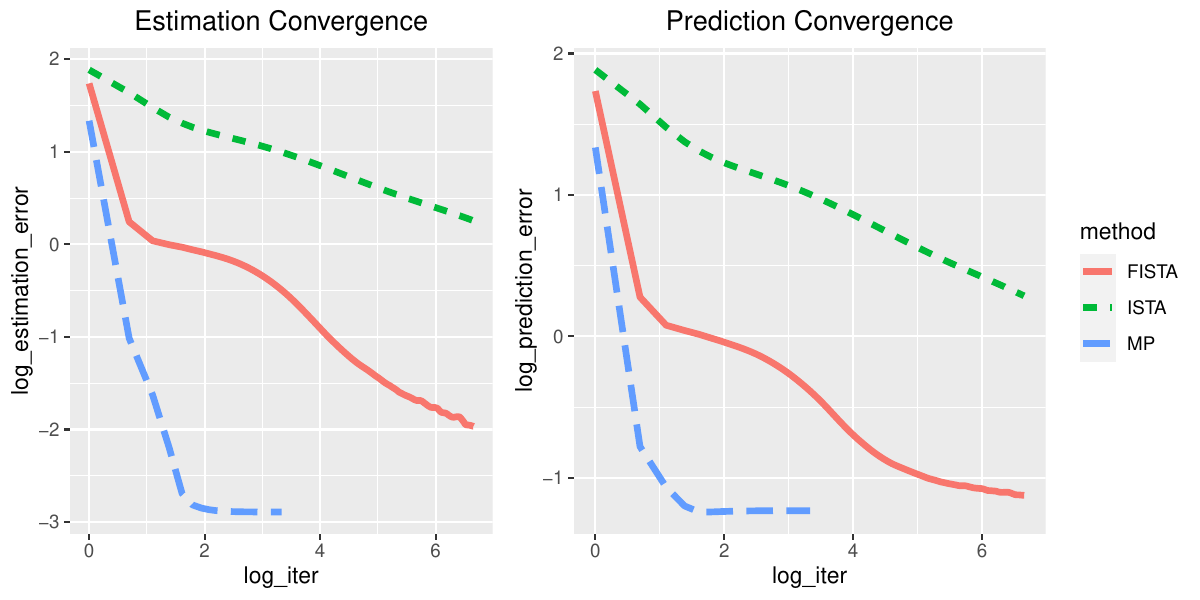}
		\caption{Comparison in iterations between message passing variational inference under proposed prior vs. proximal (ISTA) and accelerated proximal gradient descent (FISTA) under $\ell_1$ regularized trend filter problem. The  hyperparameter $\tau$ of FFS is assigned with a fixed value to avoid model selection for fair comparison. }
		\label{fig:convergence}
	\end{figure}

		Due to the fact that the variational family~\eqref{eq:SMF_joint} captures the temporal dependence, the computation of $q(\+\theta_\cdot)$ given the rest densities can be carried out in a message-passing manner. The proposed VI algorithm converges faster than the proximal gradient descent algorithm for $\ell_1$ regularized trend filtering problem without increasing the complexity of computation and storage per iteration due to the fact that message-passing utilizes the banded (block tri-diagonal) structure of the second-order moments, which can be inverted at an $O(Td^3)$ cost. Figure~\ref{fig:convergence} provides a simulation example by comparing our convergence in iterations between the proposed VI approach and the proximal and accelerated proximal gradient descent for a trend-filtering problem with $\ell_1$ regularization for $d=1$.
	
 \subsection{Computation for tensor data}\label{sec:tensor_computation}
      
            For tensor data, the shrinkage priors are similarly applied on the transitions $\+u^{(m)}_{it} \mid \+u^{(m)}_{i(t-1)} \sim \m N(\+u^{(m)}_{i(t-1)},\tau^{(m)2}_{i}\lambda^{(m)2}_{it})$, and the SMF family is similarly defined:
        \begin{equation*}
                q(\m U, \beta, \+H,\+\Lambda,\+\tau, \+\sigma_0) = \prod_{m=1}^{m}\prod_{i=1}^{n_{m}} \left[q(\+u^{(m)}_{i\cdot})q(\tau^{(m)}_i)\prod_{t=1}^{T-1}\left\{q(\lambda^{(m)}_{it})q(\eta^{(m)}_{it}) \right\}q(\sigma^{(m)}_{0i}) \right]q(\beta).
        \end{equation*}
        
        Then the extension from matrix to tensor is based on the tensor unfolding technique, as in \cite{hoff2011hierarchical} and \cite{zhou2013tensor}. Tensor unfolding allows the components of the mean to be expressed as inner products of a targeted vector and component-wise products of some other vectors, which is fortunate to be compatible with the alternating updatings of the SMF variational family. Note that when we assume       $\m M_t$ defined as equation~\eqref{eq:tensor_data_generate}, we have the mode-$m$ unfolding:
        \begin{equation*}
                \m M_{t,(m)} =\left\{ \+U_{t}^{(M)} \odot \ldots \+U_{t}^{(m+1)} \odot \+U_{t}^{(m-1)} \odot \ldots \+U_{t}^{(1)} \right\} \+U_{t}^{(m)^{\prime}},
        \end{equation*}
        where $\odot$ is the Khatri-Rao product: $\+A \odot \+B = [\+a_1\otimes \+b_1,...,\+a_K \otimes \+b_K]$ for $\+A=[\+a_1,...,\+a_K] \in \mathbb{R}^{I \times K}$ and $\+B=[\+b_1,...,\+b_K] \in \mathbb{R}^{J \times K}$.     Therefore, the components of $\m M_{t,(m)}$ equal inner products between row vectors of $ \+U_{t}^{(M)} \odot \ldots \+U_{t}^{(m+1)} \odot \+U_{t}^{(m-1)} \odot \ldots \+U_{t}^{(1)}$ and $\+U_{t}^{(m)}$.  In particular, for $\{i_{m} \in [n_m]\}_{m}$, we have 
        \begin{equation*}
                \left\{\+u_{i_M t}^{(M)} \circ \ldots \+u_{i_{m+1}t}^{(m+1)} \circ \+u_{i_{m-1}t}^{(m-1)} \circ \ldots \circ\+u_{i_{1}t}^{(1)}\right\}'\+u^{(m)}_{jt} = \m [M_{t,(m)}]_{i,j}
        \end{equation*}
        with $i=(i_M-1)(\prod_{j=1, j \ne M}^{M-1}n_j) +...+ (i_{m+1}-1)(\prod_{j=1}^{m-1}n_j) + (i_{m-1}-1)(\prod_{j=1}^{m-2}n_j) +...+ i_1$, where $\circ$ is the Hadamard (element-wise) product. In the CAVI algorithm for the SMF family, when all the targeted row vectors follow Gaussian distributions, calculating the first and second moments of $\+u_{i_M t}^{(M)} \circ \ldots \+u_{i_{m+1}t}^{(m+1)} \circ \+u_{i_{m-1}t}^{(m-1)} \circ \ldots \circ\+u_{i_{1}t}^{(1)}$ is necessary to update a new mean and covariance for the distribution of  $\+u_{i_{m} t}^{(m)}$. For the component-wise product above, the first moment is straightforward, and the following lemma can be used to calculate the second moment sequentially:
        \begin{lemma}\label{lem:4.1}
                Suppose $\+a \sim \m N(\+\mu,\+\Sigma)$ and  $\+\Xi$ is a positive definite matrix. Let $\+D(\+a) = diag(\+a)$, then we have 
                \begin{equation*}
                        \E(\+D(\+a)\+\Xi \+D(\+a))  = \+D(\+\mu)\+\Xi \+D(\+\mu) + \+\Xi \circ \+\Sigma,
                \end{equation*}
        \end{lemma}
        
        With the lemma, given $\E(\+a\+a')$ and $\+b \sim \m N(\+\mu_b,\+\Sigma_b)$, we can first calculate $\E((\+b\circ \+a)(\+b\circ \+a)') = \E(\+D(\+b) \+a\+a'\+D(\+b))$. Then $\E((\+c \circ \+b\circ \+a)(\+c \circ \+b\circ \+a)')$ can also be calculated sequentially with another normal distributed vector $\+c$. A similar technique can also be applied when additional auxiliary variables are estimated using additional MF factorization. For example, to estimate the variance of the noise of a Gaussian tensor model $\sigma$, commonly used inverse-Gamma conjugate updates can be applied together within the MF framework, then the necessary term $\E_q \{\sum_{t=1}^T(\m Y_t- \m M_t )^2\}$ can also be computed through the above strategy.
 
 \subsection{Proof of Theorem~\ref{thm:lower_minimax}}\label{proof_thm:lower_minimax}
	\begin{proof}
		For $\varTheta^a=[\m U^{a'}, \m V^{a'}]'$ and $\varTheta^b= [\m U^{b'}, \m V^{b'}]'$, let 
		\begin{equation*}
		d^2(\varTheta^a, \varTheta^b) = \sum_{t=1}^T\|\+U^a_t\+V^{a'}_t - \+U^b_t\+V^{b'}_t\|_F^2,
		\end{equation*}
		and
		\begin{equation*}
		d_0^2(\+U^a_t\+V^{a'}_t, \+U^b_t\+V^{b'}_t) = \|\+U^a_t\+V^{a'}_t - \+U^b_t\+V^{b'}_t\|_F^2.
		\end{equation*}
		
		The hypothesis set is constructed such that all elements are sufficiently distinct and the cardinality of the set can be constrained.

		\subsection{Asymmetric case:}	
		To obtain the final bound for the asymmetric case, we use an alternative strategy. Initially, we fix $\m V$ and select a suitable parameter space for $\m U$ to obtain a rate of the lower bound; we then fix $\m U$ and construct a hypothesis space for $\m V$ to obtain a second lower bound. The final rate of the lower bound should be the sum of the two.
		
		As a first step, we should first obtain a sparse Varshamov-Gilbert bound in Lemma~\ref{lemma VG} under Hamming distance for the low-rank subset construction:

		\begin{lemma}[Lemma 4.10 in \cite{massart2007concentration}]\label{lemma VG}
			Let $\Omega =\{0,1\}^n$ and $1 \leq n_0 \leq n/4$. Then there exists a subset $\{\+w^{(1)},...,\+w^{(M)}\} \subset \Omega$ such that 
			\begin{enumerate}
				\item $\|\+w^{(i)}\|_0 =n_0$ for all $1 \leq i \leq M$;
				\item $\|\+w^{(i)}-\+w^{(j)}\|_0 \geq n_0/2$ for $0\leq i \ne j \leq M $;
				\item $\log M \geq cn_0 \log(n/n_0)$ with $c \geq 0.233$.
			\end{enumerate}
		\end{lemma}
		
		We construct the following case according to Lemma~\ref{lemma VG} (the construction holds under $n-d+1\geq 4$). For each $\+w$,  we can construct an $n \times d$ matrix and a $p \times d$ matrix as follows:
		\begin{equation*}
		\+U^{w}=\begin{bmatrix}
		\epsilon \+w &\+0 \\
		\+0  & \+I_{d-1}
		\end{bmatrix} \quad  \mbox{with}  \quad  w \in \Omega_M;   \quad 
		\+V^{0}=\begin{bmatrix}
		\+1_{p-d+1} &\+0 \\
		\+0  & \+I_{d-1}
		\end{bmatrix},
		\end{equation*}
		which gives
		\begin{align*}
		\+U^{w}\+V^{0'} = \begin{bmatrix}
		\epsilon \+w \+1' &\+0 \\
		\+0  & \+I_{d-1}
		\end{bmatrix} .
		\end{align*}

		The effect of this construction is that: for different $\+w_1,\+w_2 \in \Omega_M$,  since $n_0 /2 \leq  \|\+w_1-\+w_2\|_0 \leq 2n_0$ and $\|\+V^{0}\|_F \leq \sqrt{p}$, we have
		\begin{align*}
		d_0(\+U^{w_1}\+V^{0'}, \+U^{w_2}\+V^{0'}) \leq \|\+U^{w_1}-\+U^{w_2}\|_F \|\+V^{0}\|_F = \sqrt{n}\epsilon \|\+w_1-\+w_2\|_2\leq \sqrt{{2pn_0}}\epsilon.
		\end{align*} 
		In addition,  note that
		$$d^2_0(\+U^{w_1}\+V^{0'}, \+U^{w_2}\+V^{0'}) = (p-d-1) \epsilon^2\sum_{i=1}^{n_0} ( w_{1i}-w_{2i})^2 \geq \frac{n_0 (p-d-1) \epsilon^2}{2}.$$

		Let 
		\begin{equation*}
		\+U^{0}=\begin{bmatrix}
		\+0 &\+0 \\
		\+0  & \+I_{d-1}
		\end{bmatrix} 
		\end{equation*}
		$\m U^0= [\+U^0,...,\+U^0]$, $\m V^0=[\+V^0,...,\+V^0]$ and $\varTheta^0 = [\m U^{0'}, \m V^{0'}]'.$
		
		We  need the above  Varshamov-Gilbert Bound~\ref{lemma VG} again to introduce another binary coding: 
		
		Let $\Omega_r =\{\+\phi^{(1)},...,\+\phi^{(M_0)}\}  \subset \{0,1\}^{T} $, such that $\|\+\phi^{(i)}\|_0=t_0$  for all $1\leq i \leq M_0$  and $\|\+\phi^{(i)}-\+\phi^{(j)}\|_0 \geq t_0/2$ for $0\leq i <j \leq M_0 $ with $\log M_0 \geq ct_0 \log( T/t_0)$ with $c \geq 0.233$. 
		
		Then we have the following construction:
		\begin{equation*}
		\begin{aligned}
		\Xi_\epsilon =\{\varTheta^{(w,\phi)}:&\\
		&  \+U^{(w,\phi)}_t =\+U^{w^{(i)}}, \quad  \quad \mbox{if}\quad  \phi_t=1,\\
		&\+U^{(w,\phi)}_t =\+U^{0}, \quad  \quad  \mbox{if}\quad  \phi_t=0, \\
		& \+V_t^{(w,\phi)} =\+V_0, \\
		& \+w^{(i)} \in \Omega_M, \, \forall i=1,...,t_0, \, \+w^{(i)} \,\mbox{is chosen with replacement}, \quad \+\phi \in \Omega_r\},
		\end{aligned}
		\end{equation*}
		For example, when $\phi =(0,1,0,1,0,1,...)$, $\+U^{(w,\phi)}_1$,..,$\+U^{(w,\phi)}_T$ is:
		$$
		\begin{bmatrix}\+U^{0} \\ \+V^0 \end{bmatrix},   
		\begin{bmatrix}\+U^{w^{(1)}} \\ \+V^0 \end{bmatrix},
		\begin{bmatrix} \+U^{0}, \\ \+V^0 \end{bmatrix},
		\begin{bmatrix} \+U^{w^{(2)}},\\ \+V^0 \end{bmatrix},
		\begin{bmatrix} 	\+U^{0}, \\ \+V^0 \end{bmatrix},
		\begin{bmatrix} \+U^{w^{(3)}} \\ \+V^0 \end{bmatrix},...
		$$
		For $\varTheta^{(w_1,\phi_1)}, \varTheta^{(w_2,\phi_2)} \in \Xi_\epsilon$, we have 
		\begin{equation*}
		d^2(\varTheta^{(w_1,\phi_1)}, \varTheta^{(w_2,\phi_2)}) = \sum_{t=1}^T d_0^2(\+U_t^{(w_1,\phi_1)}\+V^{0'}, \+U^{(w_2,\phi_2)}_t\+V^{0'}) \geq \frac{t_0 n_0(p-d+1)}{4} \epsilon^2 .
		\end{equation*}
		We have $$|\Xi_\epsilon| = M_0 M^{t_0}, \quad \mbox{with} \quad \log(|\Xi_\epsilon|) \geq c t_0 n_0 \log((n-d-1)/n_0) +c_0 t_0 \log(T/t_0). $$
		
		In addition, the KL divergence between any elements $\varTheta \in \Xi_\epsilon$ and $\varTheta^0$ can be upper bounded:
		\begin{equation*}
		D_{KL}(P_{\varTheta} \,||\, P_{\varTheta^0})  \leq C_0 d^2(\varTheta,\varTheta^0) \leq   C_0 t_0 n_0 (p-d-1)\epsilon^2,
		\end{equation*}
		for some constant $C_0>0$.

		Based on the construction, we have  for $\varTheta \in \Xi_\epsilon $, $\sum_{t=2}^{T} \ind\{D\+u_{it}\ne 0\} \leq 2 t_0$ for $i \in [n]$ and $\sum_{i=1}^{n} \ind\{D\+u_{it}\ne 0\} \leq 2 n_0$ for $t=2,...,T$. Therefore we have
		\begin{equation*}
		\sum_{i=1}^{n} \sum_{t=2}^{T}\ind\{D\+u_{it}\ne 0\} \leq 4 t_0 n_0.
		\end{equation*}
		We use the following Lemma~\ref{tsybakov} to show final proof that the minimax rate holds:
		\begin{lemma}[Theorem 2.5 in \cite{tsybakov2008introduction}]\label{tsybakov}
			Suppose $M \geq 2$ and $(\Theta,d)$ contains elements $\theta_0,...,\theta_M$ such that $d(\theta_i,\theta_j)\geq 2s >0$ for any $0\leq i\leq j \leq M$ and $\sum_{i=1}^{M}D_{KL}(P_{\theta_i},P_0)/M \leq \alpha \log M$ with $0<\alpha<1/8$. 
			Then we have
			$$\inf_{\hat{\theta}} \sup_{\theta \in \Theta} P_\theta (d(\hat{\theta},\theta)\geq s) \geq \frac{\sqrt{M}}{1+\sqrt{M}} \left(1-2\alpha-\sqrt{\frac{2\alpha}{\log M}} \right).$$
		\end{lemma}

		We consider different cases such that by choosing different $t_0$, $n_0$, the constraint $4 t_0 n_0 \leq s_u$ and the KL is upper bounded by the log cardinality (up to constant factor) can both be satisfied:
		\begin{equation}
		\begin{cases}
		4 t_0 n_0 &\leq s_u ;\\
		t_0 n_0 (p-d-1)\epsilon^2 &\lesssim t_0 n_0 \log((n-d-1)/n_0) + t_0 \log(T/t_0).
		\end{cases}
		\end{equation}
		
		{\bf Sparse rate ($s_u \leq T$):\\}
		When $s_u \leq T$.        Denote $n_0 =1$ and $t_0 =s_u/4$, which satisfies $4 n_0 t_0 \leq s_u$.  Then it's enough to set 
		\begin{equation*}
		s_u  (p-d-1) \epsilon^2 \lesssim    s_u \log (4T/s_u)+   s_u \log \left(n-d-1 \right).
		\end{equation*}

		{\bf Sparse rate ($s_u>T$):\\}
		First, suppose $(n-d-1)T/4>s_u>T$, we assume $s_u/T$ is an integer for simplicity.   
		Let   $n_0 =s_u/T$ and $t_0 =T/4$, which satisfies $4n_0 t_0 \leq s_u$.
		To adopt the above Lemma~\ref{tsybakov}, it suffices to show 
		\begin{equation*}
		n_0t_0(p-d-1) \epsilon^2 \leq  \alpha \log  (M_0M^{T/4})  =  \alpha \log (|\Xi_\epsilon|), 
		\end{equation*}
		with $\alpha < 1/8$. Then we only need to set 
		\begin{equation*}
		s_u  (p-d-1) \epsilon^2 \lesssim    T \log 4+  s_u \log \left(\frac{(n-d-1)T}{s_u} \right).
		\end{equation*}

		By taking $\epsilon =c^* \sqrt{\log (nT/s_u)/(p-d-1)}$ for some constant $c^*>0$ such that the above inequality is satisfied, we have
		\begin{equation*}
		\frac{t_0 n_0(p-d+1) \epsilon^2}{4} \gtrsim s_u \log(Tn/s_u).
		\end{equation*}

		By taking $\epsilon =c^* \sqrt{ \log (Tn/s_u)/(p-d-1)}$ for some constant $c^*>0$, we have
		\begin{equation*}
		\frac{t_0 n_0(p-d+1) \epsilon^2}{4} \gtrsim s_u \log(Tn/s_u).
		\end{equation*}

		{\bf Dense rate:\\}
		If $s_u \geq (n-d-1)T/4$, the difference between latent vectors is dense instead. Then we assign  $n_0 =(n-d-1)/2$ and $t_0 =T/2$, which satisfies the constraint. In this case,
		we need to set 
		\begin{equation*}
		(p-d-1)T  n \epsilon^2 \lesssim    T +   T (n-d-1).
		\end{equation*}
		
		By taking $\epsilon =c^* \sqrt{1/(p-d-1)}$ for some constant $c^*>0$ such that the above inequality is satisfied, we have
		\begin{equation*}
		\frac{t_0 n_0(p-d+1) \epsilon^2}{4} \gtrsim nT.
		\end{equation*}

		{\bf Initial estimation rate:\\}
		When the sparse rate is too small, then the above construction if not optimal, since we at least need to estimate the first matrix well without any sparse structures. Therefore, we consider we consider $T$ copies of the same matrix.    Note that the sparse constraint on the differences of the matrix is automatically satisfied when all matrices are the same. By constructing the following subset
		\begin{equation}
		\begin{aligned}
		\Xi_\epsilon =\{\varTheta^{(w)}:  \+U^{(w)}_t =\+U^{w}, \+V^{(w)}_t =\+V_0, \quad \forall t =[T], \,\+w \in \Omega_M\}.
		\end{aligned}
		\end{equation}
		the KL divergence between any elements $\varTheta\in \Xi_\epsilon$ and $\varTheta^0$ can be upper bounded:
		\begin{equation}
		D_{KL}(P_{\varTheta} \,||\, P_{\varTheta^0}) \lesssim T(p-d-1) \|\+w\|_2^2 \epsilon^2 \leq  T n_0 (p-d-1)\epsilon^2.
		\end{equation}
		Let $n_0$ be the largest possible integer less than or equal to $(n-d-1)$,
		then it suffices to let 
		\begin{equation*}
		T (n-d-1) (p-d-1) \epsilon^2 \lesssim n-d-1.
		\end{equation*}
		Therefore,  we can choose $\epsilon=\sqrt{1/((p-d-1)T)}$.  Then we have
		$$ d^2(\varTheta^{(w_1)}, \varTheta^{(w_2)}) \gtrsim T n_0(p-d+1) \epsilon^2 \gtrsim n.$$
		
		Finally, based on Markov's inequality, by combining all the above cases, we have
		
		\begin{align}
		\begin{aligned}
		\inf_{\hat{\varTheta}}\sup_{ \varTheta\in \mbox{DSF}(s_u,s_v)} \E_{\varTheta} \left[\frac{1}{Tnp} \sum_{t=1}^{T}\|\hat{\+U}_t\hat{\+V}_t'-\+U^*_t\+V^{*'}_t\|_F^2 \right] 
		\gtrsim \frac{1}{npT} \left\{ {s_u \log \left(\frac{Tn}{s_u}\right)}+ n\right\}.
		\end{aligned}
		\end{align}
		
		By the alternating technique: fixing $\m U$ and constructing a similar hypothesis set for $\m V$, we can also obtain the other rate of the lower bound 
		
		\begin{align}
		\begin{aligned}
		\inf_{\hat{\varTheta}}\sup_{ \varTheta\in \mbox{DSF}(s_u,s_v)} \E_{\varTheta} \left[\frac{1}{Tnp} \sum_{t=1}^{T}\|\hat{\+U}_t\hat{\+V}_t'-\+U^*_t\+V^{*'}_t\|_F^2 \right] 
		\gtrsim \frac{1}{npT} \left\{ {s_v \log \left(\frac{Tp}{s_v}\right)}+ p\right\}.
		\end{aligned}
		\end{align}
		Then the final conclusion is obtained.

		\subsection{Symmetric case}	
		In the symmetric case, we adopt a different hypothesis construction.	Let $\Omega_M =\{\+w^{(1)},...,\+w^{(M)}\}  \subset \{0,1\}^{(n-d+1)/2} $ constructed based on the above Lemma~\ref{lemma VG} (the construction holds under $n-d+1\geq 8$). For each $\+w$, we can construct a $n \times d$ matrix as follows:
		\begin{equation*}
		\+U^{w}=\begin{bmatrix}
		\+u^{w} &\+0 \\
		\+0  & \+I_{d-1}
		\end{bmatrix} \quad  \mbox{with} \quad \+u^w =\begin{bmatrix}
		1 \\...\\1 \\\epsilon \+w
		\end{bmatrix} \in \mathbb{R}^{n-d+1}, w \in \Omega_M
		\end{equation*}
		where the first $(n-d+1)/2$ components for $\+v^w$ are all ones.

		The effect of this construction is that: for different $\+w_1,\+w_2 \in \Omega_M$,  since $n_0 /2 \leq  \|\+w_1-\+w_2\|_0 \leq 2n_0$ and $\|\+U^w\|_F \leq \sqrt{n}$, we have
		\begin{align*}
		d_0(\+U^{w_1}, \+U^{w_2}) \leq\|\+U^{w_1}\+U^{'w_1}-\+U^{w_2}\+U^{'w_2} \|_F \leq \|\+U^{w_1}(\+U^{'w_1}-\+U^{'w_2} )\|_F+\|(\+U^{w_1}-\+U^{w_2})\+U^{'w_2} \|_F \\
		\leq 2\sqrt{n}\|\+U^{w_1}-\+U^{w_2}\|_F  = 2\sqrt{n}\|\+u^{w_1}-\+u^{w_2}\|_2\leq 2\sqrt{{2nn_0}}\epsilon.
		\end{align*} 
		In addition, consider $A:=\{i+(n-d+1)/2 : w_{1i} \ne 0\}$, $B:= \{j+(n-d+1)/2:w_{2j} \ne 0\}$, $C:=A \cap B$, where $w_{1i},w_{2j}$ are $i$ and $j$th component of $w_1$ and $w_2$. We have $|C| \leq n_0/2$, $|A-C|\geq n_0/2$ and $|B-C|\geq n_0/2$.
		$$d^2_0(\+U^{w_1}, \+U^{w_2}) = \sum_{i\ne j}^{n-d+1}( u^{w_1}_{i} u^{w_1}_{j}- u^{w_{2}}_i u^{w_2}_j  )^2.$$
		By only considering the sum for $i  \in \{1,...,(n-d+1)/2\}, j\in A-C$ where $u_i^{w_1}=u_i^{w_2} = 1$ $u^{w_1}_{j} =\epsilon $, $u^{w_2}_j =0$ and $i \ne j$,  we have
		\begin{align*}
		d^2_0(\+U^{w_1}, \+U^{w_2})  \geq \sum_{i=1}^{(n-d+1)/2}\sum_{j \in A-C}(   \epsilon {w_1}_{j}-  \epsilon {w_2}_j  )^2\geq \frac{n_0(n-d+1)}{4} \epsilon^2.
		\end{align*}

		Let 
		\begin{equation*}
		\+U^{0}=\begin{bmatrix}
		\+0 &\+0 \\
		\+0  & \+I_{d-1}
		\end{bmatrix} \quad  \mbox{with} \quad \+v^{w_0} =\begin{bmatrix}
		1 \\...\\1 \\ \+w_0
		\end{bmatrix} \in \mathbb{R}^{n-d+1}
		\end{equation*}
		and $\m U^0= [\+U^0,...,\+U^0]$.
		We  need the above  Varshamov-Gilbert Bound~\ref{lemma VG} again to introduce another binary coding: Let $\Omega_r =\{\+\phi^{(1)},...,\+\phi^{(M_0)}\}  \subset \{0,1\}^{T} $, such that $\|\+\phi^{(i)}\|_0=t_0$  for all $1\leq i \leq M_0$  and $\|\+\phi^{(i)}-\+\phi^{(j)}\|_0 \geq t_0/2$ for $0\leq i <j \leq M_0 $ with $\log M_0 \geq ct_0 \log( T/t_0)$ with $c \geq 0.233$. 
		
		Then we have the following construction:
		\begin{equation*}
		\begin{aligned}
		\Xi_\epsilon =\{\m U^{(w,\phi)}:&\\
		&  \+U^{(w,\phi)}_t =\+U^{w^{(i)}}, \quad  \quad \mbox{if}\quad  \phi_t=1,\\
		&\+U^{(w,\phi)}_t =\+U^{0}, \quad  \quad  \mbox{if}\quad  \phi_t=0, \\
		& \+w^{(i)} \in \Omega_M, \, \forall i=1,...,t_0, \, \+w^{(i)} \,\mbox{is chosen with replacement}, \quad \+\phi \in \Omega_r\},
		\end{aligned}
		\end{equation*}
		For example, when $\phi =(0,1,0,1,0,1,...)$, $\+U^{(w,\phi)}_1$,..,$\+U^{(w,\phi)}_T$ is:
		$$
		\+U^{0},\+U^{w^{(1)}},\+U^{0},\+U^{w^{(2)}},\+U^{0},\+U^{w^{(3)}},...
		$$
		
		for $\m U^a, \m U^b \in \Xi_\epsilon$, we have 
		\begin{equation*}
		d^2(\m U^a, \m U^b) = \sum_{t=1}^T d_0^2(\+U^a_t, \+U^b_t) \geq \frac{t_0 n_0(n-d+1)}{8} \epsilon^2 .
		\end{equation*}
		We have $|\Xi_\epsilon| = M_0 M^{t_0} $. 
		
		In addition, the KL divergence between any elements $\m U \in \Xi_\epsilon$ and $\m U^0$ can be upper bounded:
		\begin{equation*}
		D_{kl}(P_{\m U} \,||\, P_{\m U^0})  \leq C_0 d^2(\m U,\m V) \leq 4 C_0 t_0 n_0 n\epsilon^2,
		\end{equation*}
		for some constant $C_0>0$ ($C_0=1$ for the binary case, $C_0=1/(2\sigma^2)$ for the Gaussian case).

		Based on the construction, we have  for $\m U\in \Xi_\epsilon $, $\sum_{t=2}^{T} \ind\{D\+u_{it}\ne 0\} \leq 2 t_0$ for $i \in [n]$ and $\sum_{i=1}^{n} \ind\{D\+u_{it}\ne 0\} \leq 2 n_0$ for $t=2,...,T$. Therefore we have
		\begin{equation*}
		\sum_{i=1}^{n} \sum_{t=2}^{T}\ind\{D\+u_{it}\ne 0\} \leq 4 t_0 n_0.
		\end{equation*}
		
		We use the following Lemma~\ref{tsybakov} to show finally prove that the minimax rate holds:

		{\bf Sparse rate ($s\leq T$):\\}
		When $s \leq T$.        Denote $n_0 =1$ and $t_0 =s/4$, which satisfies $4 n_0 t_0 \leq s$.  Then it's enough to set 
		\begin{equation*}
		s  n \epsilon^2 \leq   \alpha c s \log (4T/s)+ \alpha c  s \log \left(\frac{n-d-1}{2} \right)/8
		\end{equation*}
		with $c=0.233/C_0$. 
		
		By taking $\epsilon =c^* \sqrt{s \log (Tn/s)/n}$ for some constant $c^*>0$ such that the above inequality is satisfied, we have
		\begin{equation*}
		\frac{t_0 n_0(n-d+1) \epsilon^2}{8} \gtrsim s \log(Tn/s).
		\end{equation*}
		
		{\bf Sparse rate ($s>T$):\\}
		First, suppose $(n-d-1)T/8>s>T$, we assume $s/T$ is an integer for simplicity since it doesn't affect the final rate.   
		
		Denote $n_0 =s/T$ and $t_0 =T/4$, which satisfies $4n_0 t_0 \leq s$. Since  $s/T \leq (n-d-1)/8$, the construction holds given.
		To adopt the above Lemma~\ref{tsybakov}, it suffices to show 
		\begin{equation*}
		2 C_0 s  n \epsilon^2/4 \leq  \alpha \log  (M_0M^{T/4})  =  \alpha \log (|\Xi_\epsilon|),  
		\end{equation*}
		with $\alpha < 1/8$. Then we need to set 
		\begin{equation*}
		s  n \epsilon^2 \leq   \alpha c T \log 4+ \alpha c  s \log \left(\frac{(n-d-1)T}{2s} \right)/2
		\end{equation*}
		with $c=0.233/C_0$. 
		
		By taking $\epsilon =c^* \sqrt{\log (nT/s)/n}$ for some constant $c^*>0$ such that the above inequality is satisfied, we have
		\begin{equation*}
		\frac{t_0 n_0(n-d+1) \epsilon^2}{8} \gtrsim s \log(Tn/s).
		\end{equation*}
		
		{\bf Dense rate:\\}
		If $s\geq (n-d-1)T/8$, the difference between latent vectors is dense instead. Then we assign  $n_0 =(n-d-1)/8$ and $t_0 =T/4$, which satisfies the constraint. In this case,
		we need to set 
		\begin{equation*}
		(n-d-1)T/32  n \epsilon^2 \leq   \alpha c T \log 4+ \alpha c  T (n-d-1)\log 4/4
		\end{equation*}
		with $c=0.233/C_0$. 
		
		By taking $\epsilon =c^* \sqrt{1/n}$ for some constant $c^*>0$ such that the above inequality is satisfied, we have
		\begin{equation*}
		\frac{t_0 n_0(n-d+1) \epsilon^2}{8} \gtrsim nT.
		\end{equation*}

		{\bf Initial estimation rate:\\}
		When the sparse rate is too small, then the above construction if not optimal, since we at least need to estimate the first matrix well without any sparse structures. Therefore, we consider we consider $T$ copies of the same matrix.    Note that the sparse constraint on the differences of the matrix is automatically satisfied when all matrices are the same. By constructing the following subset
		\begin{equation}\label{eq:construct2}
		\begin{aligned}
		\Xi_\epsilon =\{\m U^{(w)}:  \+U^{(w)}_t =\+U^{w}, \quad \forall t =[T], \,\+w \in \Omega_M\}.
		\end{aligned}
		\end{equation}
		the KL divergence between any elements $\m U\in \Xi_\epsilon$ and $\m U^0$ can be upper bounded:
		\begin{equation}\label{eq:KL_2}
		D_{KL}(P_{\m U} \,||\, P_{\m U^0}) \leq  C_0d_0^2(\+U_t, \+V_t)  \leq 4 T n (n-d-1)\epsilon^2.
		\end{equation}
		Let $n_0$ be the largest possible value $(n-d-1)/8$,
		then it suffices to let 
		\begin{equation*}
		4 T n (n-d-1) \epsilon^2 \leq  \frac{c (n-d-1) \log 4}{16} \leq \alpha \log  (M) .
		\end{equation*}
		Therefore, based on the above equation, we need to choose $\epsilon=\sqrt{1/(nT)}$.  Then we have
		$$ \frac{t_0 n_0(n-d+1) \epsilon^2}{8} \gtrsim n.$$
		
		Finally, based on Markov's inequality, by combining all the above cases the final conclusion holds.

	\end{proof}

	\subsection{Proof of Theorem~\ref{thm:posterior}}\label{sec:proof_thm1}

	Denote the $\epsilon$ ball for KL divergence neighborhood centered at $\varTheta^*$ as
	\begin{equation*}
	B_n(\varTheta^*;\epsilon) = \left\{\varTheta\in \Xi: \int p_{\varTheta^*}\log(\frac{p_{\varTheta^*}}{p_{\varTheta}}) d\mu \leq npT\epsilon^2, \int p_{\varTheta^*}\log^2(\frac{p_{\varTheta^*}}{p_{\varTheta}}) d\mu \leq npT\epsilon^2\right\},
	\end{equation*}
	where $\mu$ is the Lebesgue measure.

	Based on Assumption $1$, we have $\max\{D_{KL}(p_{\varTheta},p_{\varTheta^*}),V_2(p_{\varTheta},p_{\varTheta^*})\}\lesssim \sum_{t=1}^{T} \|\+U_{t}\+V'_t-\+U_t^* \+V_t^{*'}\|_F^2$.  Hence we only need to lower bound the prior probability of the set $E_0: =\{\sum_{i,j,t}(\+u_{it}'\+v_{jt}-\+u^{*'}_{it}\+v^{*}_{jt})^2\leq npT\epsilon^2\} \supset \{\max_{i,j,t}(\+u_{it}'\+v_{jt}-\+u^{*'}_{it}\+v^{*}_{jt})^2 \leq \epsilon^2\} $. Given $i, j, t$ we have
	\begin{equation*}
	\begin{aligned}
	|\+u_{it}'\+v_{jt}-\+u^{*'}_{it}\+v^{*}_{jt}| &\leq |(\+u'_{it}-\+u^{*'}_{it})\+v^{*}_{jt}|+|\+u'_{it}(\+v_{jt}-\+v^{*}_{jt})|= |(\+u'_{it}-\+u^{*'}_{it})\+v^{*}_{jt}|+|(\+u'_{it}-\+u^{*'}_{it}+\+u^{*'}_{it})(\+v_{jt}-\+v^{*}_{jt})| \\
	&\leq  \|\+u_{it}-\+u^{*}_{it}\|_2\|\+v^*_{jt}\|_2 + \|\+v_{jt}-\+v^{*}_{jt}\|_2\|\+u^*_{it}\|_2 + \|\+u_{it}-\+u^{*}_{it}\|_2\| \|\+v_{jt}-\+v^{*}_{jt}\|_2
	\end{aligned}
	\end{equation*}
	Note that $\max\{\+u^*_{it},\+v^*_{jt}\} \leq C$ for a constant $C$ based on assumption. Then when $\epsilon=o(1)$,  $\max_{it} \|\+u_{it}-\+u^{*}_{it}\|_2 \leq \epsilon/3C $ and $\max_{jt} \|\+v_{jt}-\+v^{*}_{jt}\|_2 \leq \epsilon/3C $, we have
	\begin{equation*}
	\max_{i,j,t}|\+u_{it}'\+v_{jt}-\+u^{*'}_{it}\+v^{*}_{jt}| \leq \frac{2\epsilon}{3}+\frac{\epsilon^2}{9C^2} \leq \epsilon, 
	\end{equation*}
	for large enough $n,p,T$. We first show the prior concentration for $\max_{it} \|\+u_{it}-\+u^{*}_{it}\|_2 \leq \epsilon/3C $, the prior concentration for $\max_{jt} \|\+v_{jt}-\+v^{*}_{jt}\|_2 \leq \epsilon/3C $ will be similarly derived.
	
	Let $\epsilon_u = \sqrt{(s_u+n)\log (npT)/(npT)}$, then $npT\epsilon_u^2 =(s_u+n)\log (npT)$, suppose for subjects $i$ the sparsity is $s_{ui}$ such that $\sum_{i=1}^{n} s_{ui} =s_u$.
	
	Denote $E_0 = \{\max_{it} \|\+u_{it}-\+u^{*}_{it}\|_2 \leq C_0\epsilon/3C\} $ for some large enough constant $C_0>0$,    $E_1 = \{\max_{i,j,t} |(U_{ijt}-U_{ij1})-(U^{*}_{ijt}-U^*_{ij1})| \leq \epsilon_{ui} \}$, $E_2 = \{\max_{i,j} |U_{ij1}-U^*_{ij1}| \leq \epsilon_{u}\}$ with $ \epsilon_{ui}= \sqrt{(s_{ui}+1) \log(npT)/\{3C\sqrt{d}npT\}}$.  Note that $\sum_{i=1}^{n} \epsilon_{ui}^2 \lesssim \epsilon^2_{u}$, so  we have $E_1\bigcap E_2 \subset E_0 $ as long as $C_0$ is a large enough constant, which implies that
	\begin{equation*}
	\begin{aligned}
	\Pi(E_0) \geq \Pi\left(E_1\right) \Pi\left(E_2\right) 
	=  \prod_{i,j}\Pi\left(\sup_{t \geq 2}|\tilde U_{ijt}-\tilde U^{*}_{ijt}|\leq {\epsilon_{ui}}\right)\prod_{i,j}\Pi\left(|U_{ij1}-U^{*}_{ij1}|\leq {\epsilon_{u}}\right),
	\end{aligned}
	\end{equation*}
	where $\tilde U_{ijt} = U_{ijt}-U_{ij1}$ for all $i,j,t$. Note that we have by triangular inequality:
	\begin{equation*}
	|\tilde U_{ijt}-\tilde U^{*}_{ijt}| \leq \sum_{t_0=2}^{t} |\tilde U_{ijt_0}-\tilde U_{ij(t_0-1)}- \left(\tilde U^{*}_{ijt_0}-\tilde U^{*}_{ij(t_0-1)}\right)|.
	\end{equation*}
	Therefore, based on Lemma~\ref{lem:horse_net} and $\epsilon_{ui} = O(\sqrt{\max\{s_{ui},1\} \log(npT)/npT})$, and the choice of $\tau_u$, we have
	\begin{equation*}
	\begin{aligned}
	&\Pi\left(\sup_{t \geq 2}|\tilde U_{ijt}-\tilde U^{*}_{ijt}| \leq {\epsilon_{ui}}\right) \\
	&\geq \Pi\left(\sum_{t=2}^{T} |\tilde U_{ijt}-\tilde U_{ij(t-1)}- \left(\tilde U^{*}_{ijt}-\tilde U^{*}_{ij(t-1)}\right)|\leq  C_2\sqrt{ \frac{\max\{s_{ui},1\} \log(npT)}{npT}}\right) \\
	&\gtrsim e^{-K (s_i+1) \log(npT)},
	\end{aligned}
	\end{equation*}
	for some constant $C_2,K>0$, $i \in [n]$ and $j \in [d]$, since each $\tilde U_{ijt}-\tilde U_{ij(t-1)}$ is assigned with a component of the shrinkage prior. This implies that
	\begin{equation*}
	\Pi(E_1) \gtrsim e^{-K(s_u+n) \log (npT)}.
	\end{equation*}

	Let $\sigma_{0}^*=1$, we consider the event on the initial variance $E_3 =\{|\sigma_{0i}^2-\sigma_0^{*2}|\leq \sigma_{0}^{*2}/2\}$. Note that the density of Inverse-Gamma$(a_{\sigma_0},b_{\sigma_0})$ is lower bounded by a constant within the constrained space $[\sigma_0^{*2}/2,3\sigma_0^{*2}/2]$. Therefore we have the prior concentration:
	\begin{align*}
	\Pi(E_3) = \prod_{i=1}^n\Pi\left(|\sigma_{0i}^2-\sigma_0^{*2}|\leq \sigma_{0}^{*2}/2\right) \gtrsim e^{-K_2 n},
	\end{align*}
	for some constant $K_2>0$.
	
	Then for the initial error concentration $\Pi(E_2)$, by the mean-zero Gaussian of $U_{ij1}$ for all $i,j$, we have the concentration:
	\begin{equation}\label{eq:static}
	\begin{aligned}
	\Pi(E_2) = \Pi(E_2 \mid E_3)\Pi(E_3) &=\prod_{i,j}\Pi\left(|U_{ij1}-U^{*}_{ij1}|\leq {\epsilon_{u}}\mid E_3 \right )\Pi(E_3) \\
	&\gtrsim \frac{1}{(\sqrt{2\pi}\sigma_0^*/\sqrt{2})^{nd}}\exp(-\sum_{i,j} \frac{U^{*2}_{ij1}}{2\sigma_0^{*2}}) (2\epsilon_{u})^{nd} e^{-K_2 n}\\
	& \gtrsim \exp \left[-K_3 \left\{\frac{\|\+U^*_1\|_F^2}{2 \sigma_0^{*2}}+nd+nd \log( \frac{1}{\epsilon_{u}}) \right\}\right].
	\end{aligned}
	\end{equation}
	for some constant $K_3>0$. Note that $\sigma_0^{*}$ is a constant and $\|\+U^*_1\|_F^2 = O(n)$. We have $\log \Pi(E_2) \gtrsim -n \log( {1/\epsilon_u}) $ and $\log \Pi(E_0) \gtrsim -(s_u+n) \log( {1/\epsilon_u}) $. 
	
	Similarly, by the same technique, we can show that $\log \Pi(\max_{jt} \|\+v_{jt}-\+v^{*}_{jt}\|_2 \leq \epsilon_v) \gtrsim -(s_v+p) \log( {1/\epsilon_v}) $ for $\epsilon_v= \sqrt{(s_v+p) \log(npT)/(npT)}$. Finally, by choosing $\epsilon_{n,p,T} = \sqrt{\epsilon^2_u+\epsilon^2_v} $, we have
	\begin{equation*}
	\begin{aligned}
	\log \Pi(\max_{it} \|\+u_{it}-\+v^{*}_{it}\|_2 \leq \epsilon_{n,p,T}, \, \max_{jt} \|\+v_{jt}-\+v^{*}_{jt}\|_2 \leq \epsilon_{n,p,T} )
	\gtrsim -(s_u+s_v+n+p) \log(npT).
	\end{aligned}
	\end{equation*}
	Note that for the symmetric case, the prior concentration still holds by only considering the concentration for $\m U$: $\log \Pi(\max_{it} \|\+u_{it}-\+v^{*}_{it}\|_2 \leq \epsilon_{n,p,T}) \gtrsim - (s_u+n) \log(n^2T) $.  
	
	 Finally, as shown in Theorem 3.2 in \cite{bhattacharya2019bayesian}, under the condition that 
	\begin{equation*}
	\Pi(B_n(\varTheta^*;\epsilon_{n,p,T})) \geq e^{-Tnp\epsilon_{n,p,T}^2}, 
	\end{equation*}
	we can obtain the convergence of the $\alpha$-divergence at the rate:
	\begin{equation*}
	D_\alpha(p_{\varTheta},p_{\varTheta^*}) = \frac{1}{\alpha-1} \log \int (p_{\varTheta^*})^\alpha (p_{\varTheta})^{1-\alpha} d\mu.
	\end{equation*}
	
	Therefore, the conclusion holds that the $\alpha$-divergence is lower bound by the loss function under Assumption 2. 
	
		\subsection{Proof of Corollary~\ref{thm:dynamical_comparison}}\label{sec:proof_dynamic_comparison}

					We first show that the static rate can be achieved marginally under the same prior for any time. Given a time point $t$, the goal is to show the rate of estimating $\|\hat{\+U}_t \hat{\+U}'_t- \+U^*_t \+U^{*'}_t\|^2_F$ under the proposed prior. Note that the prior for each component of $\+u_{it}$ is $\m N(0,\sigma_{it}^2)$ conditional on $\sigma_{0i},  \lambda_{it_0}, t_0=1,...,t$, with $\sigma_{it}^2= \sigma_{0i}^2+\sum_{t_0=1}^{t-1} \lambda^2_{it0} \tau_u^2$. We can consider the above prior in estimating $t_0$ copies of $ \+U^*_t$: for time point $1,..,t$, the true latent matrices are $\+U^*_t,\+U^*_t,...,\+U^*_t$, where the sparsity of the transitions is zero. Note that the prior concentration of the proposed prior on the truth is $(n+s) \log(nT)$, while now $s=0$. This implies that the prior concentration for $\|\hat{\+U}_t \hat{\+U}'_t- \+U^*_t \+U^{*'}_t\|^2_F$ is $n \log (nT)$, by the same argument with proof of the dynamic rate, we have  for large enough $n$, we have $\|\hat{\+U}_t \hat{\+U}'_t- \+U^*_t \+U^{*'}_t\|^2_F \lesssim n \log (nT)$ for any $t$.

	To prove the conclusion, we start with $t=1$, corresponding to the truth $\+U_1^*,\+U^*_2$. 	For any $t_0=[T]$, we denote 
	\begin{equation*}
	\+O_{t_0} = \argmin_{\+O \in \mathbb{O}^{d \times d}} \|\hat{\+U}^o_{t_0} - \+U^*_{t_0} \+O \|_F.
	\end{equation*}
	First, note that $\hat{\+U_1^o} = \hat{\+U}_1$, consider the common matrix $\+O_1$, we have
	\begin{align*}
	\|\hat{\+U}^o_2 \hat{\+O}_{21} - \+U^*_2\+O_1\|_F &\leq \|\hat{\+U}^o_2 \hat{\+O}_{21}- \hat{\+U}^o_1\|_F +  \|\hat{\+U}^o_1-\+U^*_1\+O_1\|_F +\|\+U^*_1\+O_1-\+U^*_2\+O_1\|_F \\
	&\leq \|\hat{\+U}^o_2 \hat{\+O}_{21}- \hat{\+U}^o_1\|_F + \|\hat{\+U}^o_1-\+U^*_1\+O_1\|_F+\|\+U^*_1-\+U^*_2\|_F,
	\end{align*}
	and
	\begin{align*}
	\|\hat{\+U}^o_2 \hat{\+O}_{21}-\hat{\+U}_1^o\|_F &\leq \|\hat{\+U}^o_2 \+O_2'-\hat{\+U}_1^o \+O_1'\|_F  \\
	&\leq \|\hat{\+U}^o_2 \+O_2'-\+U^*_2\|_F  + \|\+U_2^*-\+U_1^*\|_F +\|\hat{\+U}_1^o \+O_1'- \+U^*_1\|_F \\
	&= \|\hat{\+U}^o_2 -\+U^*_2\+O_2\|_F  + \|\+U_2^*-\+U_1^*\|_F +\|\hat{\+U}_1^o- \+U^*_1 \+O_1\|_F.
	\end{align*}
	The above two inequalities imply that
	\begin{equation*}
	\|\hat{\+U}^o_2 \hat{\+O}_{21} - \+U^*_2\+O_1\|_F \leq 2\|\hat{\+U}^o_1-\+U^*_1\+O_1\|_F + \|\hat{\+U}^o_2-\+U^*_2\+O_2\|_F +2\|\+U^*_2-\+U^*_1\|_F .
	\end{equation*}

	Therefore, similarly, for $k \geq 3$, we have
	\begin{equation*}
	\begin{aligned}
	\|\hat{\+U}^o_k \hat{\+O}_{k1} - \+U^*_{k}\+O_1\|_F &\leq \|\hat{\+U}^o_k -\+U^*_k\+O_k\|_F + 2\|\+U^*_k-\+U^*_1\|_F +2\|\hat{\+U}^o_1- \+U^*_1 \+O_1\|_F,
	\end{aligned}
	\end{equation*}
	which implies that
		\begin{equation*}
	\begin{aligned}
	\|\hat{\+U}^o_k \hat{\+O}_{k1} - \+U^*_{k}\+O_1\|^2_F &\leq 2\|\hat{\+U}^o_k -\+U^*_k\+O_k\|^2_F + 16\|\+U^*_k-\+U^*_1\|^2_F +16\|\hat{\+U}^o_1- \+U^*_1 \+O_1\|^2_F.
	\end{aligned}
	\end{equation*}
	Therefore, by induction, we have
			\begin{equation*}
	\begin{aligned}
	\sum_{k_0=1}^{k}\|\hat{\+U}^o_{k_0} \hat{\+O}_{k_01} - \+U^*_{k_0}\+O_1\|^2_F &\leq 2\sum_{k_0=1}^{k}\|\hat{\+U}^o_{k_0} -\+U^*_{k_0}\+O_{k_0}\|^2_F + 16\sum_{k_0=1}^{k}\|\+U^*_{k_0}-\+U^*_1\|^2_F +16 k\|\hat{\+U}_1- \+U^*_1 \+O_1\|^2_F.
	\end{aligned}
	\end{equation*}
	
		For large enough $n$, we have $\|\hat{\+U}_t \hat{\+U}'_t- \+U^*_t \+U^{*'}_t\|_F$. By the eigenvalue condition, we have $\|\+U^{*\dagger}_t\| \lesssim 1/\sqrt{n}$ where $\+U^{*\dagger}_t$ is the psedo-inverse of $\+U^{*}_t$. The above two conclusion implies that $\|\+U^{*\dagger}_t\| \sqrt{\|\hat{\+U}_t \hat{\+U}'_t- \+U^*_t \+U^{*'}_t\|_F } \leq 1/2$ for large enough $n$. 
	Therefore, by Lemma~\ref{lem:perturbation}, we have:
	
	\begin{equation*}
	\|\hat{\+U}^o_t- \+U^*_t \+O_t \|_F \lesssim   \|\hat{\+U}_t \+U'_t- \+U^*_t \+U^{*'}_t\|_F/\sqrt{n} \lesssim \sqrt{\log nT}.
	\end{equation*}


For the rest of the windows (e.g., $\+U^*_{k+1},...,\+U^*_{2k-1}$),  the proof follows the same technique, with the common orthogonal transformation $\+O_{k}$.
Then when the average error is considered, note that 
	\begin{equation*}
	\begin{aligned}
\sum_{t=1}^{\bar t} \sum_{k_0=1}^{k}\|\hat{\+U}^o_{(t-1)k+k_0} -\+U^*_{(t-1)k+k_0}\+O_{(t-1)k+1}\|_F^2 \\
\lesssim \sum_{t=1}^{\bar t} \sum_{k_0=1}^{k}\|\hat{\+U}^o_{(t-1)k+k_0} -\+U^*_{(t-1)k+k_0}\+O_{(t-1)k+k_0}\|^2_F\\
 +\sum_{t=1}^{\bar t} \sum_{k_0=1}^{k}\|\+U^*_{(t-1)k+k_0}-\+U^*_{(t-1)k+1}\|^2_F +\sum_{t=1}^{\bar t}   k\|\hat{\+U}^o_{(t-1)k+1}- \+U^*_{(t-1)k+1} \+O_{(t-1)k+1}\|^2_F.
	\end{aligned}
	\end{equation*}
	First, by the estimation error rate, we have  \begin{align*}
	    \sum_{t=1}^{\bar t}    k\|\hat{\+U}^o_{(t-1)k+1}- \+U^*_{(t-1)k+1} \+O_{(t-1)k+1}\|^2_F \\\lesssim \min \{\sum_{t=1}^{T} k \|\hat{\+U}_{t}- \+U^*_{t} \+O_{t}\|^2_F,T \log(nT)\} \lesssim \min\{k(s+n) \log(nT)/n, T \log(nT)\};
	\end{align*}
	and
	\begin{align*}
	\sum_{t=1}^{\bar t} \sum_{k_0=1}^{k}\|\hat{\+U}^o_{(t-1)k+k_0} -\+U^*_{(t-1)k+k_0}\+O_{(t-1)k+k_0}\|^2_F \\
	\lesssim \sum_{t=1}^{T} 2 \|\hat{\+U}_{t}- \+U^*_{t} \+O_{t}\|^2_F \lesssim (s+n) \log(nT)/n.
		\end{align*}
	Suppose the true transitions are $s_1,...s_{\bar t}, t=1,...,\bar t$ in each window, then $\sum_{t=1}^{\bar t} s_{t} = s$, then we have for any $t,k,k_0>1$,
	$\|\+U^*_{(t-1)k+k_0}-\+U^*_{(t-1)k+1}\|_F \lesssim \min\{\sqrt{s_t k},\sqrt{s},\sqrt{n} \}$.
	Therefore,  we have
	$$ \sum_{t=1}^{\bar t} \sum_{k_0=1}^{k}\|\+U^*_{(t-1)k+k_0}-\+U^*_{(t-1)k+1}\|^2_F \lesssim \min \{s k(k-1),Tn,sT \}.$$
	
	Therefore, the final conclusion holds by aggregating the corresponding three terms.

	\subsection{An improved rate for Corollary~\ref{thm:dynamical_comparison}}\label{sec:improved_rate}
	
	 The rate in Corollary~\ref{thm:dynamical_comparison} is not optimal in some special cases. Consider that there is only one node change $O(T)$ times with $T>n$, then the best rate of $k$ could only approach $O(\sqrt{n})\leq O(\sqrt{T})$. This does not match our intuition that when almost all nodes do not change, it should be no problem to compare all latent positions across all time. To handle this case, we consider the following more stringent assumption:

 \begin{assumption}[Active nodes constraint]\label{asm:active_node}
 There is a $s_{0,k} = o(n)$, such that
\begin{equation}
    \max_{k_0=1}^k \max_{t=1}^T \|\+U^*_{(t-1)k+k_0} -\+U^*_{(t-1)k+1}\|_{2,0} \leq s_{0,k}.
\end{equation}
\end{assumption}

The Assumption~\ref{asm:active_node} assumes that within a window for $k_0=1,...,k$, only at most $2s_{0,k}$ of $n$ nodes transit over time, which are therefore considered as `active nodes', while the rest of nodes remain static. However, the active nodes can transit over all time points.

\begin{corollary}\label{thm:active_node}
Under the assumptions in \ref{thm:dynamical_comparison}, if we further assume that Assumption~\ref{asm:active_node} holds, then we have
	\begin{equation*}
		  \inf_{\bar {\+O}_{t,k} \in \mathbb{O}^{d \times d}}\sum_{k_0=1}^{k}\|\hat{\+U}^o_{(t-1)k+k_0} -\+U^*_{(t-1)k+k_0} \bar{\+O}_{t,k}\|_F^2\lesssim k \log nT+ s_{0,k}(k-1),
		\end{equation*}
		
	and	the average error for all windows satisfies
		\begin{equation*}
		\begin{aligned}
		\frac{1}{nT}\sum_{t=0}^{\bar t-1}\inf_{\bar {\+O}_{t,k} \in \mathbb{O}^{d \times d}}\sum_{k_0=1}^{k}\|\hat{\+U}^o_{(t-1)k+k_0} -\+U^*_{(t-1)k+k_0} \bar{\+O}_{t}\|_F^2\lesssim \frac{  (s+n) \log(nT)}{n^2T}\\
		+\min \left\{\frac{ k(s+n) \log(nT)}{n^2T},  \frac{\log(nT)}{n} \right\}+\min\left\{\frac{s k (k-1)}{nT},\frac{s_{0,k}}{n} \right\}.
		\end{aligned}
		\end{equation*}
\end{corollary}
According to Corollary~\ref{thm:dynamical_comparison}, as long as $s_{0,k}=o(n)$ (which is possible when $s=o(nT)$, therefore improved upon Theorem~\ref{thm:dynamical_comparison}), then one can compare all the latent space for time points of length $k=O(T) $ simultaneously, which can be intuitively explained that when the majority of nodes do not move across all time, there is no need to adjust the axis within all the time period.
	
	\begin{proof}
	    	Under Assumption~\ref{thm:active_node}, we can improve the following bound: for any $t,k,k_0>1$,
	$\|\+X^*_{(t-1)k+k_0}-\+X^*_{(t-1)k+1}\|_F \lesssim \sqrt{s_t k},\sqrt{s_{0,k}} \}$ in the final step in the proof in subsection~\ref{sec:proof_dynamic_comparison}.
	Therefore,  we have
	$$ \sum_{t=0}^{\bar t} \sum_{k_0=1}^{k}\|\+X^*_{(t-1)k+k_0}-\+X^*_{(t-1)k+1}\|^2_F \lesssim \min \{s k(k-1),s_{0,k}T \},$$ and the rest follows similarly.

	\end{proof}

		\subsection{Proof of Theorem~\ref{thm:single_cluster}}\label{sec:proof_thm:single_cluster}

	For any $t \in [T]$, denote 
	\begin{equation*}
	\+O_t = \argmin_{\+O \in \mathbb{O}^{d \times d}} \|\hat{\+U}_t- \+U^*_t \+O \|_F.
	\end{equation*}
	Note that $\+U^*_t \+O$ has the same cluster configuration with $\+U^*_t$. Note that for large enough $n$, we have $\|\hat{\+U}_t \hat{\+U}'_t- \+U^*_t \+U^{*'}_t\|_F \lesssim \sqrt{n \log nT}$ by the posterior concentration for the static latent space model. By assumption 2, we have $\|\+U^{*\dagger}_t\| \lesssim 1/\sqrt{n}$ where $\+U^{*\dagger}_t$ is the psedo-inverse of $\+U^{*}_t$. The above two conclusion implies that $\|\+U^{*\dagger}_t\| \sqrt{\|\hat{\+U}_t \+U'_t- \+U^*_t \+U^{*'}_t\|_F } \leq 1/2$ for large enough $n$. 
	Therefore, by Lemma~\ref{lem:perturbation}, we have:
	
	\begin{equation*}
	\|\hat{\+U}_t- \+U^*_t \+O_t \|_F \lesssim   \|\hat{\+U}_t \+U'_t- \+U^*_t \+U^{*'}_t\|_F/\sqrt{n} \lesssim \sqrt{\log nT}.
	\end{equation*}
	Note that the smallest cluster size is greater than $16 \|\hat{\+U}_t- \+U^*_t \+O_t \|_F^2 /\delta^2$.	By lemma~\ref{lem:k_means}, the above bounds implies that by applying $K_t$ means on $\hat{\+U}_t$ to obtain $\hat{\+\Xi}_t$, we have
	\begin{equation*}
	L(\hat{\+\Xi}_t,\+\Xi_t) \leq 16 \|\hat{\+U}_t- \+U^*_t \+O_t \|_F^2 /\delta^2.
	\end{equation*}
	The above two inequalities lead to 
	\begin{equation*}
	\frac{\sum_{t=1}^{T} L(\hat{\+\Xi}_t,\+\Xi_t)}{T} \lesssim  \frac{\sum_{t=1}^{T} \|\hat{\+U}_t \hat{\+U}'_t- \+U^*_t \+U^{*'}_t\|^2_F}{n T \delta ^2}  \lesssim \frac{(s+n) \log (nT)}{nT  \delta^2}.
	\end{equation*}
	Therefore, the final conclusion holds.

\subsection{Theoretical results for sparse networks}\label{sec:beta_unknown}

       \begin{proof}
Based on the constant smoothness of logistic link function, we still have\\ $\max\{D_{KL}(p_{\varTheta},p_{\varTheta^*}),V_2(p_{\varTheta},p_{\varTheta^*})\}\lesssim \sum_{t=1}^{T} \|\+U_{t}\+U'_t+\beta\+1\+1'-\+U_t \+U_t^{*'}-\beta^*\+1\+1'\|_F^2 \leq  \sum_{t=1}^{T} 2\|\+U_{t}\+U'_t -\+U_t \+U_t^{*'}\|_F^2 +2 (\beta-\beta^*)^2$.  Based on the proof in Section \ref{sec:proof_thm1}, we only need to lower bound the prior probability of the set $E_0:=E_1 \cap E_2$ with $E_1:=\{\max_{i,j,t}(\+u_{it}'\+u_{jt}-\+u^{*'}_{it}\+u^{*}_{jt})^2 \leq \epsilon^2\}$ and $E_2 = \{|\beta-\beta^*|\leq \epsilon\}$. The proof in Section \ref{sec:proof_thm1} gives us  $\log \Pi(E_1) \gtrsim -(s_u+n)\log(nT)$. 
and the Assumption~\ref{asm:sparsity_prior} gives us  $\log \Pi(E_2) \gtrsim -(s_u+n)\log(nT)$. In particular, suppose we use a $\m N(0,\sigma^2_\beta)$ prior for $\beta$ for a constant $\sigma_\beta$, then the prior concentration shows that 
 	\begin{equation}
	\begin{aligned}
	\Pi(E_2)  &= \Pi\left(|\beta-\beta^*|\leq \epsilon  \right) \\
	&\gtrsim \frac{1}{(\sqrt{2\pi}\sigma_\beta)}\exp(- \frac{\beta^{*2} }{2\sigma_\beta^{2}}) (2\epsilon)   \\
	& \gtrsim \exp \left[-K \left\{\frac{(\beta^*)^2}{2 \sigma_\beta^{2}}+ \log( \frac{1}{\epsilon})+\log (\sigma_\beta) \right\}\right].
	\end{aligned}
	\end{equation}
Then for a constant $\sigma_\beta$ and $\beta^* \ll \log(1/\epsilon_{n,T})$, we can see the right-hand side of the above equation is lower bounded by $e^{-n^2T\epsilon_{n,T}^2}.$
 
 Therefore, we have  $\log \Pi(E_0) \gtrsim -(s_u+n)\log(nT)$. Hence, the prior concentration still holds. We still have with high posterior probability $D_{\frac{1}{2}}(P_{\m{ U}, \beta},P_{\m U^*, \beta^*}) \lesssim n^2T\epsilon_{n,T}^2$. 



Based on Lemma~\ref{lem:lower of divergence} in the appendix, we have for $a,b \rightarrow -\infty$ such that $p_a,p_b \rightarrow 0$, then we have
 		\begin{align*}
		D_{\frac{1}{2}}(P_a,P_b)\gtrsim (\sqrt{p_a}-\sqrt{p_b})^2,
		\end{align*}
  and
\begin{equation*}
		D_{\frac{1}{2}}(P_a,P_b) \gtrsim  \exp \left\{a \wedge b \right\}(b-a)^2.
\end{equation*}

Denote $p^*_{ijt} = 1/\{1+\exp(-\beta^*-\+u_{it}^{*'}\+u_{jt}^{*})\}$, $\hat{p}_{ijt} = 1/\{1+\exp(-\hat{\beta}-\hat {\+u}_{it}^{'}\hat {\+u}_{jt})\}$, $p^*_{\beta} = 1/\{1+\exp(-\beta^*)\}$ and $\hat p_{\hat \beta} = 1/\{1+\exp(-\hat \beta)\}$. Due to the uniform boundedness of $\+u_{it}^*$ and $\hat{\+u}_{it}$, we have there are uniform constants $c,C$ such that
\begin{equation*}
c p^*_{\beta}   \leq p^*_{ijt} \leq Cp^*_{\beta}; \quad c \hat p_{\hat \beta} \leq \hat{p}_{ijt}  \leq C  \hat p_{\hat \beta}.
\end{equation*}
Then we have 
\begin{equation*}
\begin{aligned}
    \frac{1}{n^2T}D_{\frac{1}{2}}(P_{\m{\hat{U}}, \hat \beta},P_{\m U^*, \beta^*}) = \frac{1}{n^2T}\sum_{i\ne j=1}^n\sum_{t =1}^T D_{\frac{1}{2}}(p^*_{ijt},\hat{p}_{ijt})\\
    \gtrsim  \frac{1}{n^2T}\sum_{i\ne j=1}^n\sum_{t =1}^T (\sqrt{p^*_{ijt}}-\sqrt{\hat{p}_{ijt}})^2\\
    \gtrsim (C \sqrt{p^*_{\beta} \vee \hat{p}_{ \hat \beta}}-c\sqrt{p^*_{\beta} \wedge \hat{p}_{\hat \beta}} )^2.
    \end{aligned}
\end{equation*}
If $\hat{\beta} \ll \beta^* $, which means $\hat p_{\hat{\beta}} \ll p_{\beta^*}$, then we have $\epsilon_{n,T}^2 \gtrsim D_{\frac{1}{2}}(P_{ \hat{\m U}, \hat \beta},P_{\m U^*, \beta^*})/ (n^2T) \gtrsim (c\sqrt{\hat p_{\hat{\beta}}}-C\sqrt{p_{\beta^*}})^2 \gtrsim p_{\beta^*}$, which causes contradiction with Condition~\ref{asm:sparsity_level}. Therefore, we must have $\hat{\beta} \gtrsim \beta^*$, then
 \begin{align*}
     \epsilon_{n,T}^2 \gtrsim \frac{1}{n^2T}D_{\frac{1}{2}}(P_{\m{\hat{U}}, \hat \beta},P_{\m U^*, \beta^*})\\
     \gtrsim \frac{1}{n^2T}\sum_{i\ne j=1}^n\sum_{t =1}^T \exp \left\{\hat \beta \wedge \beta^* \right\}(\hat \beta-\beta^*+\hat {\+u}_{it}' \hat {\+u}_{jt}-\+u_{it}^{*'}\+u_{jt}^{*'})^2 \\
     \gtrsim \frac{1}{n^2T}\exp(\beta^*) \sum_{i\ne j=1}^n\sum_{t =1}^T (\hat \beta-\beta^*+\hat {\+u}_{it}' \hat{\+u}_{jt}-\+u_{it}^{*'}\+u_{jt}^{*'})^2 .
 \end{align*}

   \end{proof}

	\subsection{Auxiliary Lemmas}
	Consider the prior concentration for dynamic networks with sparsity on the differences between two consecutive vectors of the same subject:  
	\begin{align}\label{shrinkage}
	\begin{aligned}
	\lambda_t &\sim \mbox{Ca}^+(0,1), \quad t \in [T], \quad \tau \sim g \\
	\beta_t &\sim \mathcal{N}(0,\lambda_{t}^2 \tau^2), \quad t \in [T].
	\end{aligned}
	\end{align}
	In the following Lemma, we reformulate the form of prior concentration for shrinkage prior with additional variable $n$ to fit the network and factor problems. Note that the additional variable $n$ here stands for $np$ for factor models and $n^2$ for network models. 
	\begin{lemma}[$\ell_1$ concentration for proposed shrinkage prior with additional variable $n$]\label{lem:horse_net}
		Suppose $\+\beta^* \in \mathbb{R}^{T}$ with $S=\{j: \beta^*_{j}\ne 0\}$ and $|S| \leq s$, $1\leq s\leq T$. Denote $\delta=\sqrt{s\log (nT)/(nT)}$ with $\log T =o (n)$, $n \geq 1$ and  $\max_{t=1}^T |\beta_t| =O(n^\alpha T^\alpha)$ for some constant $\alpha>0$. Under the above shrinkage prior~\eqref{shrinkage} with $g$ satisfying the following property:
		\begin{equation}\label{eq:global_property}
		\log	\left\{g(\tau^\ast<\tau<2\tau^\ast)  \right\}\gtrsim -s_i\log(nT), \quad \mbox{with} \quad \tau^\ast = \frac{s^{\frac{1}{2}}(\log (nT))^{\frac{1}{2}}}{n^{\frac{3}{2}}T^\frac{5}{2}}.
		\end{equation}
		then for some constant $c_0>0$, we have
		\begin{equation}\label{eq:hs_net}
		\Pi(\|\+\beta-\+\beta^*\|_1 \leq \delta) \geq e^{-K s\log (nT)},
		\end{equation}
		where $K>0$ is a constant.
	\end{lemma}

	\begin{proof}
		$\log T =o (n)$ guarantees $\delta =o(1)$ for any $s=[T]$.

		First,  we have
		\begin{equation*}
		Pr(\|\+\beta-\+\beta^*\|_1 <\delta) \geq \prod_{j \in S} Pr(|\beta_j-\beta^*_{j}|< \frac{\delta}{2s})\prod_{j \in S^c} Pr(|\beta_j-\beta^*_{j}|< \frac{\delta}{2T}),
		\end{equation*}
		
		Denote \begin{equation}
		\tau^\ast = \frac{s^{\frac{1}{2}}(\log (nT))^{\frac{1}{2}}}{n^{\frac{3}{2}}T^\frac{5}{2}},
		\end{equation}
		and consider the event $E_\tau = \{\tau:\tau \in [\tau^\ast,2\tau^\ast]\}$.
		
		For the non-signal part. For $j \in S^c$, by the Chernoff bound for a Gaussian random variable, we have
		\begin{equation*}
		Pr(|\beta_j|> \frac{\delta}{2T} \mid \lambda_j,\tau) \leq  2e^{-\delta^2/(8T^2\lambda_j^2\tau^2)}.
		\end{equation*}
		Then we have
		\begin{equation*}
		\begin{aligned}
		Pr(|\beta_j|< \frac{\delta}{2T} \mid \tau) &\geq \int_{\lambda_j} \left\{1-2e^{-\delta^2/(8T^2\lambda_j^2\tau^2)}\right\} f(\lambda_j ) d\lambda_j \\
		&= 1- \int_{\lambda_j} 2e^{-\delta^2/(8T^2\lambda_j^2\tau^2)} f(\lambda_j ) d\lambda_j,
		\end{aligned}
		\end{equation*}
		with $f(\lambda_j ) = 1/(1+\lambda_j^2) <\lambda_j^{-2}$. Then, we have
		\begin{equation*}
		\begin{aligned}
		\int_{\lambda_j} 2e^{-\delta^2/(8T^2\lambda_j^2\tau^2)} f(\lambda_j) d\lambda_j &<  2 \int_{\lambda_j} e^{-\delta^2/(8T^2\lambda_j^2\tau^2)} \lambda_j^{-2}  d\lambda_j \\
		& = 2 \frac{\Gamma(1/2)}{\{\delta^2/(8T^2\tau^2)\}^{1/2}} \\
		& = C \frac{T\tau}{\sqrt{s\log (nT)/(nT)}}\\
		& {\leq} C' \frac{1}{nT},
		\end{aligned}
		\end{equation*}
		where in the final step we use $\tau < 2\tau^*$.
		
		Moreover, for the signal part, let $\delta_0 =\delta/(2s)$, we have
		\begin{equation*}
		\begin{aligned}
		Pr(|\beta_j-\beta^*_j| < \delta_0 \mid \tau) &=\int_{\lambda_j} \int_{|\beta_j-\beta^*_j|<\delta_0} (\frac{2}{\pi^3})^{1/2} \exp\left\{-\beta_j^2/(2\lambda_j^2 \tau^2)\right\} \frac{1}{\lambda_j \tau (1+\lambda_j^2)}  d\lambda_j d \beta_j\\
		&\geq (\frac{2}{\pi^3})^{1/2} \int_{|\beta_j-\beta^*_j|<\delta_0}\int_{n^\alpha T^\alpha/\tau^\ast}^{2 n^\alpha T^\alpha/\tau^\ast} \frac{1}{\sqrt{2\pi}} \exp\left\{-\beta_j^2/(2\lambda_j^2 \tau^2)\right\} \frac{1}{\lambda_j \tau (1+\lambda_j^2)}  d\lambda_j d \beta_j\\
		&\stackrel{(i)}{\geq} (\frac{2}{\pi^3})^{1/2}   \int_{|\beta_j-\beta^*_j|<\delta_0}\exp\left\{-\beta_j^2/(4n^{2\alpha} T^{2\alpha})\right\} \int_{n^\alpha T^\alpha/\tau^*}^{n^\alpha T^\alpha/\tau^*}  \frac{1}{4 n^\alpha T^\alpha ( 1+\lambda_j^2)}d\lambda_j d \beta_j \\
		& \geq (\frac{2}{\pi^3})^{1/2} \frac{\tau}{2(4n^{2\alpha}T^{2\alpha}+\tau^2)} \int_{|\beta_j-\beta^*_j|<\delta_0} \exp\left\{-\beta_j^2/(4n^{2\alpha }T^{2\alpha})\right\}d \beta_j\\
		& \stackrel{(ii)}{\geq} K \delta_0 \tau^* n^{-2\alpha}T^{-2\alpha}  \\
		& {\geq } K' \frac{\sqrt{\log (nT)}}{\sqrt{nTs}} n^{-2\alpha} T^{-2\alpha} \frac{s^{\frac{1}{2}}(\log (nT))^{\frac{1}{2}}}{n^{\frac{3}{2}}T^\frac{5}{2}} \stackrel{(iii)}{\geq} K'' (nT)^{-M}
		\end{aligned}
		\end{equation*}
		where $(i)$ holds based on $n^\alpha T^\alpha<\tau \lambda_j<4 n^\alpha T^\alpha$; $(ii)$ is due to $\tau<1$ and $\max_t (|\beta_t|) = O(n^\alpha T^\alpha)$; $(iii)$ is because $s<p$; $M,K,K',K''>0$ are some constants.
		
		
		Therefore, we under Assumption~\ref{asm:global_prior}, we have
		\begin{equation*}
		\Pi(\|\+\beta-\+\beta^*\|_1 <\delta ) \geq	\Pi(\|\+\beta-\+\beta^*\|_1 <\delta \mid E_\tau)\Pi(E_\tau) \geq (1-C'/(nT))^{T-s} K'' e^{-M s\log (nT)}   \geq e^{-K^* s\log (nT)},
		\end{equation*}
		where $K^*$ is a positive constant.
		
	\end{proof}
	
	\subsection{Proof of Lemma~\ref{lem:4.1}}
	\begin{proof}
		\begin{equation*}
		[\E(\+D(\+a)\+\Xi \+D(\+a))]_{ij} =\E( a_{i}  \Xi_{ij} a_j ) = \begin{cases}
		\mu_{i}  \Xi_{ij} \mu_{j} + \Xi_{ij} \Sigma_{ij}  \quad \mbox{if} \quad i \ne j,\\
		\mu_{i} \Xi_{ij} \mu_{i} + \Xi_{ij}\Sigma_{ii} \quad \mbox{if} \quad i = j.
		\end{cases}
		\end{equation*}
	\end{proof}

	\begin{prop}\label{proposition_lemma_8}
		The equation~\eqref{eq:global_property} is satisfied for $\tau \sim \mbox{Ca}^+(0,1)$ and $\tau^2 \sim \mbox{Gamma} (a_\tau,b_\tau)$ with positive constants $a_\tau,b_\tau$.
	\end{prop}
	
	\begin{proof}
		Given $\tau \sim \mbox{Ca}^+(0,1)$,   for any $s=[T]$,  by mean-value theorem, we have
		\begin{equation*}
		g(\tau^* <\tau <2\tau^*) =\int_{\tau^{*}}^{2\tau^{*}} \frac{1}{1+\tau^{2}} d\tau \geq \tau^* \geq c_1e^{-c_2 s\log(nT)},
		\end{equation*}
		where the final inequality follows by $\log(\tau^*) \gtrsim -s\log(nT)$, and $c_1,c_2$ are positive constants.
		Given the $\Gamma(a_\tau,b_\tau)$ prior for $\tau^2$ with constants $a_\tau,b_\tau$, for any $s=[T]$,  by mean-value theorem,
		\begin{equation*}
		\begin{aligned}
		g(\tau^* <\tau <2\tau^*) = \int_{\tau^{*2}}^{4\tau^{*2}} \frac{b_\tau^{a_\tau}}{\Gamma(a_\tau)} \tau^{2(a_\tau-1)} e^{-b_\tau \tau^2}  d\tau^2 \geq c_1 \tau^{*(2a_\tau-2)}e^{-c_2\tau^{*2}} \tau^{*2}
		\\	\geq c_1 e^{-c_2 \tau^{*2}+2 a_\tau \log \tau^*}  \geq c_1 e^{-c_3 s\log(nT)},
		\end{aligned}
		\end{equation*}
		where the final inequality follows by $-\tau^{*2} \gtrsim -s\log(nT)$ and $\log(\tau^*) \gtrsim -s\log(nT)$, and $c_1,c_2,c_3$ are positive constants.
	\end{proof}
	
		\begin{lemma}[Perturbation bound for Procrustes, Theorem 1 in \cite{arias2020perturbation}] \label{lem:perturbation}
		Consider two matrices $\+X, \+Y \in \mathbb{R}^{n \times d}$, with $\+X$ having full rank, and set $\epsilon^2 =\|\+Y\+Y^T-\+X\+X^T\|_F$. Then if $\|\+X^{\dagger}\|\epsilon \leq 1/\sqrt{2}$, we have
		\begin{equation}
		\min_{\+O \in \mathbb{O}^{d \times d}} \|\+Y-\+X\+O\|_F \leq (1+\sqrt{2})\| \+X^{\dagger}\|\epsilon^2,
		\end{equation} 
		where $ \+X^{\dagger}$ is the pseudo-inverse of $\+X$.
	\end{lemma}

	\begin{lemma}[$K$-means error bound, adapted from Lemma 5.3 in  \cite{lei2015consistency}] \label{lem:k_means}
		For any two matrices $\hat{\+U},\+U\in \mathbb{R}^{n \times d}$ such that $\+U=\+\Theta \+X$ with $\+\Theta \in \mathbb{M}_{n,K}$, $\+X \in \mathbb{R}^{K \times d}$, let $(\hat{\+\Theta},\hat{\+X})$ be the solution to the $K$-means problem and $\bar{\+U}=\hat{\+\Theta}\hat{\+X}$. Then for any $\delta_k \leq \min_{l \ne k} \|\+u_{l}-\+u_{k} \|_2$, define $S_k= \{i \in G_k(\+\Theta): \|\bar{\+u}_i-\+u_{i} \|_2 \geq \delta_k/2\}$, then
		\begin{equation*}
		\sum_{k=1}^{K} |S_k|\delta_k^2 \leq {16 \|\hat{\+U}-\+U\|_F^2}.
		\end{equation*}
	\end{lemma}
	Moreover, if
	\begin{equation*}
	16 \|\hat{\+U}-\+U\|_F^2/\delta_k^2 <n_k \quad \text{for all } k,
	\end{equation*}
	then there exists a $K \times K$ permutation matrix $J$ such that $\hat{\+\Theta}_{G^*}=\+\Theta_{G^*} J$, where $G = \cap_{k=1}^K(G_k-S_k)$.
	 {
	\begin{lemma}[Upper bound for binary KL divergence]\label{lem:binary_KL}
		Let $p_a=1/(1+\exp(-a))$ and $p_b=1/(1+\exp (-b))$. Define $P_a$ and $P_b$ as the Bernoulli measures with probability $p_a$ and $p_b$. Then we have
		\begin{equation*}
		D_{KL}(P_{a} \,||\, P_{b}) + D_{KL}(P_{b} \,||\, P_{a}) \leq (p_a \vee p_b)(a-b)^2.
		\end{equation*}
	\end{lemma}
	}
	\begin{proof}
		\begin{equation*}
		\begin{aligned}
		D_{KL}(P_{a} \,||\, P_{b}) + D_{KL}(P_{b} \,||\, P_{a}) &=(p_a-p_b) \log \frac{p_a}{p_b}+(p_b-p_a) \log \frac{1-p_a}{1-p_b}\\
		=(p_a-p_b) \log\left(\frac{p_a}{1-p_a} \frac{1-p_b}{p_b}\right) 
		&= \left\{\frac{1}{1+\exp(-a)}-\frac{1}{1+\exp(-b)}\right\} \log(e^{a} e^{-b}) \\
		& = (a-b)\left\{\frac{1}{1+\exp(-a)}-\frac{1}{1+\exp(-b)}\right\}.
		\end{aligned}
		\end{equation*}
		Without loss of generality, we can assume $a>b$, then by $\exp(x)\geq 1+x$, we have
		\begin{align*}
		\frac{1}{1+\exp(-a)}-\frac{1}{1+\exp(-b)} = \frac{e^{-b}-e^{-a}}{(1+\exp(-a))(1+\exp(-b))} \\
		\leq  \frac{1-e^{b-a}}{(1+e^{-a})(1+e^b)} \leq p_a(1-e^{b-a}) \leq p_a(a-b).
		\end{align*}
	\end{proof}


{
	\begin{lemma}[Upper bound of second order KL moment]\label{lem:second_KL}
		Let $p_a=1/(1+\exp(-a))$ and $p_b=1/(1+\exp (-b))$. Define $P_a$ and $P_b$ as the Bernoulli measures with probability $p_a$ and $p_b$. Then we have
		\begin{equation*}
		\int P_a \log^2 \left( \frac{P_a}{P_b}\right) d\mu \leq  \left[\frac{p_a}{(p_a \wedge  p_b)^2}+\frac{1-p_a}{(1-p_a\vee p_b)^2}\right] (p_a \vee p_b)^2(a-b)^2.
		\end{equation*}
	\end{lemma}
 }
	
	\begin{proof}
 Note that 
		\begin{equation*}
		\begin{aligned}
\int P_a \log^2 \left( \frac{P_a}{P_b}\right) d\mu = p_a\log^2 \left( \frac{p_a}{p_b}\right) +(1-p_a) \log^2\left( \frac{1-p_a}{1-p_b}\right).
		\end{aligned}
		\end{equation*}
  We have
\begin{align*}
    \log^2 \left( \frac{p_a}{p_b}\right)  = \log^2\left( \frac{p_a \vee p_b}{p_a \wedge p_b}-1+1\right) \leq \left(\frac{p_a \vee p_b - p_a \wedge p_b}{p_a \wedge p_b}\right)^2 = \left(\frac{p_a - p_b}{p_a \wedge p_b}\right)^2.
\end{align*}
  Similarly,
\begin{align*}
    \log^2 \left( \frac{1-p_a}{1-p_b}\right)  = \log^2\left( \frac{(1-p_a) \vee (1-p_b)}{(1-p_a) \wedge (1-p_b)}-1+1\right) \leq \left(\frac{p_a - p_b}{1-p_a \vee p_b}\right)^2.
\end{align*}
  
	For the $(p_a-p_b)^2$ term,  by $\exp(x)\geq 1+x$, we have
		\begin{align*}
		\frac{1}{1+\exp(-a \vee b)}-\frac{1}{1+\exp(-a \wedge b)} = \frac{e^{-a \wedge b}-e^{-a \vee b}}{(1+\exp(-a \vee b))(1+\exp(-a \wedge b))} \\
		\leq  \frac{1-e^{a \vee b-a \wedge b}}{(1+e^{a \wedge b})(1+e^{-a \vee b})} \leq (p_a \vee p_b)(1-e^{a \vee b-a \wedge b}) \leq (p_a \vee p_b) (a \wedge b-a \vee b).
		\end{align*}
	\end{proof}

{	\begin{lemma}[Lower bound of the $1/2$ divergence] \label{lem:lower of divergence}Let $p_a=1/(1+\exp(-a))$ and $p_b=1/(1+\exp (-b))$. Define $P_a$ and $P_b$ as the Bernoulli measures with probability $p_a$ and $p_b$.
 \begin{enumerate}
     \item  Suppose that there exist constants $c,C>0$ such that $c<a,b<C$, then we have
		\begin{equation*}
		D_{\frac{1}{2}}(P_a,P_b) \gtrsim  (b-a)^2.
		\end{equation*}
\item   Suppose that $a,b \rightarrow -\infty$ such that $p_a,p_b \rightarrow 0$, then we have
		\begin{equation*}
		D_{\frac{1}{2}}(P_a,P_b) \gtrsim  \exp \left\{a \wedge b \right\}(b-a)^2.
		\end{equation*}
  \end{enumerate}
  	\end{lemma}
   }

	\begin{proof}
 		\begin{align*}
		D_{\frac{1}{2}}(p_a,p_b)=-2\log(1-h^2(p_a,p_b)) \geq 2h^2(p_a,p_b) =  \left[(\sqrt{p_a}-\sqrt{p_b})^2+(\sqrt{1-p_a}-\sqrt{1-p_b})^2\right].
		\end{align*}
	For the first conclusion,  since $a,b$ are bounded, $p_a,p_b$ are bounded away form $0$ and $1$, and $(\sqrt{p_a}+\sqrt{p_b}),(\sqrt{1-p_a}+\sqrt{1-p_b})$ are bounded from $0$ as well. Hence,
		\begin{align*}
		D_{\frac{1}{2}}(p_a,p_b) \gtrsim \left[(\sqrt{p_a}-\sqrt{p_b})^2(\sqrt{p_a}+\sqrt{p_b})^2+(\sqrt{1-p_a}-\sqrt{1-p_b})^2(\sqrt{1-p_a}+\sqrt{1-p_b})^2\right] \\
		\gtrsim (p_a-p_b)^2 \stackrel{(i)}{=} \left\{\frac{\exp(x)}{(1+\exp(x))^2} \right\}^2(a-b)^2 \gtrsim (a-b)^2.
		\end{align*}
		where $(i)$ is because the mean value theorem and $a<x<b$ is bounded.

  For the second conclusion, when the probabilities $p_a,p_b$ are converging to zeros, for the term $(\sqrt{p_a}-\sqrt{p_b})^2$, by the mean value theorem of function $\sqrt{p_x}$ with respect to $x$, we have
  \begin{equation*}
     (\sqrt{p_a}-\sqrt{p_b})^2 \geq  \left( \frac{\sqrt{\exp(x)} \sqrt{1+\exp(x)}}{2 (1+\exp(x))^2} \right)^2 (a-b)^2 \stackrel{(i)}{\gtrsim} e^{a \wedge b }(a-b)^2,
  \end{equation*}
  where $(i)$ is because $\exp(x)$ is the order of $\exp \{a \wedge b \}$ for $a\wedge b<x<a \vee b$. For the term $(\sqrt{1-p_a}-\sqrt{1-p_b})^2$, note that $\sqrt{1-p_a}+\sqrt{1-p_b}$ is still bound away from $0$, we have 
  \begin{equation*}
      (\sqrt{1-p_a}-\sqrt{1-p_b})^2 \gtrsim (p_a-p_b)^2 = \left\{\frac{\exp(x)}{(1+\exp(x))^2} \right\}^2(a-b)^2 \gtrsim e^{(2a) \wedge (2b) } (a-b)^2,
  \end{equation*}
  for $a<x<b$. Finally, $\exp(a \wedge b )(a-b)^2$ dominates when the sum of the two lower bounds is taken into account.
  
	\end{proof}
{ \putbib}
	\end{bibunit}

\end{document}

%% file: CP_standalone.tex
\newcommand{\Depth}{2}
\newcommand{\Height}{2}
\newcommand{\Width}{2}
\newcommand{\xx}{1}
\newcommand{\yy}{1}
\newcommand{\zz}{1}
\newcommand{\xd}{4.5}
\newcommand{\yd}{-0.3}
\newcommand{\ye}{1.1}
\newcommand{\zd}{1}

\begin{tikzpicture}
\coordinate (O) at (0,0,0);
\coordinate (A) at (0,\Width,0);
\coordinate (B) at (0,\Width,\Height);
\coordinate (C) at (0,0,\Height);
\coordinate (D) at (\Depth,0,0);
\coordinate (E) at (\Depth,\Width,0);
\coordinate (F) at (\Depth,\Width,\Height);
\coordinate (G) at (\Depth,0,\Height);
\draw[black!60!black,fill=black!5] (O) -- (C) -- (G) -- (D) -- cycle;
\draw[black!60!black,fill=black!5] (O) -- (A) -- (E) -- (D) -- cycle;
\draw[black!60!black,fill=black!5] (O) -- (A) -- (B) -- (C) -- cycle;
\draw[black!60!black,fill=black!5,opacity=0.8] (D) -- (E) -- (F) -- (G) -- cycle;
\draw[black!60!black,fill=black!5,opacity=0.6] (C) -- (B) -- (F) -- (G) -- cycle;
\draw[black!60!black,fill=black!5,opacity=0.8] (A) -- (B) -- (F) -- (E) -- cycle;

\coordinate (O) at (0+\xx,0+\yy,0+\zz);
\coordinate (A) at (0+\xx,0.25\Width+\yy,0+\zz);
\coordinate (B) at (0+\xx,0.25\Width+\yy,0.25\Height+\zz);
\coordinate (C) at (0+\xx,0+\yy,0.25\Height+\zz);
\coordinate (D) at (0.25\Depth+\xx,0+\yy,0+\zz);
\coordinate (E) at (0.25\Depth+\xx,0.25\Width+\yy,0+\zz);
\coordinate (F) at (0.25\Depth+\xx,0.25\Width+\yy,0.25\Height+\zz);
\coordinate (G) at (0.25\Depth+\xx,0+\yy,0.25\Height+\zz);
\draw[black!80!black,fill=black!10] (O) -- (C) -- (G) -- (D) -- cycle;
\draw[black!80!black,fill=black!10] (O) -- (A) -- (E) -- (D) -- cycle;
\draw[black!80!black,fill=black!10] (O) -- (A) -- (B) -- (C) -- cycle;
\draw[black!40!black,fill=black!10,opacity=0.8] (D) -- (E) -- (F) -- (G) -- cycle;
\draw[black!40!black,fill=black!10,opacity=0.6] (C) -- (B) -- (F) -- (G) -- cycle;
\draw[black!40!black,fill=black!10,opacity=0.8] (A) -- (B) -- (F) -- (E) -- cycle;
\draw (0.2,-1,0) node {\scriptsize{\color{black}$n_2$}};
\draw (-0.2,1,2) node[rotate = 90] {\scriptsize{\color{black}$n_1$}};
\draw (2.3,0,1.4) node[rotate = 45] {\scriptsize{\color{black}$n_3$}};
\draw [draw=gray!75,thick,->] (0.8,0.1,1) -- (1.2,0.8,1) node [right] {{\color{black!50!black}$M_{i_1 i_2 i_3,t}$}};
\draw (0.8,0,1) node {\scriptsize{\color{black}$(i_1,i_2,i_3)$-th}};
\draw (1.2,-1.2,1) node {\color{black}$\mathcal{M}_t \in\mathbb{R}^{n_1 \times n_2 \times n_3}$};

\draw (2.8,0.5,0) node[rotate = 0]{{\color{black}\LARGE{$=$}}};

\coordinate (O) at (0+\xd,0+\yd,0+\zd);
\coordinate (A) at (0+\xd,\Width+\yd,0+\zd);
\coordinate (B) at (0+\xd,\Width+\yd,0.25\Height+\zd);
\coordinate (C) at (0+\xd,0+\yd,0.25\Height+\zd);
\coordinate (D) at (0.5\Depth+\xd,0+\yd,0+\zd);
\coordinate (E) at (0.5\Depth+\xd,\Width+\yd,0+\zd);
\coordinate (F) at (0.5\Depth+\xd,\Width+\yd,0.25\Height+\zd);
\coordinate (G) at (0.5\Depth+\xd,0+\yd,0.25\Height+\zd);
\draw[black!60,fill=black!25] (O) -- (C) -- (G) -- (D) -- cycle;
\draw[black!60,fill=black!25] (O) -- (A) -- (E) -- (D) -- cycle;
\draw[black!60,fill=black!25] (O) -- (A) -- (B) -- (C) -- cycle;
\draw[black!60,fill=black!25,opacity=0.8] (D) -- (E) -- (F) -- (G) -- cycle;
\draw[black!60,fill=black!25,opacity=0.6] (C) -- (B) -- (F) -- (G) -- cycle;
\draw[black!60,fill=black!25,opacity=0.8] (A) -- (B) -- (F) -- (E) -- cycle;

\coordinate (O) at (0+\xd,0+\yy-0.1,0+\zd);
\coordinate (A) at (0+\xd,0.25\Width+\yy-0.1,0+\zd);
\coordinate (B) at (0+\xd,0.25\Width+\yy-0.1,0.25\Height+\zd);
\coordinate (C) at (0+\xd,0+\yy-0.1,0.25\Height+\zd);
\coordinate (D) at (0.5\Depth+\xd,0+\yy-0.1,0+\zd);
\coordinate (E) at (0.5\Depth+\xd,0.25\Width+\yy-0.1,0+\zd);
\coordinate (F) at (0.5\Depth+\xd,0.25\Width+\yy-0.1,0.25\Height+\zd);
\coordinate (G) at (0.5\Depth+\xd,0+\yy-0.1,0.25\Height+\zd);
\draw[black!80!black,fill=black!10] (O) -- (C) -- (G) -- (D) -- cycle;
\draw[black!80!black,fill=black!10] (O) -- (A) -- (E) -- (D) -- cycle;
\draw[black!80!black,fill=black!10] (O) -- (A) -- (B) -- (C) -- cycle;
\draw[black!40!black,fill=black!10,opacity=0.8] (D) -- (E) -- (F) -- (G) -- cycle;
\draw[black!40!black,fill=black!10,opacity=0.6] (C) -- (B) -- (F) -- (G) -- cycle;
\draw[black!40!black,fill=black!10,opacity=0.8] (A) -- (B) -- (F) -- (E) -- cycle;

\draw[black!60] (0.5\Depth+\xd,\Width+\yd,0.25\Height+\zd) -- (0.5\Depth+\xd,0+\yd,0.25\Height+\zd);
\draw (\xd+0.2,\yy-1.9,\zz) node {\color{black}$\boldsymbol U_t^{(1)}\in\mathbb{R}^{n_1 \times d}$};
\draw (\xd-0.5,\yy-0.1,\zz) node {\color{black!50!black}$\boldsymbol{u}^{(1)}_{i_1t}$};

\coordinate (O) at (0+\xd+1.2,0+\ye,0+\zd);
\coordinate (A) at (0+\xd+1.2,0.5\Width+\ye,0+\zd);
\coordinate (B) at (0+\xd+1.2,0.5\Width+\ye,0.25\Height+\zd);
\coordinate (C) at (0+\xd+1.2,0+\ye,0.25\Height+\zd);
\coordinate (D) at (\Depth+\xd+1.2,0+\ye,0+\zd);
\coordinate (E) at (\Depth+\xd+1.2,0.5\Width+\ye,0+\zd);
\coordinate (F) at (\Depth+\xd+1.2,0.5\Width+\ye,0.25\Height+\zd);
\coordinate (G) at (\Depth+\xd+1.2,0+\ye,0.25\Height+\zd);
\draw[black!60,fill=black!15] (O) -- (C) -- (G) -- (D) -- cycle;
\draw[black!60,fill=black!15] (O) -- (A) -- (E) -- (D) -- cycle;
\draw[black!60,fill=black!15] (O) -- (A) -- (B) -- (C) -- cycle;
\draw[black!60,fill=black!15,opacity=0.8] (D) -- (E) -- (F) -- (G) -- cycle;
\draw[black!60,fill=black!15,opacity=0.6] (C) -- (B) -- (F) -- (G) -- cycle;
\draw[black!60,fill=black!15,opacity=0.8] (A) -- (B) -- (F) -- (E) -- cycle;

\coordinate (O) at (0+\xd+2.5,0+\ye,0+\zd);
\coordinate (A) at (0+\xd+2.5,0.5\Width+\ye,0+\zd);
\coordinate (B) at (0+\xd+2.5,0.5\Width+\ye,0.25\Height+\zd);
\coordinate (C) at (0+\xd+2.5,0+\ye,0.25\Height+\zd);
\coordinate (D) at (0.25\Depth+\xd+2.5,0+\ye,0+\zd);
\coordinate (E) at (0.25\Depth+\xd+2.5,0.5\Width+\ye,0+\zd);
\coordinate (F) at (0.25\Depth+\xd+2.5,0.5\Width+\ye,0.25\Height+\zd);
\coordinate (G) at (0.25\Depth+\xd+2.5,0+\ye,0.25\Height+\zd);
\draw[black!80!black,fill=black!10] (O) -- (C) -- (G) -- (D) -- cycle;
\draw[black!80!black,fill=black!10] (O) -- (A) -- (E) -- (D) -- cycle;
\draw[black!80!black,fill=black!10] (O) -- (A) -- (B) -- (C) -- cycle;
\draw[black!40!black,fill=black!10,opacity=0.8] (D) -- (E) -- (F) -- (G) -- cycle;
\draw[black!40!black,fill=black!10,opacity=0.6] (C) -- (B) -- (F) -- (G) -- cycle;
\draw[black!40!black,fill=black!10,opacity=0.8] (A) -- (B) -- (F) -- (E) -- cycle;

\draw[black!60] (0+\xd+1.2,0.5\Width+\ye,0.25\Height+\zd) -- (\Depth+\xd+1.2,0.5\Width+\ye,0.25\Height+\zd);
\draw (\xd+2.2,\yy-1,\zz) node {\color{black}$\boldsymbol U^{(2)}_t \in\mathbb{R}^{n_2 \times d}$};
\draw (\xd+2.6,\yy-0.4,\zz) node {\color{black!50!black}$\boldsymbol{u}^{(2)}_{i_2 t}$};

\coordinate (O) at (0+\xd+1.2,0+\ye+1.5,0+\zd);
\coordinate (A) at (0+\xd+1.2,0.25\Width+\ye+1.5,0+\zd);
\coordinate (B) at (0+\xd+1.2,0.25\Width+\ye+1.5,\Height+\zd);
\coordinate (C) at (0+\xd+1.2,0+\ye+1.5,\Height+\zd);
\coordinate (D) at (0.5\Depth+\xd+1.2,0+\ye+1.5,0+\zd);
\coordinate (E) at (0.5\Depth+\xd+1.2,0.25\Width+\ye+1.5,0+\zd);
\coordinate (F) at (0.5\Depth+\xd+1.2,0.25\Width+\ye+1.5,\Height+\zd);
\coordinate (G) at (0.5\Depth+\xd+1.2,0+\ye+1.5,\Height+\zd);
\draw[black!60,fill=black!15] (O) -- (C) -- (G) -- (D) -- cycle;
\draw[black!60,fill=black!15] (O) -- (A) -- (E) -- (D) -- cycle;
\draw[black!60,fill=black!15] (O) -- (A) -- (B) -- (C) -- cycle;
\draw[black!60,fill=black!15,opacity=0.8] (D) -- (E) -- (F) -- (G) -- cycle;
\draw[black!60,fill=black!15,opacity=0.6] (C) -- (B) -- (F) -- (G) -- cycle;
\draw[black!60,fill=black!15,opacity=0.8] (A) -- (B) -- (F) -- (E) -- cycle;

\coordinate (O) at (0+\xd+0.8,0+\yy+1.2,0+\zd);
\coordinate (A) at (0+\xd+0.8,0.25\Width+\yy+1.2,0+\zd);
\coordinate (B) at (0+\xd+0.8,0.25\Width+\yy+1.2,0.25\Height+\zd);
\coordinate (C) at (0+\xd+0.8,0+\yy+1.2,0.25\Height+\zd);
\coordinate (D) at (0.5\Depth+\xd+0.8,0+\yy+1.2,0+\zd);
\coordinate (E) at (0.5\Depth+\xd+0.8,0.25\Width+\yy+1.2,0+\zd);
\coordinate (F) at (0.5\Depth+\xd+0.8,0.25\Width+\yy+1.2,0.25\Height+\zd);
\coordinate (G) at (0.5\Depth+\xd+0.8,0+\yy+1.2,0.25\Height+\zd);
\draw[black!80!black,fill=black!10] (O) -- (C) -- (G) -- (D) -- cycle;
\draw[black!80!black,fill=black!10] (O) -- (A) -- (E) -- (D) -- cycle;
\draw[black!80!black,fill=black!10] (O) -- (A) -- (B) -- (C) -- cycle;
\draw[black!40!black,fill=black!10,opacity=0.8] (D) -- (E) -- (F) -- (G) -- cycle;
\draw[black!40!black,fill=black!10,opacity=0.6] (C) -- (B) -- (F) -- (G) -- cycle;
\draw[black!40!black,fill=black!10,opacity=0.8] (A) -- (B) -- (F) -- (E) -- cycle;

\draw[black!60] (0.5\Depth+\xd+1.2,0.25\Width+\ye+1.5,0+\zd) -- (0.5\Depth+\xd+1.2,0.25\Width+\ye+1.5,\Height+\zd);
\draw (\xd+2.6,\yy+1.3,\zz) node {\color{black}$\boldsymbol U_t^{(3)}\in\mathbb{R}^{n_3\times d}$};
\draw (\xd,\yy+1.4,\zz) node {\color{black!50!black}$\boldsymbol{u}^{(3)}_{i_3 t}$};

\end{tikzpicture}

%% file: CP2_standalone.tex
\newcommand{\Depth}{2}
\newcommand{\Height}{2}
\newcommand{\Width}{2}
\newcommand{\xx}{1}
\newcommand{\yy}{1}
\newcommand{\zz}{1}
\newcommand{\xd}{4.5}
\newcommand{\yd}{-0.3}
\newcommand{\ye}{1.1}
\newcommand{\zd}{1}

\begin{tikzpicture}

\coordinate (O) at (0,0,0);
\coordinate (A) at (0,\Width,0);
\coordinate (B) at (0,\Width,\Height);
\coordinate (C) at (0,0,\Height);
\coordinate (D) at (\Depth,0,0);
\coordinate (E) at (\Depth,\Width,0);
\coordinate (F) at (\Depth,\Width,\Height);
\coordinate (G) at (\Depth,0,\Height);
\draw[black!60!black,fill=black!5] (O) -- (C) -- (G) -- (D) -- cycle;
\draw[black!60!black,fill=black!5] (O) -- (A) -- (E) -- (D) -- cycle;
\draw[black!60!black,fill=black!5] (O) -- (A) -- (B) -- (C) -- cycle;
\draw[black!60!black,fill=black!5,opacity=0.8] (D) -- (E) -- (F) -- (G) -- cycle;
\draw[black!60!black,fill=black!5,opacity=0.6] (C) -- (B) -- (F) -- (G) -- cycle;
\draw[black!60!black,fill=black!5,opacity=0.8] (A) -- (B) -- (F) -- (E) -- cycle;

\coordinate (O) at (0,0,0+\zz);
\coordinate (A) at (0,\Width,0+\zz);
\coordinate (B) at (0,\Width,0.25\Height+\zz);
\coordinate (C) at (0,0,0.25\Height+\zz);
\coordinate (D) at (\Depth,0,0+\zz);
\coordinate (E) at (\Depth,\Width,0+\zz);
\coordinate (F) at (\Depth,\Width,0.25\Height+\zz);
\coordinate (G) at (\Depth,0,0.25\Height+\zz);

\draw[black!60!black,fill=black!50,opacity=0.2] (O) -- (C) -- (G) -- (D) -- cycle;
\draw[black!60!black,fill=black!50,opacity=0.2] (O) -- (A) -- (E) -- (D) -- cycle;
\draw[black!60!black,fill=black!50,opacity=0.2] (O) -- (A) -- (B) -- (C) -- cycle;
\draw[black!60!black,fill=black!50,opacity=0.9] (D) -- (E) -- (F) -- (G) -- cycle;
\draw[black!60!black,fill=black!50,opacity=0.2] (C) -- (B) -- (F) -- (G) -- cycle;
\draw[black!60!black,fill=black!50,opacity=0.9] (A) -- (B) -- (F) -- (E) -- cycle;

\coordinate (O) at (0+\xx,0+\yy,0+\zz);
\coordinate (A) at (0+\xx,0.25\Width+\yy,0+\zz);
\coordinate (B) at (0+\xx,0.25\Width+\yy,0.25\Height+\zz);
\coordinate (C) at (0+\xx,0+\yy,0.25\Height+\zz);
\coordinate (D) at (0.25\Depth+\xx,0+\yy,0+\zz);
\coordinate (E) at (0.25\Depth+\xx,0.25\Width+\yy,0+\zz);
\coordinate (F) at (0.25\Depth+\xx,0.25\Width+\yy,0.25\Height+\zz);
\coordinate (G) at (0.25\Depth+\xx,0+\yy,0.25\Height+\zz);

\draw[black!80!black,fill=black!50] (O) -- (C) -- (G) -- (D) -- cycle;
\draw[black!80!black,fill=black!50] (O) -- (A) -- (E) -- (D) -- cycle;
\draw[black!80!black,fill=black!50] (O) -- (A) -- (B) -- (C) -- cycle;
\draw[black!40!black,fill=black!50,opacity=0.8] (D) -- (E) -- (F) -- (G) -- cycle;
\draw[black!40!black,fill=black!50,opacity=0.6] (C) -- (B) -- (F) -- (G) -- cycle;
\draw[black!40!black,fill=black!50,opacity=0.8] (A) -- (B) -- (F) -- (E) -- cycle;
\draw (0.2,-1,0) node {\scriptsize{\color{black}$n_2$}};
\draw (-0.2,1,2) node[rotate = 90] {\scriptsize{\color{black}$n_1$}};
\draw (2,-0.1,1) node[rotate = 45] {\scriptsize{\color{black}$i_3$}};
\draw (2.3,-0.2,1.4) node[rotate = 45] {\scriptsize{\color{black}$n_3$}};
\draw [draw=gray!75,thick,->] (0.8,0.1,1) -- (1,0.8,1) node [right] {{\color{black!50!black}$M_{i_1 i_2 i_3,t+1}$}};
\draw (0.8,-0.2) node {\scriptsize{\color{black}$(i_1,i_2,i_3)$-th}};
\draw (1,-1.2,1) node {\color{black}$\mathcal{M}_{t+1} \in\mathbb{R}^{n_1 \times n_2 \times n_3}$};

\draw (2.8,0.5,0) node[rotate = 0]{{\color{black}\LARGE{$=$}}};

\coordinate (O) at (0+\xd,0+\yd,0+\zd);
\coordinate (A) at (0+\xd,\Width+\yd,0+\zd);
\coordinate (B) at (0+\xd,\Width+\yd,0.25\Height+\zd);
\coordinate (C) at (0+\xd,0+\yd,0.25\Height+\zd);
\coordinate (D) at (0.5\Depth+\xd,0+\yd,0+\zd);
\coordinate (E) at (0.5\Depth+\xd,\Width+\yd,0+\zd);
\coordinate (F) at (0.5\Depth+\xd,\Width+\yd,0.25\Height+\zd);
\coordinate (G) at (0.5\Depth+\xd,0+\yd,0.25\Height+\zd);
\draw[black!60,fill=black!25] (O) -- (C) -- (G) -- (D) -- cycle;
\draw[black!60,fill=black!25] (O) -- (A) -- (E) -- (D) -- cycle;
\draw[black!60,fill=black!25] (O) -- (A) -- (B) -- (C) -- cycle;
\draw[black!60,fill=black!25,opacity=0.8] (D) -- (E) -- (F) -- (G) -- cycle;
\draw[black!60,fill=black!25,opacity=0.6] (C) -- (B) -- (F) -- (G) -- cycle;
\draw[black!60,fill=black!25,opacity=0.8] (A) -- (B) -- (F) -- (E) -- cycle;

\coordinate (O) at (0+\xd,0+\yy-0.1,0+\zd);
\coordinate (A) at (0+\xd,0.25\Width+\yy-0.1,0+\zd);
\coordinate (B) at (0+\xd,0.25\Width+\yy-0.1,0.25\Height+\zd);
\coordinate (C) at (0+\xd,0+\yy-0.1,0.25\Height+\zd);
\coordinate (D) at (0.5\Depth+\xd,0+\yy-0.1,0+\zd);
\coordinate (E) at (0.5\Depth+\xd,0.25\Width+\yy-0.1,0+\zd);
\coordinate (F) at (0.5\Depth+\xd,0.25\Width+\yy-0.1,0.25\Height+\zd);
\coordinate (G) at (0.5\Depth+\xd,0+\yy-0.1,0.25\Height+\zd);
\draw[black!80!black,fill=black!10] (O) -- (C) -- (G) -- (D) -- cycle;
\draw[black!80!black,fill=black!10] (O) -- (A) -- (E) -- (D) -- cycle;
\draw[black!80!black,fill=black!10] (O) -- (A) -- (B) -- (C) -- cycle;
\draw[black!40!black,fill=black!10,opacity=0.8] (D) -- (E) -- (F) -- (G) -- cycle;
\draw[black!40!black,fill=black!10,opacity=0.6] (C) -- (B) -- (F) -- (G) -- cycle;
\draw[black!40!black,fill=black!10,opacity=0.8] (A) -- (B) -- (F) -- (E) -- cycle;

\draw[black!60] (0.5\Depth+\xd,\Width+\yd,0.25\Height+\zd) -- (0.5\Depth+\xd,0+\yd,0.25\Height+\zd);
\draw (\xd+0.2,\yy-1.9,\zz) node {\color{black}$\boldsymbol U_{t}^{(1)}\in\mathbb{R}^{n_1 \times d}$};
\draw (\xd-0.6,\yy-0.1,\zz) node {\color{black!50!black}$\boldsymbol{u}^{(1)}_{i_1 t}$};

\coordinate (O) at (0+\xd+1.2,0+\ye,0+\zd);
\coordinate (A) at (0+\xd+1.2,0.5\Width+\ye,0+\zd);
\coordinate (B) at (0+\xd+1.2,0.5\Width+\ye,0.25\Height+\zd);
\coordinate (C) at (0+\xd+1.2,0+\ye,0.25\Height+\zd);
\coordinate (D) at (\Depth+\xd+1.2,0+\ye,0+\zd);
\coordinate (E) at (\Depth+\xd+1.2,0.5\Width+\ye,0+\zd);
\coordinate (F) at (\Depth+\xd+1.2,0.5\Width+\ye,0.25\Height+\zd);
\coordinate (G) at (\Depth+\xd+1.2,0+\ye,0.25\Height+\zd);
\draw[black!60,fill=black!15] (O) -- (C) -- (G) -- (D) -- cycle;
\draw[black!60,fill=black!15] (O) -- (A) -- (E) -- (D) -- cycle;
\draw[black!60,fill=black!15] (O) -- (A) -- (B) -- (C) -- cycle;
\draw[black!60,fill=black!15,opacity=0.8] (D) -- (E) -- (F) -- (G) -- cycle;
\draw[black!60,fill=black!15,opacity=0.6] (C) -- (B) -- (F) -- (G) -- cycle;
\draw[black!60,fill=black!15,opacity=0.8] (A) -- (B) -- (F) -- (E) -- cycle;

\coordinate (O) at (0+\xd+2.5,0+\ye,0+\zd);
\coordinate (A) at (0+\xd+2.5,0.5\Width+\ye,0+\zd);
\coordinate (B) at (0+\xd+2.5,0.5\Width+\ye,0.25\Height+\zd);
\coordinate (C) at (0+\xd+2.5,0+\ye,0.25\Height+\zd);
\coordinate (D) at (0.25\Depth+\xd+2.5,0+\ye,0+\zd);
\coordinate (E) at (0.25\Depth+\xd+2.5,0.5\Width+\ye,0+\zd);
\coordinate (F) at (0.25\Depth+\xd+2.5,0.5\Width+\ye,0.25\Height+\zd);
\coordinate (G) at (0.25\Depth+\xd+2.5,0+\ye,0.25\Height+\zd);
\draw[black!80!black,fill=black!10] (O) -- (C) -- (G) -- (D) -- cycle;
\draw[black!80!black,fill=black!10] (O) -- (A) -- (E) -- (D) -- cycle;
\draw[black!80!black,fill=black!10] (O) -- (A) -- (B) -- (C) -- cycle;
\draw[black!40!black,fill=black!10,opacity=0.8] (D) -- (E) -- (F) -- (G) -- cycle;
\draw[black!40!black,fill=black!10,opacity=0.6] (C) -- (B) -- (F) -- (G) -- cycle;
\draw[black!40!black,fill=black!10,opacity=0.8] (A) -- (B) -- (F) -- (E) -- cycle;

\draw[black!60] (0+\xd+1.2,0.5\Width+\ye,0.25\Height+\zd) -- (\Depth+\xd+1.2,0.5\Width+\ye,0.25\Height+\zd);
\draw (\xd+2.2,\yy-1,\zz) node {\color{black}$\boldsymbol U^{(2)}_t \in\mathbb{R}^{n_2 \times d}$};
\draw (\xd+2.6,\yy-0.4,\zz) node {\color{black!50!black}$\boldsymbol{u}^{(2)}_{i_2 t}$};

\coordinate (O) at (0+\xd+1.2,0+\ye+1.5,0+\zd);
\coordinate (A) at (0+\xd+1.2,0.25\Width+\ye+1.5,0+\zd);
\coordinate (B) at (0+\xd+1.2,0.25\Width+\ye+1.5,\Height+\zd);
\coordinate (C) at (0+\xd+1.2,0+\ye+1.5,\Height+\zd);
\coordinate (D) at (0.5\Depth+\xd+1.2,0+\ye+1.5,0+\zd);
\coordinate (E) at (0.5\Depth+\xd+1.2,0.25\Width+\ye+1.5,0+\zd);
\coordinate (F) at (0.5\Depth+\xd+1.2,0.25\Width+\ye+1.5,\Height+\zd);
\coordinate (G) at (0.5\Depth+\xd+1.2,0+\ye+1.5,\Height+\zd);
\draw[black!60,fill=black!15] (O) -- (C) -- (G) -- (D) -- cycle;
\draw[black!60,fill=black!15] (O) -- (A) -- (E) -- (D) -- cycle;
\draw[black!60,fill=black!15] (O) -- (A) -- (B) -- (C) -- cycle;
\draw[black!60,fill=black!15,opacity=0.8] (D) -- (E) -- (F) -- (G) -- cycle;
\draw[black!60,fill=black!15,opacity=0.6] (C) -- (B) -- (F) -- (G) -- cycle;
\draw[black!60,fill=black!15,opacity=0.8] (A) -- (B) -- (F) -- (E) -- cycle;

\coordinate (O) at (0+\xd+0.8,0+\yy+1.2,0+\zd);
\coordinate (A) at (0+\xd+0.8,0.25\Width+\yy+1.2,0+\zd);
\coordinate (B) at (0+\xd+0.8,0.25\Width+\yy+1.2,0.25\Height+\zd);
\coordinate (C) at (0+\xd+0.8,0+\yy+1.2,0.25\Height+\zd);
\coordinate (D) at (0.5\Depth+\xd+0.8,0+\yy+1.2,0+\zd);
\coordinate (E) at (0.5\Depth+\xd+0.8,0.25\Width+\yy+1.2,0+\zd);
\coordinate (F) at (0.5\Depth+\xd+0.8,0.25\Width+\yy+1.2,0.25\Height+\zd);
\coordinate (G) at (0.5\Depth+\xd+0.8,0+\yy+1.2,0.25\Height+\zd);
\draw[black!80!black,fill=black!50] (O) -- (C) -- (G) -- (D) -- cycle;
\draw[black!80!black,fill=black!50] (O) -- (A) -- (E) -- (D) -- cycle;
\draw[black!80!black,fill=black!50] (O) -- (A) -- (B) -- (C) -- cycle;
\draw[black!40!black,fill=black!50,opacity=0.8] (D) -- (E) -- (F) -- (G) -- cycle;
\draw[black!40!black,fill=black!50,opacity=0.6] (C) -- (B) -- (F) -- (G) -- cycle;
\draw[black!40!black,fill=black!50,opacity=0.8] (A) -- (B) -- (F) -- (E) -- cycle;

\draw[black!60] (0.5\Depth+\xd+1.2,0.25\Width+\ye+1.5,0+\zd) -- (0.5\Depth+\xd+1.2,0.25\Width+\ye+1.5,\Height+\zd);
\draw (\xd+2.6,\yy+1.3,\zz) node {\color{black}$\boldsymbol U^{(3)}_{t+1}\in\mathbb{R}^{n_3\times d}$};
\draw (\xd,\yy+1.4,\zz) node {\color{black!50!black}$\boldsymbol{u}^{(3)}_{i_3 (t+1)}$};

\end{tikzpicture}

%% file: resubmit.bbl
\begin{thebibliography}{}

\bibitem[Arias-Castro et~al., 2020]{arias2020perturbation}
Arias-Castro, E., Javanmard, A., and Pelletier, B. (2020).
\newblock Perturbation bounds for procrustes, classical scaling, and
  trilateration, with applications to manifold learning.
\newblock {\em The Journal of Machine Learning Research}, 21:15--1.

\bibitem[A{\ss}mann et~al., 2016]{assmann2016bayesian}
A{\ss}mann, C., Boysen-Hogrefe, J., and Pape, M. (2016).
\newblock Bayesian analysis of static and dynamic factor models: An ex-post
  approach towards the rotation problem.
\newblock {\em Journal of Econometrics}, 192(1):190--206.

\bibitem[Bai and Yin, 2008]{bai2008limit}
Bai, Z.-D. and Yin, Y.-Q. (2008).
\newblock Limit of the smallest eigenvalue of a large dimensional sample
  covariance matrix.
\newblock In {\em Advances In Statistics}, pages 108--127. World Scientific.

\bibitem[Bashir et~al., 2019]{bashir2019post}
Bashir, A., Carvalho, C.~M., Hahn, P.~R., and Jones, M.~B. (2019).
\newblock Post-processing posteriors over precision matrices to produce sparse
  graph estimates.
\newblock {\em Bayesian Analysis}, 14(4):1075--1090.

\bibitem[Bhattacharya and Dunson, 2011]{bhattacharya2011sparse}
Bhattacharya, A. and Dunson, D.~B. (2011).
\newblock Sparse {Bayesian} infinite factor models.
\newblock {\em Biometrika}, pages 291--306.

\bibitem[Bhattacharya et~al., 2019]{bhattacharya2019bayesian}
Bhattacharya, A., Pati, D., and Yang, Y. (2019).
\newblock Bayesian fractional posteriors.
\newblock {\em The Annals of Statistics}, 47(1):39--66.

\bibitem[Bishop and Nasrabadi, 2006]{bishop2006pattern}
Bishop, C.~M. and Nasrabadi, N.~M. (2006).
\newblock {\em Pattern recognition and machine learning}, volume~4.
\newblock Springer.

\bibitem[Carvalho et~al., 2009]{carvalho2009handling}
Carvalho, C.~M., Polson, N.~G., and Scott, J.~G. (2009).
\newblock Handling sparsity via the horseshoe.
\newblock In {\em Artificial Intelligence and Statistics}, pages 73--80. PMLR.

\bibitem[Chakraborty et~al., 2020]{chakraborty2020bayesian}
Chakraborty, A., Bhattacharya, A., and Mallick, B.~K. (2020).
\newblock Bayesian sparse multiple regression for simultaneous rank reduction
  and variable selection.
\newblock {\em Biometrika}, 107(1):205--221.

\bibitem[Chan et~al., 2012]{chan2012time}
Chan, J.~C., Koop, G., Leon-Gonzalez, R., and Strachan, R.~W. (2012).
\newblock Time varying dimension models.
\newblock {\em Journal of Business \& Economic Statistics}, 30(3):358--367.

\bibitem[Chatterjee, 2015]{chatterjee2015matrix}
Chatterjee, S. (2015).
\newblock Matrix estimation by universal singular value thresholding.
\newblock {\em The Annals of Statistics}, 43(1):177--214.

\bibitem[Donoho et~al., 2020]{donoho2020screenot}
Donoho, D.~L., Gavish, M., and Romanov, E. (2020).
\newblock {ScreeNOT}: Exact {MSE}-optimal singular value thresholding in
  correlated noise.
\newblock {\em arXiv preprint arXiv:2009.12297}.

\bibitem[Doukopoulos and Moustakides, 2008]{doukopoulos2008fast}
Doukopoulos, X.~G. and Moustakides, G.~V. (2008).
\newblock Fast and stable subspace tracking.
\newblock {\em IEEE Transactions on Signal Processing}, 56(4):1452--1465.

\bibitem[Durante and Dunson, 2014]{durante2014nonparametric}
Durante, D. and Dunson, D.~B. (2014).
\newblock Nonparametric {Bayes} dynamic modelling of relational data.
\newblock {\em Biometrika}, 101(4):883--898.

\bibitem[Fan and Guan, 2018]{fan2018approximate}
Fan, Z. and Guan, L. (2018).
\newblock Approximate $\ell_0$-penalized estimation of piecewise-constant
  signals on graphs.
\newblock {\em Annals of Statistics}, 46(6B):3217--3245.

\bibitem[Friel et~al., 2016]{friel2016interlocking}
Friel, N., Rastelli, R., Wyse, J., and Raftery, A.~E. (2016).
\newblock Interlocking directorates in {Irish} companies using a latent space
  model for bipartite networks.
\newblock {\em Proceedings of the National Academy of Sciences},
  113(24):6629--6634.

\bibitem[Fr{\"u}hwirth-Schnatter, 2023]{fruhwirth2023generalized}
Fr{\"u}hwirth-Schnatter, S. (2023).
\newblock Generalized cumulative shrinkage process priors with applications to
  sparse {Bayesian} factor analysis.
\newblock {\em Philosophical Transactions of the Royal Society A},
  381(2247):20220148.

\bibitem[Fr{\"u}hwirth-Schnatter and Wagner, 2010]{fruhwirth2010stochastic}
Fr{\"u}hwirth-Schnatter, S. and Wagner, H. (2010).
\newblock Stochastic model specification search for {Gaussian} and partial
  {non-Gaussian} state space models.
\newblock {\em Journal of Econometrics}, 154(1):85--100.

\bibitem[Gavish and Donoho, 2014]{gavish2014optimal}
Gavish, M. and Donoho, D.~L. (2014).
\newblock The optimal hard threshold for singular values is $4/\sqrt{3}$.
\newblock {\em IEEE Transactions on Information Theory}, 60(8):5040--5053.

\bibitem[Gibler, 2008]{gibler2008international}
Gibler, D.~M. (2008).
\newblock {\em International military alliances, 1648-2008}.
\newblock CQ Press.

\bibitem[Gil et~al., 2013]{gil2013renyi}
Gil, M., Alajaji, F., and Linder, T. (2013).
\newblock R{\'e}nyi divergence measures for commonly used univariate continuous
  distributions.
\newblock {\em Information Sciences}, 249:124--131.

\bibitem[Gower and Dijksterhuis, 2004]{gower2004procrustes}
Gower, J.~C. and Dijksterhuis, G.~B. (2004).
\newblock {\em Procrustes problems}, volume~30.
\newblock OUP Oxford.

\bibitem[Hahn and Carvalho, 2015]{hahn2015decoupling}
Hahn, P.~R. and Carvalho, C.~M. (2015).
\newblock Decoupling shrinkage and selection in {Bayesian} linear models: a
  posterior summary perspective.
\newblock {\em Journal of the American Statistical Association},
  110(509):435--448.

\bibitem[Hegde et~al., 2008]{hegde2008dynamic}
Hegde, S.~R., Manimaran, P., and Mande, S.~C. (2008).
\newblock Dynamic changes in protein functional linkage networks revealed by
  integration with gene expression data.
\newblock {\em PLoS computational biology}, 4(11):e1000237.

\bibitem[Hewapathirana et~al., 2020]{hewapathirana2020change}
Hewapathirana, I.~U., Lee, D., Moltchanova, E., and McLeod, J. (2020).
\newblock Change detection in noisy dynamic networks: a spectral embedding
  approach.
\newblock {\em Social Network Analysis and Mining}, 10:1--22.

\bibitem[Hoff, 2007]{hoff2007model}
Hoff, P.~D. (2007).
\newblock Model averaging and dimension selection for the singular value
  decomposition.
\newblock {\em Journal of the American Statistical Association},
  102(478):674--685.

\bibitem[Hoff, 2011]{hoff2011hierarchical}
Hoff, P.~D. (2011).
\newblock Hierarchical multilinear models for multiway data.
\newblock {\em Computational Statistics \& Data Analysis}, 55(1):530--543.

\bibitem[Hoff, 2015]{hoff2015multilinear}
Hoff, P.~D. (2015).
\newblock Multilinear tensor regression for longitudinal relational data.
\newblock {\em The Annals of Applied Statistics}, 9(3):1169.

\bibitem[Hoff et~al., 2002]{hoff2002latent}
Hoff, P.~D., Raftery, A.~E., and Handcock, M.~S. (2002).
\newblock Latent space approaches to social network analysis.
\newblock {\em Journal of the American Statistical Association},
  97(460):1090--1098.

\bibitem[Huang et~al., 2016]{huang2016variational}
Huang, X., Wang, J., and Liang, F. (2016).
\newblock A variational algorithm for {Bayesian} variable selection.
\newblock {\em arXiv preprint arXiv:1602.07640}.

\bibitem[Jaakkola and Jordan, 2000]{jaakkola2000bayesian}
Jaakkola, T.~S. and Jordan, M.~I. (2000).
\newblock {Bayesian} parameter estimation via variational methods.
\newblock {\em Statistics and Computing}, 10(1):25--37.

\bibitem[Kalli and Griffin, 2014]{kalli2014time}
Kalli, M. and Griffin, J.~E. (2014).
\newblock Time-varying sparsity in dynamic regression models.
\newblock {\em Journal of Econometrics}, 178(2):779--793.

\bibitem[Kolda and Bader, 2009]{kolda2009tensor}
Kolda, T.~G. and Bader, B.~W. (2009).
\newblock Tensor decompositions and applications.
\newblock {\em SIAM review}, 51(3):455--500.

\bibitem[Kowal et~al., 2019]{kowal2019dynamic}
Kowal, D.~R., Matteson, D.~S., and Ruppert, D. (2019).
\newblock Dynamic shrinkage processes.
\newblock {\em Journal of the Royal Statistical Society: Series B (Statistical
  Methodology)}, 81(4):781--804.

\bibitem[Lee et~al., 2004]{lee2004voters}
Lee, D.~S., Moretti, E., and Butler, M.~J. (2004).
\newblock Do voters affect or elect policies? evidence from the {US} house.
\newblock {\em The Quarterly Journal of Economics}, 119(3):807--859.

\bibitem[Lee and Lee, 2021]{lee2021post}
Lee, K. and Lee, J. (2021).
\newblock Post-processed posteriors for sparse covariances and its application
  to global minimum variance portfolio.
\newblock {\em arXiv preprint arXiv:2108.09462}.

\bibitem[Legramanti et~al., 2020]{legramanti2020bayesian}
Legramanti, S., Durante, D., and Dunson, D.~B. (2020).
\newblock Bayesian cumulative shrinkage for infinite factorizations.
\newblock {\em Biometrika}, 107(3):745--752.

\bibitem[Lei and Rinaldo, 2015]{lei2015consistency}
Lei, J. and Rinaldo, A. (2015).
\newblock Consistency of spectral clustering in stochastic block models.
\newblock {\em The Annals of Statistics}, 43(1):215--237.

\bibitem[Loyal, 2024]{loyal2024fast}
Loyal, J.~D. (2024).
\newblock Fast variational inference of latent space models for dynamic
  networks using bayesian p-splines.
\newblock {\em arXiv preprint arXiv:2401.09715}.

\bibitem[Loyal and Chen, 2021]{loyal2021eigenmodel}
Loyal, J.~D. and Chen, Y. (2021).
\newblock An eigenmodel for dynamic multilayer networks.
\newblock {\em arXiv preprint arXiv:2103.12831}.

\bibitem[L{\"u}tkepohl, 2013]{lutkepohl2013vector}
L{\"u}tkepohl, H. (2013).
\newblock Vector autoregressive models.
\newblock In {\em Handbook of Research Methods and Applications in Empirical
  Macroeconomics}, pages 139--164. Edward Elgar Publishing.

\bibitem[Ma et~al., 2020]{ma2020universal}
Ma, Z., Ma, Z., and Yuan, H. (2020).
\newblock Universal latent space model fitting for large networks with edge
  covariates.
\newblock {\em The Journal of Machine Learning Research}, 21(4):1--67.

\bibitem[Massart, 2007]{massart2007concentration}
Massart, P. (2007).
\newblock {\em Concentration inequalities and model selection}, volume~6.
\newblock Springer.

\bibitem[Matias and Miele, 2017]{matias2017statistical}
Matias, C. and Miele, V. (2017).
\newblock Statistical clustering of temporal networks through a dynamic
  stochastic block model.
\newblock {\em Journal of the Royal Statistical Society Series B},
  79(4):1119--1141.

\bibitem[Nakajima and West, 2013]{nakajima2013bayesian}
Nakajima, J. and West, M. (2013).
\newblock Bayesian analysis of latent threshold dynamic models.
\newblock {\em Journal of Business \& Economic Statistics}, 31(2):151--164.

\bibitem[Neville et~al., 2014]{neville2014mean}
Neville, S.~E., Ormerod, J.~T., and Wand, M. (2014).
\newblock Mean field variational {Bayes} for continuous sparse signal
  shrinkage: pitfalls and remedies.
\newblock {\em Electronic Journal of Statistics}, 8(1):1113--1151.

\bibitem[Ng et~al., 2001]{ng2001spectral}
Ng, A., Jordan, M., and Weiss, Y. (2001).
\newblock On spectral clustering: Analysis and an algorithm.
\newblock {\em Advances in Neural Information Processing Systems}, 14.

\bibitem[Padilla et~al., 2022]{padilla2022change}
Padilla, O. H.~M., Yu, Y., and Priebe, C.~E. (2022).
\newblock Change point localization in dependent dynamic nonparametric random
  dot product graphs.
\newblock {\em The Journal of Machine Learning Research}, 23(1):10661--10719.

\bibitem[Padilla et~al., 2021]{padilla2021optimal}
Padilla, O. H.~M., Yu, Y., Wang, D., and Rinaldo, A. (2021).
\newblock Optimal nonparametric multivariate change point detection and
  localization.
\newblock {\em IEEE Transactions on Information Theory}, 68(3):1922--1944.

\bibitem[Papastamoulis and Ntzoufras, 2022]{papastamoulis2022identifiability}
Papastamoulis, P. and Ntzoufras, I. (2022).
\newblock On the identifiability of {Bayesian} factor analytic models.
\newblock {\em Statistics and Computing}, 32(2):1--29.

\bibitem[Park and Sohn, 2020]{park2020detecting}
Park, J.~H. and Sohn, Y. (2020).
\newblock Detecting structural changes in longitudinal network data.
\newblock {\em Bayesian Analysis}, 15(1):133--157.

\bibitem[Ray and Szab{\'o}, 2021]{ray2021variational}
Ray, K. and Szab{\'o}, B. (2021).
\newblock Variational {Bayes} for high-dimensional linear regression with
  sparse priors.
\newblock {\em Journal of the American Statistical Association}, pages 1--12.

\bibitem[Sarkar and Moore, 2005]{sarkar2005dynamic}
Sarkar, P. and Moore, A.~W. (2005).
\newblock Dynamic social network analysis using latent space models.
\newblock {\em Acm Sigkdd Explorations Newsletter}, 7(2):31--40.

\bibitem[Schiavon et~al., 2022]{schiavon2022generalized}
Schiavon, L., Canale, A., and Dunson, D.~B. (2022).
\newblock Generalized infinite factorization models.
\newblock {\em Biometrika}, 109(3):817--835.

\bibitem[Sewell and Chen, 2015]{sewell2015latent}
Sewell, D.~K. and Chen, Y. (2015).
\newblock Latent space models for dynamic networks.
\newblock {\em Journal of the American Statistical Association},
  110(512):1646--1657.

\bibitem[Sewell and Chen, 2017]{sewell2017latent}
Sewell, D.~K. and Chen, Y. (2017).
\newblock Latent space approaches to community detection in dynamic networks.
\newblock {\em Bayesian analysis}, 12(2):351--377.

\bibitem[Stock and Watson, 2011]{stock2011dynamic}
Stock, J.~H. and Watson, M. (2011).
\newblock Dynamic factor models.
\newblock {\em Oxford Handbooks Online}.

\bibitem[Stock and Watson, 2016]{stock2016dynamic}
Stock, J.~H. and Watson, M.~W. (2016).
\newblock Dynamic factor models, factor-augmented vector autoregressions, and
  structural vector autoregressions in macroeconomics.
\newblock In {\em Handbook of macroeconomics}, volume~2, pages 415--525.
  Elsevier.

\bibitem[Sun et~al., 2014]{sun2014collaborative}
Sun, J.~Z., Parthasarathy, D., and Varshney, K.~R. (2014).
\newblock Collaborative {Kalman} filtering for dynamic matrix factorization.
\newblock {\em IEEE Transactions on Signal Processing}, 62(14):3499--3509.

\bibitem[Tan et~al., 2016]{tan2016short}
Tan, H., Wu, Y., Shen, B., Jin, P.~J., and Ran, B. (2016).
\newblock Short-term traffic prediction based on dynamic tensor completion.
\newblock {\em IEEE Transactions on Intelligent Transportation Systems},
  17(8):2123--2133.

\bibitem[Tsybakov, 2008]{tsybakov2008introduction}
Tsybakov, A.~B. (2008).
\newblock {\em Introduction to nonparametric estimation}.
\newblock Springer Science \& Business Media.

\bibitem[{United States of America}, 1933]{montevideo1933}
{United States of America} (1933).
\newblock United states of america - convention on rights and duties of states
  adopted by the seventh international conference of american states.
\newblock Signed at Montevideo, December 26th, 1933 [1936] LNTSer 9; 165 LNTS
  19.

\bibitem[Vershynin, 2010]{vershynin2010introduction}
Vershynin, R. (2010).
\newblock Introduction to the non-asymptotic analysis of random matrices.
\newblock {\em arXiv preprint arXiv:1011.3027}.

\bibitem[Von~Luxburg, 2007]{von2007tutorial}
Von~Luxburg, U. (2007).
\newblock A tutorial on spectral clustering.
\newblock {\em Statistics and computing}, 17(4):395--416.

\bibitem[Walker and Hjort, 2001]{walker2001bayesian}
Walker, S. and Hjort, N.~L. (2001).
\newblock On {Bayesian} consistency.
\newblock {\em Journal of the Royal Statistical Society: Series B (Statistical
  Methodology)}, 63(4):811--821.

\bibitem[Wang et~al., 2021]{wang2021optimal}
Wang, D., Yu, Y., and Rinaldo, A. (2021).
\newblock Optimal change point detection and localization in sparse dynamic
  networks.
\newblock {\em The Annals of Statistics}, 49(1):203--232.

\bibitem[Xu et~al., 2022]{xuchangepoints2022}
Xu, H., Padilla, O., Wang, D., and Li, M. (2022).
\newblock {\em {changepoints:} A Collection of Change-Point Detection Methods}.
\newblock R package version 1.1.0.

\bibitem[Zhang et~al., 2022a]{zhang2022directed}
Zhang, J., He, X., and Wang, J. (2022a).
\newblock Directed community detection with network embedding.
\newblock {\em Journal of the American Statistical Association},
  117(540):1809--1819.

\bibitem[Zhang et~al., 2018]{zhang2018local}
Zhang, R., Lin, C.~D., and Ranjan, P. (2018).
\newblock Local {Gaussian} process model for large-scale dynamic computer
  experiments.
\newblock {\em Journal of Computational and Graphical Statistics},
  27(4):798--807.

\bibitem[Zhang et~al., 2022b]{zhang2022joint}
Zhang, X., Xu, G., and Zhu, J. (2022b).
\newblock Joint latent space models for network data with high-dimensional node
  variables.
\newblock {\em Biometrika}, 109(3):707--720.

\bibitem[Zhang et~al., 2021]{zhang2021dynamic}
Zhang, Y., Bi, X., Tang, N., and Qu, A. (2021).
\newblock Dynamic tensor recommender systems.
\newblock {\em The Journal of Machine Learning Research}, 22(65):1--35.

\bibitem[Zhao et~al., 2022]{zhao2022structured}
Zhao, P., Bhattacharya, A., Pati, D., and Mallick, B.~K. (2022).
\newblock Structured optimal variational inference for dynamic latent space
  models.
\newblock {\em arXiv preprint arXiv:2209.15117}.

\bibitem[Zhou et~al., 2013]{zhou2013tensor}
Zhou, H., Li, L., and Zhu, H. (2013).
\newblock Tensor regression with applications in neuroimaging data analysis.
\newblock {\em Journal of the American Statistical Association},
  108(502):540--552.

\bibitem[Zhu et~al., 2016]{zhu2016scalable}
Zhu, L., Guo, D., Yin, J., Ver~Steeg, G., and Galstyan, A. (2016).
\newblock Scalable temporal latent space inference for link prediction in
  dynamic social networks.
\newblock {\em IEEE Transactions on Knowledge and Data Engineering},
  28(10):2765--2777.

\bibitem[Zou and Li, 2017]{zou2017modeling}
Zou, N. and Li, J. (2017).
\newblock Modeling and change detection of dynamic network data by a network
  state space model.
\newblock {\em IISE Transactions}, 49(1):45--57.

\end{thebibliography}


\begin{thebibliography}{}

\bibitem[Arias-Castro et~al., 2020]{arias2020perturbation}
Arias-Castro, E., Javanmard, A., and Pelletier, B. (2020).
\newblock Perturbation bounds for procrustes, classical scaling, and
  trilateration, with applications to manifold learning.
\newblock {\em The Journal of Machine Learning Research}, 21:15--1.

\bibitem[Bhattacharya et~al., 2019]{bhattacharya2019bayesian}
Bhattacharya, A., Pati, D., and Yang, Y. (2019).
\newblock Bayesian fractional posteriors.
\newblock {\em The Annals of Statistics}, 47(1):39--66.

\bibitem[Hoff, 2011]{hoff2011hierarchical}
Hoff, P.~D. (2011).
\newblock Hierarchical multilinear models for multiway data.
\newblock {\em Computational Statistics \& Data Analysis}, 55(1):530--543.

\bibitem[Jaakkola and Jordan, 2000]{jaakkola2000bayesian}
Jaakkola, T.~S. and Jordan, M.~I. (2000).
\newblock {Bayesian} parameter estimation via variational methods.
\newblock {\em Statistics and Computing}, 10(1):25--37.

\bibitem[Kolda and Bader, 2009]{kolda2009tensor}
Kolda, T.~G. and Bader, B.~W. (2009).
\newblock Tensor decompositions and applications.
\newblock {\em SIAM review}, 51(3):455--500.

\bibitem[Lei and Rinaldo, 2015]{lei2015consistency}
Lei, J. and Rinaldo, A. (2015).
\newblock Consistency of spectral clustering in stochastic block models.
\newblock {\em The Annals of Statistics}, 43(1):215--237.

\bibitem[Massart, 2007]{massart2007concentration}
Massart, P. (2007).
\newblock {\em Concentration inequalities and model selection}, volume~6.
\newblock Springer.

\bibitem[Tsybakov, 2008]{tsybakov2008introduction}
Tsybakov, A.~B. (2008).
\newblock {\em Introduction to nonparametric estimation}.
\newblock Springer Science \& Business Media.

\bibitem[Zhou et~al., 2013]{zhou2013tensor}
Zhou, H., Li, L., and Zhu, H. (2013).
\newblock Tensor regression with applications in neuroimaging data analysis.
\newblock {\em Journal of the American Statistical Association},
  108(502):540--552.

\end{thebibliography}
